%% file: medima-main.tex
\documentclass[times,twocolumn,final]{elsarticle}

\usepackage{medima}
\usepackage{framed,multirow}
\usepackage{booktabs}
\usepackage{amssymb}
\usepackage{latexsym}
\usepackage{array, makecell}
\usepackage{bm}
\usepackage{xspace}
\usepackage{nicematrix}
\usepackage{subfigure}
\usepackage[T1]{fontenc}
\usepackage{graphicx}
\usepackage{multirow}
\usepackage{tabularx}
\usepackage{rotating}
\usepackage{url}
\usepackage{xcolor}
\usepackage[colorlinks = true,
            linkcolor = tealblue,
            urlcolor  = tealblue,
            citecolor = tealblue,
            anchorcolor = tealblue]{hyperref}

\DeclareMathOperator*{\argmin}{arg\,min\xspace}

\definecolor{newcolor1}{rgb}{.0, .502, .675}
\definecolor{candyAppleRed}{RGB}{255 8 0}

\newif\ifcb
\usepackage{marginnote}
\cbtrue
\makeatletter
\AtBeginDocument{\def\@citecolor{newcolor1}}
\AtBeginDocument{\def\@linkcolor{newcolor1}}
\AtBeginDocument{\def\@anchorcolor{newcolor1}}
\AtBeginDocument{\def\@filecolor{newcolor1}}
\AtBeginDocument{\def\@urlcolor{newcolor1}}
\AtBeginDocument{\def\@menucolor{newcolor1}}
\AtBeginDocument{\def\@pagecolor{newcolor1}}
\makeatother
\newcommand\sbullet[1][.5]{\mathbin{\vcenter{\hbox{\scalebox{#1}{$\bullet$}}}}}
\newcolumntype{?}{!{\vrule width 1pt}}
\usepackage{color, colortbl}
\definecolor{Gray}{gray}{0.9}
\definecolor{newcolor}{rgb}{.8,.349,.1}

\journal{}

\begin{document}

\verso{Chen, Liu, Wei \textit{et~al.}}

\begin{frontmatter}

\title{A survey on deep learning in medical image registration: new technologies, uncertainty, evaluation metrics, and beyond}

\author[1]{Junyu \snm{Chen}\corref{cor1}\fnref{fn1}}

\author[2]{Yihao \snm{Liu}\fnref{fn1}}
\author[2]{Shuwen \snm{Wei}\fnref{fn1}}
\fntext[fn1]{Contributed equally to this work.}
\author[2]{Zhangxing \snm{Bian}}
\author[1]{Shalini \snm{Subramanian}}
\author[2]{Aaron \snm{Carass}}
\author[2]{Jerry L. \snm{Prince}}
\author[1]{Yong \snm{Du}}
\address[1]{Department of Radiology and Radiological Science, Johns Hopkins School of Medicine, MD, USA}
\address[2]{Department of Electrical and Computer Engineering, Johns Hopkins University, MD, USA}
\cortext[cor1]{Corresponding author. E-mail address: 
 jchen245@jhmi.edu.}
\received{xxxx}
\finalform{xxxx}
\accepted{xxxx}
\availableonline{xxxx}
\communicated{xxxx}

\begin{abstract}
Deep learning technologies have dramatically reshaped the field of medical image registration over the past decade.
The initial developments, such as {regression-based and U-Net-based networks}, established the foundation for deep learning in image registration.
Subsequent progress has been made in various aspects of deep learning-based registration, including similarity measures, deformation regularizations, {network architectures}, and uncertainty estimation.
These advancements have not only enriched the field of image registration but have also facilitated its application in a wide range of tasks, including atlas construction, multi-atlas segmentation, motion estimation, and 2D-3D registration.
In this paper, we present a comprehensive overview of the most recent advancements in deep learning-based image registration.
We begin with a concise introduction to the core concepts of deep learning-based image registration. Then, we delve into innovative network architectures, loss functions specific to registration, and methods for estimating registration uncertainty. 
Additionally, this paper explores appropriate evaluation metrics for assessing the performance of deep learning models in registration tasks.
Finally, we highlight the practical applications of these novel techniques in medical imaging and discuss the future prospects of deep learning-based image registration.
\end{abstract}

\begin{keyword}
\KWD Image Registration\sep Deep Neural Networks\sep Medical Imaging
\end{keyword}

\end{frontmatter}



\section{Introduction}
\label{sec:introduction}
\input{introduction.tex}

\section{Fundamentals of Learning-based Image Registration}
\label{sec:Fund_Img_Reg}
\input{fundamentals.tex}

\section{Loss Functions}
\label{sec:loss}
\input{loss.tex}
\section{Network Architectures}
\label{sec:net_arch}
\input{network_arch.tex}

\section{Uncertainty in Learning-based Registration}
\label{sec:uncertainty}
\input{uncertainty.tex}

\section{Registration Evaluation Metrics}
\label{sec:Eval_Metric}
\input{evaluation.tex}

\section{Benchmark Datasets for Medical Image Registration}
\label{sec:Datasets}
\input{datasets.tex}

\section{Applications of Medical Image Registration}
\label{sec:Application}

\input{application.tex}

\section{Challenges and Future Perspectives}
\label{sec:future_persp}
\input{future_perspective}

\section{Conclusion}
In this survey, we presented a thorough examination of deep learning for medical image registration. In contrast to existing review papers, which might not fully capture the most recent advancements and tend to be systematic in nature with a limited focus on technical aspects, our comprehensive survey analyzed over 250 papers with an emphasis on the most recent technological advancements. Beginning with a review of the fundamentals of learning-based image registration, our investigation incorporated widely-used and novel loss functions, as well as network architectures for image registration. We also thoroughly investigated the estimation methods of registration uncertainty and appropriate metrics of registration accuracy and regularity. Furthermore, we provided insights into potential clinical applications, future perspectives, and challenges, aiming to guide future research in this rapidly evolving field.

\section*{Acknowledgments}
Junyu Chen and Yong Du were supported by grants from the National Institutes of Health~(NIH), United States, U01-CA140204~(PI: Y.~Du), R01-EB031023~(PI: Y.~Du), and U01-EB031798~(PI: G.~Sgouros).
Yihao Liu, Shuwen Wei, Zhangxing Bian, Aaron Carass, and Jerry L. Prince were supported by the NIH from National Eye Institute grants R01-EY024655~(PI:~J.L.~Prince) and R01-EY032284~(PI:~J.L.~Prince), as well as the National Science Foundation grant 1819326~(Co-PI: S.~Scott, Co-PI: A.~Carass).
This work was also made possible by a 2023 Johns Hopkins Discovery grant~(Co-PI: J.~Chen, Co-PI: A.~Carass).
{The authors thank Lianrui Zuo for valuable discussions and insights.}

The views expressed in written conference materials or publications and by speakers and moderators do not necessarily reflect the official policies of the NIH; nor does mention by trade names, commercial practices, or organizations imply endorsement by the U.S. Government.

\bibliographystyle{model2-names.bst}
\biboptions{authoryear}
\bibliography{references}

\end{document}

%% file: introduction.tex
Medical image registration involves estimating the optimal spatial transformation to align the structures of interest in a pair of fixed and moving images.
The choice of spatial transformation depends on the specific application and can be categorized as either rigid/affine or non-rigid/deformable.
In rigid/affine registration, all spatial coordinates are transformed using the same rigid/affine matrix.
On the other hand, non-rigid/deformable registration employs independent transformations for individual local regions of spatial coordinates.
Both types of registration are of great importance to many medical imaging tasks.
Rigid registration is commonly used when the rigid body assumption holds.
For example, it is used to align a structural scan---\emph{e.g.}, magnetic resonance image~(MRI) or computed tomography~(CT)---with a functional scan---\emph{e.g.}, functional magnetic resonance image~(fMRI) or positron emission tomography~(PET)---of the same patient for attenuation correction~\citep{hofmann2008mri} or interpretation of functional activities~\citep{studholme2000accurate}.
On the other hand, deformable image registration~(DIR) is often used in cases where more complex, spatially varying deformations are needed.
Examples of such applications include constructing deformable templates for a patient
cohort~\hbox{\citep{christensen1996deformable, ganser2004deformable}}
or registering atlases to a patient image for multi-atlas
segmentation~\citep{reed2009automatic, cabezas2011review,
aljabar2009multi}.

\begin{figure*}
\centering
\begin{tabular}{ccc}
\raisebox{1.5em}{\includegraphics[width = 0.46\textwidth]{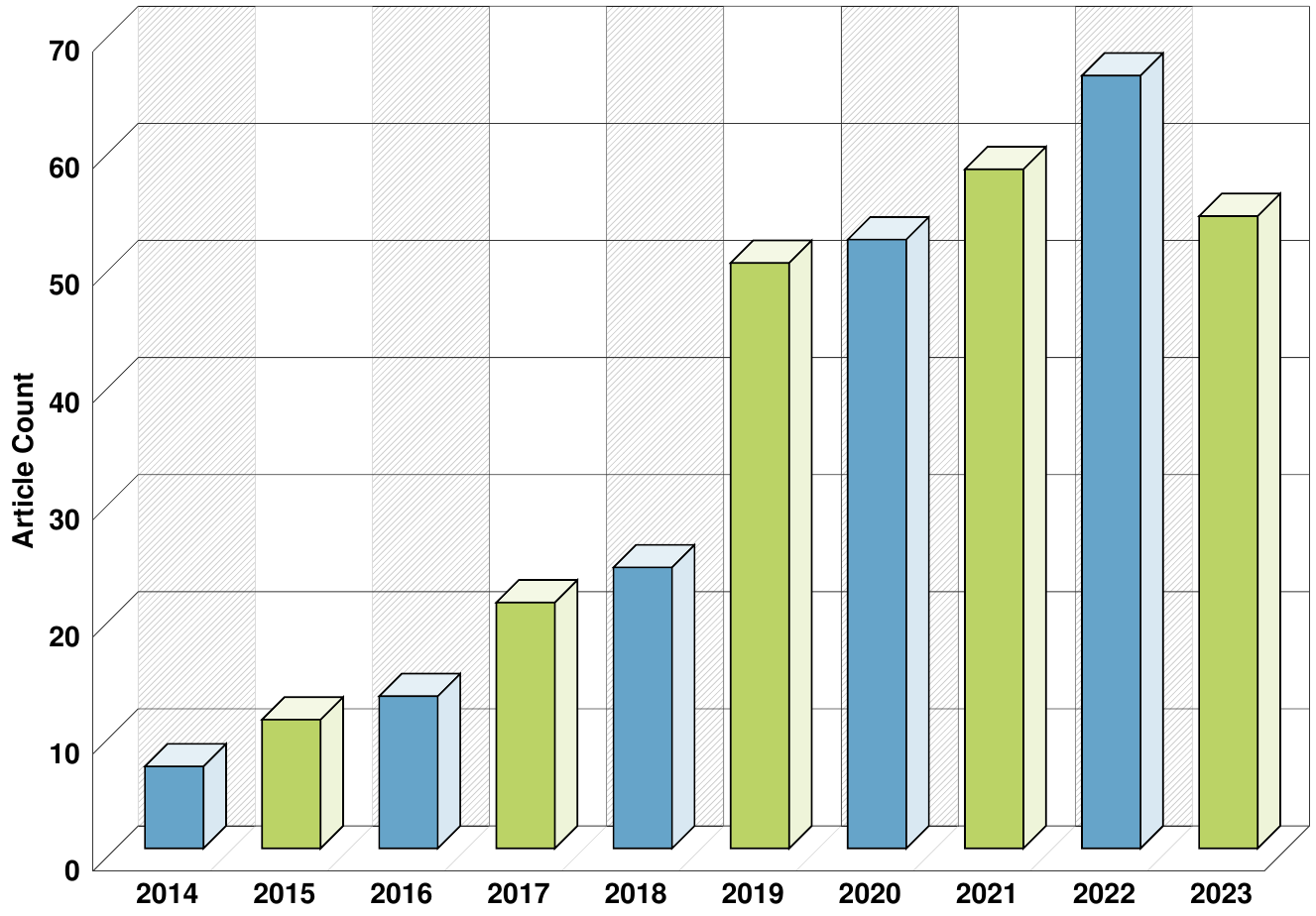}} && \includegraphics[width = 0.46\textwidth]{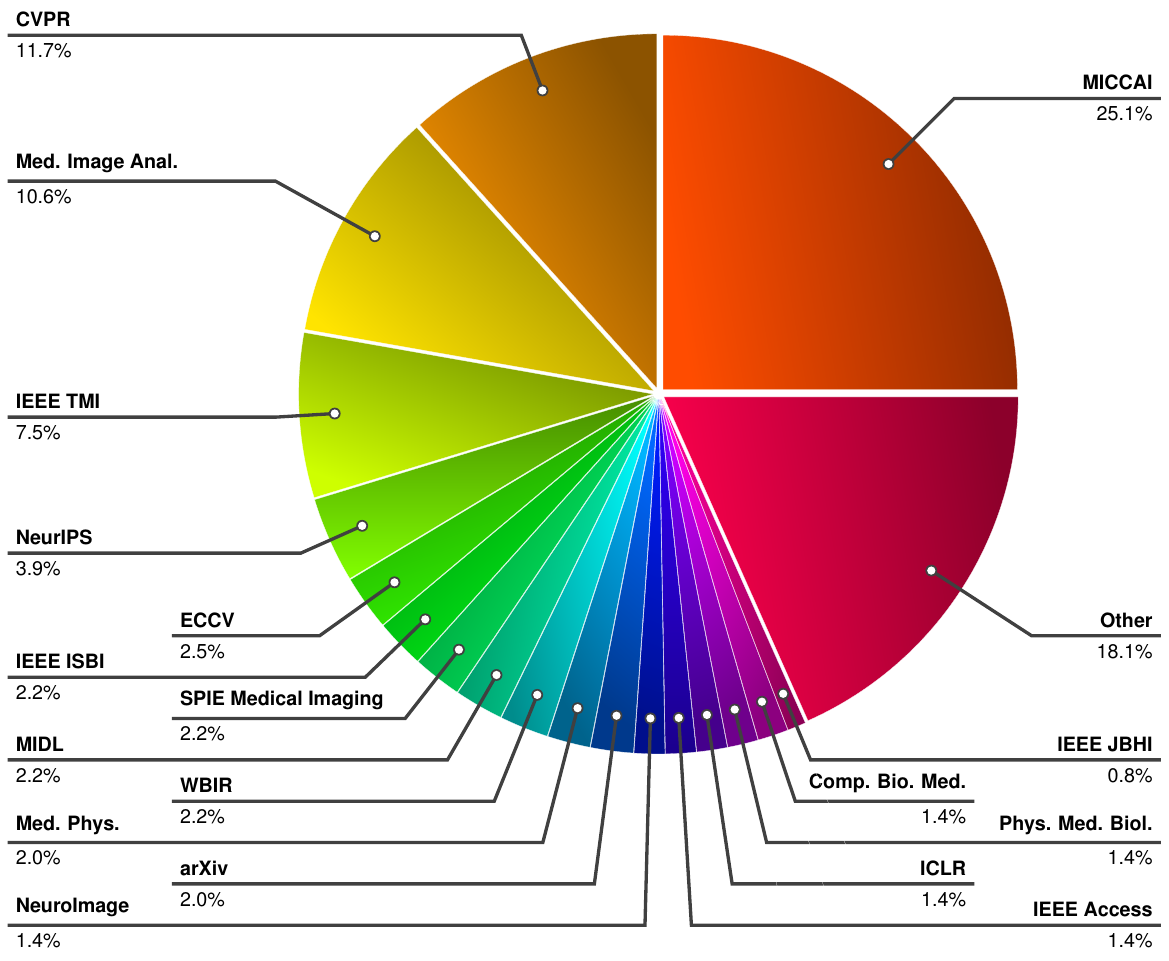}\\
\end{tabular}
\caption{Statistics of the articles investigated in this survey paper.
The left panel displays a histogram of the number of papers by year;
the vast majority of the surveyed papers were proposed within the last
five years. The right panel illustrates the sources of the
investigated articles, demonstrating that our survey draws from
sources associated with the field of medical image analysis.}
\label{fig:paper_count}
\end{figure*}

Traditionally, image registration has been accomplished by iteratively solving an optimization problem (\emph{e.g.}, demons~\citep{vercauteren2009diffeomorphic}, LDDMM~\citep{beg2005computing}, SyN~\citep{avants2008symmetric}, DARTEL~\citep{ashburner2007fast}, and Elastix~\citep{klein2009elastix}.
{These methods are well-established and supported by strong mathematical theory but come with notable limitations. Firstly, they are computationally intensive and tend to be slow, as  they involve solving a distinct optimization problem for every pair of moving and fixed images, leading to redundancy across optimizations. Secondly, the objective function related to transformation parameters (such as displacement fields or control points) typically exhibits non-linearity, creating a non-convex optimization dilemma. To address this, attempts have been made to simplify the problem, for instance, by linearizing the objective function~\citep{sun2013efficient}, convexifying the optimization~\citep{esser2010primal, heinrich2014non, rajchl2016fast}, or adopting discretized registration techniques~\citep{glocker2008dense, glocker2011deformable,
heinrich2012globally, heinrich2014non}. However, these approaches often increase computational costs and might lead to a less realistic problem, given the inherently non-convex nature of image registration~\citep{sotiras2013deformable}.} Several review papers have covered traditional medical image registration methods extensively~\citep{maintz1998survey, hill2001medical, shams2010survey, fluck2011survey, sotiras2013deformable, oliveira2014medical, viergever2016survey}.
Interested readers can refer to these references for more information on these methods. 

In the last decade, deep learning-based methods have shown promise in improving the accuracy and efficiency of image registration.
Unlike traditional methods, deep learning-based methods train a general network by optimizing a global objective function on a
training dataset. Then in the testing phase, the trained network is directly applied to each image pair with fixed network weights without further optimization. 
{Deep learning-based methods offer a threefold advantage:
First, the diversity within the training dataset acts as an implicit regularization, effectively smoothing the loss landscape by averaging out noisy or misleading gradients that may lead to suboptimal local minima.
Secondly, the optimization of non-convex problems is highly dependent on the initial starting point. The use of pretrained weights (i.e., transfer learning), combined with advanced optimization algorithms, enhances deep learning methods' capacity to locate global minima.
Lastly, the capability to process image pairs through a single forward pass during inference, avoiding iterative optimization, results in a significant speed advantage over traditional optimization-based methods.}
{Initially, image registration network architectures were mostly encoder-based, serving either as feature extractors to replace hand-crafted features in optimization-based registration methods~\citep{wu2013unsupervised} or as regressors for estimating transformation parameters from local image patches~\citep{miao2016cnn}.
However, following the success of U-Net~\citep{ronneberger2015u} in medical imaging, learning-based deformable registration methods began to incorporate an encoder-decoder architecture within a supervised learning framework~\citep{yang2017quicksilver, rohe2017svf, sokooti2017nonrigid, uzunova2017training}, often requiring ground truth deformation fields for direct supervision.
Concurrently, the advent of spatial transformer networks~\citep{jaderberg2015spatial} heralded a shift towards unsupervised and end-to-end learning for deformable registration employing the encoder-decoder framework, which has now become the dominant approach~\citep{vos2017end, li2018non,
balakrishnan2019voxelmorph, kim2021cyclemorph, chen2022deformer}.
On the other hand, learning-based rigid/affine registration methods continue to adopt encoder-only networks~\citep{miao2016cnn, hu2018weakly, de2019deep, chen2021learning, chen2022deformer, mok2022affine}, with the output being the rigid or affine parameters.
In the context of supervised learning for these deep learning-based methods, the ground truth is typically a transformation matrix for rigid/affine registration tasks, while a dense displacement field is used for deformable registration tasks.} 
While there are papers that provide general reviews of learning-based registration methods~\citep{fu2020deep, chen2021deep, xiao2021review, zou2022review}, it is important to note that these reviews may not be fully up-to-date due to the rapid advancement of the field of deep learning. Recent advancements, including learning-based similarity metrics and regularizers, novel network architectures, and innovative evaluation metrics and uncertainty estimation methods, have demonstrated promising potential for medical image registration. This paper provides a timely review of learning-based methods in medical image registration, highlighting the latest technologies that have been proposed and discussing their respective characteristics and applications. In addition, we investigate and formally define registration uncertainty for deep learning-based image registration and address the appropriate evaluation metrics for these methods that have been overlooked in previous review papers. For simplicity, we refer to deep learning-based methods as learning-based methods throughout the paper.

In this paper, we surveyed over 250 articles on learning-based
medical image registration. As depicted in Fig.~\ref{fig:paper_count},
the focus is primarily on recent advancements proposed in the last
five years. Our search covers well-established medical imaging
journals, such as Medical Image Analysis, IEEE~Transactions on Medical
Imaging, Medical Physics, and NeuroImage, as well as conference
proceedings related to medical imaging and image registration, such as
MICCAI, IPMI, WBIR, CVPR, ECCV, ICCV, and NeurIPS.
{A comprehensive list of open-sourced code from the papers reviewed in this study has been organized and is available at \url{https://bit.ly/3QgFJ9z}}.
The remainder of
the paper is organized as follows:
Section~\ref{sec:Fund_Img_Reg}~offers a brief overview of the
fundamentals of learning-based image registration.
Section~\ref{sec:loss}~explores widely-used loss functions for
learning-based registration methods which resemble objective
functions in traditional methods, and discusses other novel loss functions
enabled by deep learning. Section~\ref{sec:net_arch}~investigates
network architectures developed for medical image registration,
with a focus on recent developments.  Section~\ref{sec:uncertainty}~delves
into methods for estimating registration uncertainty in learning-based
registration.  Section~\ref{sec:Eval_Metric}~considers appropriate
evaluation metrics for learning-based methods and examines methods for
quantifying the regularity of generated deformation fields.
{Section~\ref{sec:Datasets} provides an enumeration of commonly used public benchmark datasets for medical image registration.}
Section~\ref{sec:Application}~summarizes recent applications of
learning-based registration in medical imaging. Finally,
Section~\ref{sec:future_persp}~discusses current challenges and
provides future perspectives for deep learning in medical image
registration.

%% file: fundamentals.tex
Image registration aims to estimate the optimal coordinate transformation that minimizes an energy function of the form:
\begin{equation}
\label{eqn:energy_func}
    \hat{\phi} = \operatorname*{\argmin}_{\phi}E(I_f, I_m\circ\phi) + \lambda R(\phi),
\end{equation}
where $I_f$ and $I_m$ denote the fixed and moving image, respectively, $\phi$ represents the deformation field that maps $I_m$ to $I_f$, and $R$ is a functional of $\phi$.
The first term in the energy function measures the image similarity between the fixed image and the transformed moving image. The second term enforces regularization on the deformation field, with $\lambda$ being a hyperparameter that determines the trade-off between image similarity and deformation field regularity. 
The purpose of the image similarity measure is to quantify the discrepancy between the fixed image and the transformed moving image. The regularization term is typically used in DIR, as it allows for the integration of prior knowledge about the desired characteristics of the deformation field, such as spatial smoothness. Moreover, regularization prevents the deformation field from exhibiting physically implausible behaviors, such as ``folding'' or rearranging of voxels~\citep{rohlfing2011image}.
This is particularly important for medical images because such unrealistic behavior does not accurately reflect the way that organs deform in reality and may lead to a misinterpretation of the registration results. Regularization is often not required for rigid/affine registration because the deformation field is guaranteed to be spatially uniform.

\begin{figure*}
\centering
\includegraphics[width=0.95\textwidth]{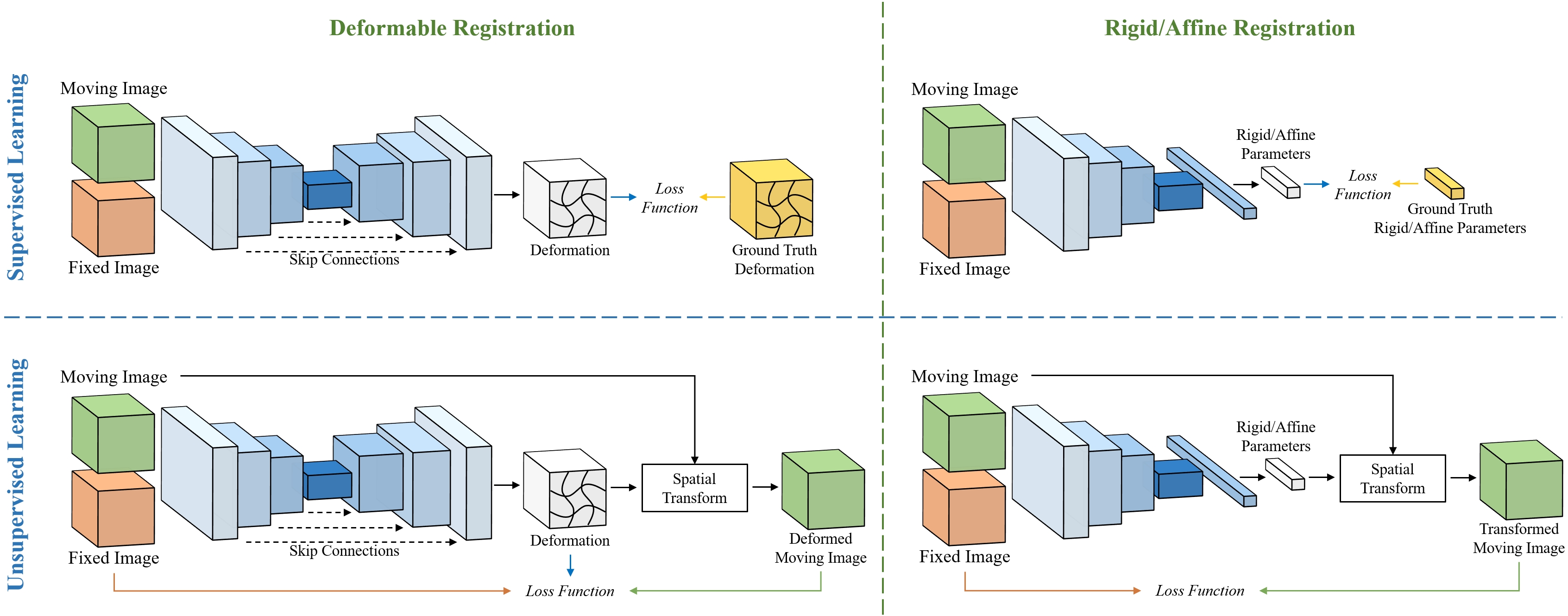}
\caption{Overview of learning-based image registration. The top panel depicts the common pipeline for supervised learning in medical image registration, which necessitates ground truth transformations. The bottom panel demonstrates the unsupervised learning pipeline, wherein the network learns to perform registration using only input images. The left panel presents the learning-based DIR pipeline, typically employing an encoder-decoder-style network architecture. The right panel exhibits the learning-based rigid/affine registration, which usually involves only an encoder.} \label{fig:img_reg}
\end{figure*}

\subsection{Supervised vs. Unsupervised Learning}
Learning-based registration methods can be broadly categorized as supervised and unsupervised.
In the machine learning paradigm, supervised learning typically refers to the use of extrinsic information during learning (such as labels) whereas unsupervised methods are concerned with discovering properties intrinsic to the data.
Both supervised and unsupervised learning-based registration methods require a training stage that uses pairs of inputs and their corresponding target outputs.
Supervised registration methods use ground truth transformations as target output during the training process.
Unsupervised methods refer to those that do not require ground truth transformations.
Yet, methods that employ landmark correspondences or anatomical label maps during their training phase are still categorized under supervised learning. This is because landmark correspondences are a sparse representation of the ground truth transformations, and matching label maps act as a surrogate for evaluating registration performance. When this extrinsic information is used alongside the image data to aim learning, these methods are referred to as semi-supervised.
In certain contexts, the term "unsupervised" might be misleading. A more precise term could be ``self-supervised'' to underscore the training aspect of deep learning. 
However, for the purposes of clarity and consistency in this discussion, we will use conventional terminology and refer to methods that do not require supervision from extrinsic information as unsupervised.

During the early stages of development, the majority of learning-based registration methods were supervised.
The ground truth transformations required for the training process are typically generated using traditional registration methods, such as \citep{yang2017quicksilver, rohe2017svf, cao2017deformable, hu2018weakly, fan2019birnet}.
However, generating ground-truth transformations this way is a time-consuming process, which is a notable drawback of such methods.
In addition, since these networks are trained to mimic the function of traditional methods, their registration performance may not surpass that of the methods they are based on.
In some cases, post-processing of the deformation fields may be required to further improve registration accuracy~\citep{yang2017quicksilver}.
Alternatively, artificial deformations can also be used as ground truth transformations in certain cases~\citep{miao2016cnn, krebs2017robust, sokooti2017nonrigid, eppenhof2018pulmonary, eppenhof2018deformable}.

More recently, the introduction of spatial transformer networks~\citep{jaderberg2015spatial} has led to a shift towards developing unsupervised methods that do not rely on ground-truth transformation~\citep{vos2017end, li2018non, de2019deep, balakrishnan2019voxelmorph, dalca2019unsupervised, mok2020fast, mok2020large, mok2021conditional, kim2021cyclemorph, chen2022transmorph, chen2021vitvnet}. These methods use the difference between the deformed moving image and the fixed image to update the network, enabling end-to-end training. By removing the reliance on ground truth transformation, these methods offer greater flexibility in modeling different properties of the deformation fields (\emph{e.g.}, smoothness, invertibility).

\subsection{Paradigm for Learning-based Registration}
Recent progress in the field of learning-based medical image registration has been focusing on exploring different ways to improve registration accuracy, such as through modifications to network architectures, loss functions, and training methods, which will be discussed in detail in subsequent sections. Despite these efforts, the fundamental principles of learning-based registration have remained unchanged.
Figure~\ref{fig:img_reg} illustrates the conventional paradigms of learning-based rigid/affine and DIR. Typically, these paradigms consist of the following components:
\begin{enumerate}
    \item Moving and fixed images as input
    \item A deep neural network
    \item The spatial transformer~(for unsupervised methods)
    \item A loss function
\end{enumerate}
The way in which moving and fixed images are inputted into deep neural networks~(DNNs) varies depending on the architecture of the network. They can either be concatenated and sent in as a single input (\emph{e.g.}, VoxelMorph~\citep{balakrishnan2019voxelmorph}) or each image can be processed separately by the DNN, with the feature maps being combined in a deeper stage (\emph{e.g.}, Quicksilver~\citep{yang2017quicksilver}). 

The architecture of DNNs can vary depending on the specific task they are designed to perform and the learning method they will undergo. For affine/rigid registration methods, DNN encoders are used for feature extraction and fully connected layers are used to output the parameters of the predicted transformation. DIR methods use DNNs with both an encoder and decoder, and the result is a deformation field of equal sizes to the input images. In the supervised setting, the network output is compared to ground truth transformations (generated from synthetic transformations or traditional image registration methods) or landmark correspondences using a loss function.
In the unsupervised setting, the predicted transformation is used by the spatial transformer~\citep{jaderberg2015spatial} to warp the moving image, and the transformed image is then evaluated against the fixed image using a loss function that incorporates an image similarity measure. 
When anatomical label maps for the fixed and moving images are available, 
the warped moving label map can also be produced by using the predicted transformation and the spatial transformer. An anatomy loss can be computed using the warped moving label map and the fixed label maps to provide extra
guidance during network training.

There is a diverse range of loss functions to choose from, depending on the learning mode. These are thoroughly discussed in Section \ref{sec:loss}.
The networks are trained by globally optimizing the loss function during the training stage using a training dataset.
The trained networks are then applied to unseen testing images for inference.

Due to the self-supervised nature of image registration, the difference between the transformed moving image and the fixed image can be further reduced at test time.
This is commonly known as \textit{instance-specific optimization}~\citep{balakrishnan2019voxelmorph, siebert2022learn, mok2022robust, heinrich2022voxelmorph++, chen2020generating}.
Specifically, the network weights can be optimized during test time to reduce the dissimilarity of each fixed and moving image pair in the test dataset and further boost the performance.
Registration networks can also be specifically designed to produce diffeomorphic transformations, which are highly desirable in DIR methods and will be discussed further in the next subsection.

\subsection{Diffeomorphic Image Registration}
\label{sec:diff_reg}
Many learning-based DIR methods follow a small deformation model~\citep{balakrishnan2019voxelmorph, kim2021cyclemorph, de2019deep, sokooti2017nonrigid, mok2021conditional, heinrich2019closing, hu2018weakly}.
In this model, $\phi$ in Eqn.~\ref{eqn:energy_func} is approximated by a displacement field, {$d$}, expressed as $\phi = id + {d}$, where the displacement is added to the identity transform, $id$.
Since $\phi$ may not be a one-to-one mapping, this model does not guarantee the invertibility of the deformation.
In some cases, the "inverse" transformation is roughly approximated by subtracting the displacement~\citep{ashburner2007fast}.
In many applications (\emph{e.g.}, \citet{avants2008symmetric, oishi2009atlas, christensen1997volumetric}), diffeomorphic image registration is highly desirable because it provides transformation invertibility and topological preservation. 
Diffeomorphic transformations are defined as smooth and continuous one-to-one mappings with a smooth and continuous inverse~(\emph{i.e.}, positive Jacobian determinants).
They are achieved mainly through two approaches: the time-dependent velocity field~\citep{beg2005computing, avants2008symmetric} or the time-stationary velocity field~\citep{arsigny2006log, ashburner2007fast, vercauteren2009diffeomorphic, hernandez2009registration} approach. 

The time-dependent velocity field approach involves integrating sufficiently smooth velocity fields that change over time. The diffeomorphism is established by using a velocity field $v^{(t)}$ at time $t$, and evolving it through~\citep{beg2005computing}:
\begin{equation}
\label{eqn:diff}
    \frac{d\phi^{(t)}}{dt}=v^{(t)}(\phi^{(t)}).
\end{equation}
The diffeomorphic transformation is achieved by starting with an identity transformation, \emph{i.e.} $\phi^{(0)}=id$, and integrating over the unit time period:
\begin{equation}
    \phi^{(1)}=\phi^{(0)}+\int^1_0v^{(t)}(\phi^{(t)})dt.
\end{equation}
{Time-varying models for learning-based registration offer the promise of capturing complex and large deformations while preserving diffeomorphic mappings that ensure biologically realistic transformations. However, such models entail certain challenges. Achieving the time-dependent diffeomorphic mapping can be done by either integrating a sequence of small, sufficiently smooth (first-order differentiable) velocity fields as proposed by the original LDDMM~\citep{beg2005computing}, or by estimating an initial velocity field and subsequently deriving the time-varying velocity fields via geodesic shooting~\citep{miller2006geodesic, zhang2019fast}. Implementing the former within a DNN framework necessitates the network to estimate a series of displacement fields at discrete time points, usually at high resolution, which leads to increased memory requirements. On the other hand, it is numerically difficult to enforce a deformation field to follow a geodesic by geodesic shooting, making end-to-end training challenging without adapting the geodesic shooting method to modern DNN libraries. Such challenges have limited the implementation of time-varying diffeomorphic mapping in current learning-based registration models. Only a handful of studies~\citep{ramon2022lddmm, pathan2018predictive, shen2019region, yang2016fast, yang2017quicksilver, han2021deep, wang2020deepflash, wang2023metamorph, wang2022geo}, have integrated this into a DNN framework. Out of these, only three studies have embedded geodesic shooting into an end-to-end learning framework for medical image registration~\citep{shen2019region, wang2023metamorph, wang2022geo}.}

The time-stationary velocity field approach considers velocity fields that remain constant throughout time.
By using this setting, the evolution of the diffeomorphism in Eqn.~\ref{eqn:diff} can be rewritten as:
\begin{equation}
\frac{d\phi^{(t)}}{dt}=v(\phi^{(t)}),
\end{equation}
where the velocity field, $v$, is now independent of time. 
\citet{dalca2019unsupervised}~were the first to use this setting in a DNN model through the \textit{scaling-and-squaring} method~\citep{arsigny2006log,ashburner2007fast}. 
This method has since become dominant in learning-based diffeomorphic registration models~\citep{mok2020fast, chen2022transmorph, mok2020large, han2023diffeomorphic, zhang2021learning, qiu2021learning, zhao2021s3reg, krebs2019learning}.
The scaling-and-squaring method considers the velocity field as a member of the Lie algebra and the deformation field as a member of the Lie group.
{The relationship between the velocity field and diffeomorphism is governed by the principle that one-parameter subgroups in the Lie group can be equivalently represented via an exponential map~\citep{arsigny2006log,vercauteren2009diffeomorphic,ashburner2007fast}, which is mathematically expressed as:}
\begin{equation}
\phi=\exp{(v)}{=\exp{(2^{-N}v)^{2^N}}},
\end{equation}
which is equivalent to integrating along the velocity field over the unit time period {(in this case, $N$ time steps)}. An alternative perspective is that the Jacobian determinant of a deformation resulting from exponentiating the velocity field is always positive, similar to how the derivative of the exponential of a real number is always positive~\citep{ashburner2007fast}. For further information on the implementation of this method, we direct interested readers to the references cited~\citep{ashburner2007fast, arsigny2006log, dalca2019unsupervised}.
It is important to note that the scaling-and-squaring method cannot guarantee a folding-free transformation in the digital domain when measured by the finite difference approximated Jacobian determinant.
This is because the scaling-and-squaring method involves bilinear or trilinear interpolation that is inconsistent with the piecewise linear transformation assumed by the finite difference based Jacobian determinant computation~\citep{liu2022finite}.

%% file: loss.tex
\begin{table*}[!t]
\caption{A compilation of unsupervised deformable image registration models (models are listed in alphabetical order). The table summarizes the models' choices of similarity and auxiliary loss functions, regularization techniques, accuracy measures, and regularity measures.}
\rowcolors{2}{cyan!10}{white}
\label{table:unsup_list}
\centering
\fontsize{6}{7.5}\selectfont 
 \begin{tabular}{ l ? c | c | c | c | c | c ? c | c ? c | c | c | c | c ? c | c | c | c | c ? c | c | c | c }
 \Xhline{1pt}
 &\multicolumn{6}{c?}{Similarity Loss} &\multicolumn{2}{c?}{Aux. Loss} &\multicolumn{5}{c?}{Regularizer}&\multicolumn{5}{c?}{Accuracy Measure}&\multicolumn{4}{c}{Regularity Measure}\\
 &\rotatebox[origin=c]{90}{MSE} & \rotatebox[origin=c]{90}{{LCC}} & \rotatebox[origin=c]{90}{Correlation} & \rotatebox[origin=c]{90}{NGF} & \rotatebox[origin=c]{90}{MI} & \rotatebox[origin=c]{90}{MIND-SSC}& \rotatebox[origin=c]{90}{Anatomy}& \rotatebox[origin=c]{90}{Landmark} & \rotatebox[origin=c]{90}{Diffusion} & \rotatebox[origin=c]{90}{Curvature} & \rotatebox[origin=c]{90}{Bending} & \rotatebox[origin=c]{90}{Jacobian} & \rotatebox[origin=c]{90}{Consistency} & \rotatebox[origin=c]{90}{TRE}& \rotatebox[origin=c]{90}{MSE}& \rotatebox[origin=c]{90}{SSIM} & \rotatebox[origin=c]{90}{Dice} & \rotatebox[origin=c]{90}{HdD} & \rotatebox[origin=c]{90}{$\%\text{ of }|J_\phi\leq0$}& \rotatebox[origin=c]{90}{$\#\text{ of }|J_\phi|\leq0$}& \rotatebox[origin=c]{90}{std.$(|J_\phi|)$}& \rotatebox[origin=c]{90}{$|\nabla J_\phi|$}\\
 
 \Xhline{1pt}
 {AC-DMiR~\citep{khor2023anatomically}}&$\sbullet[1.5]$&$\sbullet[1.5]$& & & & &$\sbullet[1.5]$& &$\sbullet[1.5]$& &$\sbullet[1.5]$&$\sbullet[1.5]$& & & & &$\sbullet[1.5]$& &$\sbullet[1.5]$& & &\\

 ADMIR~\citep{tang2020admir}                   & &$\sbullet[1.5]$& & & & & & &$\sbullet[1.5]$& & & & &$\sbullet[1.5]$& & &$\sbullet[1.5]$&$\sbullet[1.5]$& & & &\\

{AMNet~\citep{che2023amnet}} & &$\sbullet[1.5]$& & & & & & &$\sbullet[1.5]$& & &$\sbullet[1.5]$& & & & &$\sbullet[1.5]$& &$\sbullet[1.5]$& & &\\
 
 Attention-Reg~\citep{song2022cross}           & & & & & & & $\sbullet[1.5]$& &$\sbullet[1.5]$& & & & &$\sbullet[1.5]$& & &$\sbullet[1.5]$& & & &$\sbullet[1.5]$&\\

 Baum~\emph{et al.}~\cite{baum2022meta}            & & & & & & &$\sbullet[1.5]$& & & &$\sbullet[1.5]$& & &$\sbullet[1.5]$& & &$\sbullet[1.5]$& & & & &\\

 BIRNet~\citep{fan2019birnet}                  &$\sbullet[1.5]$& & & & & & & & & & & & & & & &$\sbullet[1.5]$& & & & &\\
 
 CondLapIRN~\citep{mok2021conditional}         &  &$\sbullet[1.5]$ & & & & & & & $\sbullet[1.5]$& & & & & & & &$\sbullet[1.5]$& &$\sbullet[1.5]$& &$\sbullet[1.5]$& \\

 CycleMorph~\citep{kim2021cyclemorph}          &$\sbullet[1.5]$&$\sbullet[1.5]$ & & & & & & & $\sbullet[1.5]$& & & &$\sbullet[1.5]$& &$\sbullet[1.5]$&$\sbullet[1.5]$&$\sbullet[1.5]$& &$\sbullet[1.5]$& & &\\
 
 de Vos~\emph{et al.}~\citep{de2020mutual}            & & & & &$\sbullet[1.5]$& & & & & &$\sbullet[1.5]$& & & & & &$\sbullet[1.5]$&$\sbullet[1.5]$&$\sbullet[1.5]$& & &\\ 

 Deformer~\citep{chen2022deformer}             & &$\sbullet[1.5]$& & & & & & &$\sbullet[1.5]$& & & & & & & &$\sbullet[1.5]$& &$\sbullet[1.5]$ & &$\sbullet[1.5]$&\\
 
 DiffuseMorph~\citep{kim2022diffusemorph}      & &$\sbullet[1.5]$& & & & & & & $\sbullet[1.5]$& & & & & &$\sbullet[1.5]$&$\sbullet[1.5]$&$\sbullet[1.5]$& &$\sbullet[1.5]$& & &\\

 DIRNet~\citep{de2017end}                      & & $\sbullet[1.5]$& & & & & & & & & & & & & & &$\sbullet[1.5]$& & & & & \\
 
 DLIR~\citep{de2019deep}                       & &$\sbullet[1.5]$& & & & & & & & &$\sbullet[1.5]$& & & & & &$\sbullet[1.5]$&$\sbullet[1.5]$&$\sbullet[1.5]$& &$\sbullet[1.5]$&\\

 DNVF~\citep{han2023diffeomorphic}             & &$\sbullet[1.5]$& & & & & & & $\sbullet[1.5]$& & &$\sbullet[1.5]$& & & &$\sbullet[1.5]$&$\sbullet[1.5]$& &$\sbullet[1.5]$& & &\\
 
 DTN~\citep{zhang2021learning}                 &$\sbullet[1.5]$& & & & & & & &$\sbullet[1.5]$& & & & & & & &$\sbullet[1.5]$& & & &$\sbullet[1.5]$&\\

 Dual-PRNet~\citep{hu2019dual}                 &  &$\sbullet[1.5]$& & & & & & & $\sbullet[1.5]$& & & & & & & &$\sbullet[1.5]$ & & & & & \\
 
 Dual-PRNet++~\citep{kang2022dual}              &  &$\sbullet[1.5]$& & & & & & & $\sbullet[1.5]$& & & & & & & &$\sbullet[1.5]$&$\sbullet[1.5]$&$\sbullet[1.5]$& & & \\

 FAIM~\citep{kuang2019faim}                    & &$\sbullet[1.5]$& & & & & & &$\sbullet[1.5]$& & &$\sbullet[1.5]$& & & & &$\sbullet[1.5]$& & &$\sbullet[1.5]$& &\\
 
 Fan~\emph{et al.}~\citep{fan2019adversarial}  & & & & & & & & &$\sbullet[1.5]$& & & & & & & & &$\sbullet[1.5]$& &$\sbullet[1.5]$& &\\

 Fourier-Net~\citep{jia2022fourier}            &$\sbullet[1.5]$&$\sbullet[1.5]$& & & & & & &$\sbullet[1.5]$& & & & & & & &$\sbullet[1.5]$&$\sbullet[1.5]$&$\sbullet[1.5]$& & &\\

 {FSDiffReg~\citep{qin2023fsdiffreg}} & &$\sbullet[1.5]$& & & & & & &$\sbullet[1.5]$&$\sbullet[1.5]$& & & & & & &$\sbullet[1.5]$& &$\sbullet[1.5]$& &$\sbullet[1.5]$&\\
 
 GraformerDIR~\citep{yang2022graformerdir}     & &$\sbullet[1.5]$& & & & & & &$\sbullet[1.5]$& & &$\sbullet[1.5]$& & & & &$\sbullet[1.5]$& &$\sbullet[1.5]$& & &\\

 Han~\emph{et al.}~\citep{han2022deformable}   & &$\sbullet[1.5]$& & & & & & &$\sbullet[1.5]$& & & & &$\sbullet[1.5]$& & &$\sbullet[1.5]$& &$\sbullet[1.5]$& & & \\
 
 Hering~\emph{et al.}~\citep{hering2021cnn}           &  & & &$\sbullet[1.5]$& & & &$\sbullet[1.5]$& &$\sbullet[1.5]$& & & & & & &$\sbullet[1.5]$&$\sbullet[1.5]$&$\sbullet[1.5]$& & & \\

 HyperMorph~\citep{hoopes2022hyper}       &$\sbullet[1.5]$&$\sbullet[1.5]$& & & & &$\sbullet[1.5]$& &$\sbullet[1.5]$& & & & & &$\sbullet[1.5]$& &$\sbullet[1.5]$& & & &$\sbullet[1.5]$&\\ 

 {IDIR~\citep{wolterink2022implicit}}       & &$\sbullet[1.5]$& & & & & & & & &$\sbullet[1.5]$&$\sbullet[1.5]$& &$\sbullet[1.5]$& & & & & &$\sbullet[1.5]$& &\\
 
 im2grid~\citep{liu2022coordinate}             &$\sbullet[1.5]$& & & & & & & & $\sbullet[1.5]$& & & & & & & &$\sbullet[1.5]$& &$\sbullet[1.5]$&$\sbullet[1.5]$& &\\

 Krebs~\emph{et al.}~\citep{krebs2019learning}        &  &$\sbullet[1.5]$& & & & & & & $\sbullet[1.5]$& & & & & & $\sbullet[1.5]$& &$\sbullet[1.5]$ &$\sbullet[1.5]$& & & &$\sbullet[1.5]$ \\

 LapIRN~\citep{mok2020large}                   &  &$\sbullet[1.5]$ & & & & & & & $\sbullet[1.5]$& & & & & & & &$\sbullet[1.5]$& & &$\sbullet[1.5]$&$\sbullet[1.5]$&\\

 LKU-Net~\citep{jia2022u}                      & &$\sbullet[1.5]$& & & & & $\sbullet[1.5]$& & $\sbullet[1.5]$& & & & & & & &$\sbullet[1.5]$&$\sbullet[1.5]$&$\sbullet[1.5]$ & &$\sbullet[1.5]$&\\

 Li~\emph{et al.}~\citep{li2018non}            & &$\sbullet[1.5]$& & & & & & &$\sbullet[1.5]$& & & & & & & &$\sbullet[1.5]$& & & & &\\

 Liu~\emph{et al.}~\citep{liu2019probabilistic}&$\sbullet[1.5]$& & & & & & & & & & & & & & & &$\sbullet[1.5]$& & & & &\\

 {MAIRNet~\cite{gao2024mairnet}} & & & & & &$\sbullet[1.5]$& & & &$\sbullet[1.5]$& & & & & & &$\sbullet[1.5]$& & & & &\\
 
 MIDIR~\citep{qiu2021learning}                 &  & & & &$\sbullet[1.5]$ & & & & $\sbullet[1.5]$& & & & & & & &$\sbullet[1.5]$ & &$\sbullet[1.5]$& & & $\sbullet[1.5]$ \\

 {ModeT~\citep{wang2023modet}}                 & &$\sbullet[1.5]$& & & & & & & $\sbullet[1.5]$& & & & & & & &$\sbullet[1.5]$ & &$\sbullet[1.5]$& & & \\
 
 MS-DIRNet~\citep{lei20204d}                   & &$\sbullet[1.5]$& &$\sbullet[1.5]$& & & & &$\sbullet[1.5]$& &$\sbullet[1.5]$& & &$\sbullet[1.5]$&$\sbullet[1.5]$& & & & & & &\\
 
 MS-ODENet~\citep{xu2021multi}                 &$\sbullet[1.5]$& & & & & & & & $\sbullet[1.5]$& & & & & &$\sbullet[1.5]$& &$\sbullet[1.5]$& & & & &\\

 {NICE-Trans~\citet{meng2023non}}
 & &$\sbullet[1.5]$& & & & & & & $\sbullet[1.5]$& & &$\sbullet[1.5]$& & & & &$\sbullet[1.5]$& &$\sbullet[1.5]$& & &\\
 
 NODEO~\citep{wu2022nodeo}                     & &$\sbullet[1.5]$& & & & & & & $\sbullet[1.5]$& & &$\sbullet[1.5]$& & & & &$\sbullet[1.5]$& &$\sbullet[1.5]$& & &\\

 {PC-Reg~\citep{yin2023pc}}     & &$\sbullet[1.5]$& & & & &$\sbullet[1.5]$& & $\sbullet[1.5]$& & & & & & & &$\sbullet[1.5]$&$\sbullet[1.5]$&$\sbullet[1.5]$& & &\\

 PC-SwinMorph~\citep{liu2022pc}                & &$\sbullet[1.5]$& & & & & & & $\sbullet[1.5]$& & & &$\sbullet[1.5]$& & & &$\sbullet[1.5]$& & & & &\\
 
 PDD-Net 2.5D~\citep{heinrich2020highly}       & & & & & &$\sbullet[1.5]$&$\sbullet[1.5]$& & $\sbullet[1.5]$& & & & & & & &$\sbullet[1.5]$& & & &$\sbullet[1.5]$&\\

 PDD-Net 3D~\citep{heinrich2019closing}        & & & & & &$\sbullet[1.5]$&$\sbullet[1.5]$& & $\sbullet[1.5]$& & & & & & & &$\sbullet[1.5]$& & & &$\sbullet[1.5]$&\\

 {PIViT~\citep{ma2023pivit}}        & &$\sbullet[1.5]$& & & & & & & $\sbullet[1.5]$& & & & & & & &$\sbullet[1.5]$& & &$\sbullet[1.5]$& &\\

 {R2Net~\citep{joshi2023r2net}}&$\sbullet[1.5]$& & &$\sbullet[1.5]$& & & & & & & & $\sbullet[1.5]$& & &$\sbullet[1.5]$& &$\sbullet[1.5]$& & &$\sbullet[1.5]$& &\\

 SDHNet~\citep{zhou2023self}                   & &$\sbullet[1.5]$& & & & & & & $\sbullet[1.5]$& & & & & & & &$\sbullet[1.5]$&$\sbullet[1.5]$&$\sbullet[1.5]$& & & \\

 Shao~\emph{et al.}~\citep{shao2022multi}      &$\sbullet[1.5]$& & & & & & & &$\sbullet[1.5]$& & & & & &$\sbullet[1.5]$& & & & &$\sbullet[1.5]$&$\sbullet[1.5]$&\\

{SpineRegNet~\cite{zhao2023spineregnet}} & &$\sbullet[1.5]$& & & & & & & &$\sbullet[1.5]$& & & & & & &$\sbullet[1.5]$& &$\sbullet[1.5]$& & &\\

 SVF-R2Net~\citep{joshi2022diffeomorphic}      &$\sbullet[1.5]$& & & & & & & & & & &$\sbullet[1.5]$& & &$\sbullet[1.5]$& &$\sbullet[1.5]$& & &$\sbullet[1.5]$& &\\

 SYMNet~\citep{mok2020fast}                    &  &$\sbullet[1.5]$ & & & & & & & $\sbullet[1.5]$& & &$\sbullet[1.5]$&$\sbullet[1.5]$& & & &$\sbullet[1.5]$& & &$\sbullet[1.5]$ & &\\

 SymTrans~\citep{ma2022symmetric}              &$\sbullet[1.5]$& & & & & & & &$\sbullet[1.5]$& & & & & & & &$\sbullet[1.5]$& & &$\sbullet[1.5]$& &\\

 SynthMorph~\citep{hoffmann2021synthmorph}     & & & & & & &$\sbullet[1.5]$& &$\sbullet[1.5]$& & & & & & & &$\sbullet[1.5]$& &$\sbullet[1.5]$& & &\\

 TM-DCA~\citep{chen2023deform}                 & &$\sbullet[1.5]$& & & & & $\sbullet[1.5]$& & $\sbullet[1.5]$& & & & & & & &$\sbullet[1.5]$& &$\sbullet[1.5]$& &$\sbullet[1.5]$&\\

 TM-TVF~\citep{chen2022unsupervised}   &$\sbullet[1.5]$&$\sbullet[1.5]$& & & & & $\sbullet[1.5]$& & $\sbullet[1.5]$& & & & & & &$\sbullet[1.5]$&$\sbullet[1.5]$&$\sbullet[1.5]$ &$\sbullet[1.5]$& &$\sbullet[1.5]$&\\

 {TransMatch~\citep{chen2023transmatch}}         &$\sbullet[1.5]$&$\sbullet[1.5]$& & & & & & & $\sbullet[1.5]$& & & & & & & &$\sbullet[1.5]$& &$\sbullet[1.5]$ & &$\sbullet[1.5]$&\\
 
 TransMorph~\citep{chen2022transmorph}         &$\sbullet[1.5]$&$\sbullet[1.5]$& & & & & $\sbullet[1.5]$& & $\sbullet[1.5]$& &$\sbullet[1.5]$& & & & &$\sbullet[1.5]$&$\sbullet[1.5]$& &$\sbullet[1.5]$ & & &\\

 ViT-V-Net~\citep{chen2021vitvnet}             &$\sbullet[1.5]$& & & & & & & &$\sbullet[1.5]$& & & & & & & &$\sbullet[1.5]$& &$\sbullet[1.5]$& & &\\

 VoxelMorph~\citep{balakrishnan2019voxelmorph} &$\sbullet[1.5]$& $\sbullet[1.5]$ & & & & & $\sbullet[1.5]$& &$\sbullet[1.5]$& & & & & & & &$\sbullet[1.5]$& &$\sbullet[1.5]$&$\sbullet[1.5]$& & \\

 VoxelMorph-diff~\citep{dalca2019unsupervised} &$\sbullet[1.5]$& & & & & &$\sbullet[1.5]$& & $\sbullet[1.5]$& & & & & & & &$\sbullet[1.5]$&  &$\sbullet[1.5]$&$\sbullet[1.5]$& & \\

 VoxelMorph++~\citep{heinrich2022voxelmorph++} & & & & & &$\sbullet[1.5]$& &$\sbullet[1.5]$&$\sbullet[1.5]$& & & & &$\sbullet[1.5]$& & & & & &$\sbullet[1.5]$&$\sbullet[1.5]$&\\
 
 VR-Net~\citep{jia2021learning}                &$\sbullet[1.5]$& & & & & & & &$\sbullet[1.5]$& & & & & & & &$\sbullet[1.5]$&$\sbullet[1.5]$&$\sbullet[1.5]$& & &$\sbullet[1.5]$\\

 VTN~\citep{zhao2019unsupervised}              & & &$\sbullet[1.5]$& & & & & & $\sbullet[1.5]$& & &$\sbullet[1.5]$&$\sbullet[1.5]$& & & &$\sbullet[1.5]$ & &$\sbullet[1.5]$& &$\sbullet[1.5]$& \\
 
 XMorpher~\citep{shi2022xmorpher}              & &$\sbullet[1.5]$& & & & & $\sbullet[1.5]$& & $\sbullet[1.5]$& & & &$\sbullet[1.5]$& & & &$\sbullet[1.5]$& &$\sbullet[1.5]$& & &\\

 Zhang~\emph{et al.}~\citep{zhang2020diffeomorphic}   &$\sbullet[1.5]$& & & & & & & &$\sbullet[1.5]$& & &$\sbullet[1.5]$& & & & &$\sbullet[1.5]$& &$\sbullet[1.5]$& & &\\
 
 \Xhline{1pt}
\end{tabular}
\end{table*}

%
%
Table~\ref{table:unsup_list} provides a compilation of unsupervised DIR models, summarizing the similarity and auxiliary loss functions, as well as other details.

{The topic of similarity measures is only touched upon briefly in this section and is in itself a wide area of active research, interested readers may find the review articles by~\citet{santini1999tpami} and~\citet{unnikrishnan2005wacv} to be a useful resource.}
See the text for complete details and discussion.

\subsection{Supervised Learning}
In supervised learning, where the ground truth transformation is used, the loss function is typically easy to define, with the mean square error~(MSE)~\citep{miao2016cnn, krebs2017robust, eppenhof2018deformable, rohe2017svf, cao2017deformable, fan2019birnet}, the equivalent end-point-error (EPE), and mean absolute error (MAE)~\citep{yang2017quicksilver, sokooti2017nonrigid} being the most popular choices.
{Recently,~\citet{terpstra2022loss} highlighted that a loss based on $\ell^2$ (equivalent to MSE or EPE) does not distribute errors symmetrically across deformation directions when calculated separately for each direction.
To address this, they introduced $\bot$-loss, which considers the interplay between directions.
In this method, the $x$ and $y$ directions of the deformation field are viewed as the real and imaginary components of a complex image, thereby creating a polar representation of deformations.
$\bot$-loss quantifies the phase error between the predicted and actual complex image representations of the deformation fields.
When combined with Euclidean distance, forming the $\perp$$+\ell^2$ loss, it ensures symmetric error distribution across both deformation directions. The authors demonstrated that a network trained with a combination of $\bot$-loss and $\ell^2$ loss outperforms a network trained with $\ell^2$ loss alone in terms of registration performance. However, the applicability of this method to 3D or 4D registration tasks remains to be further explored, given the challenge of representing additional dimensions within a 2D complex image framework.}

\subsection{Unsupervised \& Semi-supervised Learning}
In unsupervised learning, where there is no ground truth transformation to reference, regularization is usually used to enforce smoothness in the transformation. As a result, the loss function is often similar to the energy function used in traditional methods (\emph{i.e.}, Eqn.~\ref{eqn:energy_func}), which includes an image similarity measure and a transformation regularizer. The following subsections provide a summary of commonly used and recently proposed loss functions for image registration.

\subsection{Similarity Measure}

\noindent\textbf{Mono-modality.} The choice of image similarity measure can vary depending on each specific application. For mono-modal registration, MSE is still a popular choice and has the advantage of having a straightforward probabilistic interpretation of the Gaussian likelihood approximation~\citep{dalca2019unsupervised, chen2022transmorph, kim2021cyclemorph, balakrishnan2019voxelmorph, meng2022enhancing, jia2021learning, liu2022coordinate}.
However, a disadvantage of MSE is that it averages the difference across all voxels in the image, making it sensitive to local intensity variations within the image.
{Local correlation coefficient (LCC)} is known to be more robust to local intensity variations and has been found to be superior in brain MR registration applications~\citep{avants2008symmetric}.
{LCC} has been extended as a loss function for training learning-based models, with the local window computation often being done through convolution operations~\citep{kuang2019faim, chen2022transmorph, kim2021cyclemorph, balakrishnan2019voxelmorph, zhang2018inverse, mok2020fast, mok2020large, mok2021conditional}. One disadvantage of {LCC} is its higher computing cost in comparison to MSE, which is mainly attributable to the comparatively large convolution kernel size (typically chosen between $5\times5\times5$ and $9\times9\times9$ voxels~\citep{avants2008symmetric, balakrishnan2019voxelmorph, mok2020fast}).
{To improve the computational efficiency, Gaussian weighted LCC can be used and implemented as three 1D convolutions.
The locality is controlled by the standard deviation of the Gaussian window.
However, the computation time complexity is independent of the standard deviation, because the 1D Gaussian kernel can be approximated by a fast recursive filter~\citep{cachier2000non}.}
The structural similarity index (SSIM)~\citep{wang2004image} has also been demonstrated to be an effective loss function for mono-modal image registration~\citep{chen2020generating, mahapatra2018deformable, sandkuhler2018airlab}. SSIM takes into account luminance, contrast, and structure. It can be thought of as an extension of the {LCC}, with the structure term in SSIM being the square root of {LCC}. This allows SSIM to capture more information about the similarity of two images beyond just the degree of correlation between them.
{The mono-modality loss function can also be adapted for multi-modality registration by leveraging image synthesis techniques. \citet{liu2023geometry} introduced a network architecture that accepts two images from different modalities to estimate a deformation field. Rather than employing a loss function  designed for multi-modality images, the authors integrate an image-to-image translation network. This network aims to match the modality of the moving image with that of the fixed image, thereby facilitating the use of a mono-modality loss. To ensure that the image-to-image translation network is solely responsible for modifying intensity without compensating for spatial transformations, a specialized training scheme is implemented.}

\noindent\textbf{Multi-modality.} For multi-modal applications, traditional methods often use mutual information~(MI)~\citep{viola1997alignment}, correlation ratio~\citep{roche1998correlation}, self-similarity context~(SSC)~\citep{heinrich2013towards}, or normalized gradient fields~(NGF)~\citep{haber2006intensity} as similarity measures. Both MI and correlation ratio evaluate the relationship between the two images by calculating intensity statistics, such as intensity histograms, to measure statistical dependence.
However, the standard method for calculating intensity histograms, which involves counting, is not differentiable, so a Parzen window formulation~\hbox{\citep{thevenaz2000optimization}} is often used to allow the loss to be backpropagated during network training.
Parzen-window-based MI has been employed as a loss function in many multi-modal applications~\citep{qiu2021learning, de2020mutual, nan2020drmime, guo2019multi, hoffmann2021synthmorph}, but it can be relatively difficult to implement and also sensitive to factors such as the number of intensity bins and the smoothness of the {window} function. {Meanwhile, local MI (LMI) has been adapted for deep learning-based image registration to incorporate detailed spatial information~\cite{guo2019multi}. The calculation of MI in general often requires a vectorized approach to bypass loop operations for speed and automatic differentiation, which significantly increases memory consumption as the spatial dimension must expand to accommodate the number of bins used. Therefore, LMI is typically implemented using non-overlapping square patches to conserve memory, although overlapping patches are also feasible~\cite{guo2019multi}.} 
As far as we are aware, the correlation ratio has not been used in learning-based medical image registration.
It should be noted that these intensity-statistic-based measurements do not take into account local structural information, making them more suitable for rigid/affine registration and less suitable for deformable registration applications~\citep{pluim2000image, heinrich2013towards}.
SSC is another commonly used loss function for multi-modal applications, and it is an improvement on the modality-independent neighborhood descriptor~(MIND)~\citep{heinrich2012mind}.
Both SSC and MIND operate by calculating the descriptor between a voxel and its neighboring voxels within a given image, turning an image of any modality into a feature representation of these descriptors. 
The similarity is determined by summing the absolute differences between the descriptors of the two images.
As SSC and MIND consider local structural information, they are not limited in the same way as MI or correlation ratio, making them more useful for multi-modal deformable registration~\citep{hansen2021graphregnet, mok2021conditional2, yang2020unsupervised, xu2020adversarial, blendowski2021weakly}.
{Note that SSC and MIND generate descriptors by analyzing the spatial intensity distribution within a localized area surrounding each voxel, making them sensitive to local orientations (i.e., they are not rotationally invariant).
Consequently, when large rotations are expected between image pairs, one should calculate SSC and MIND for a warped moving image, instead of applying the warp to the SSC and MIND descriptors of the moving image.}
NGF compares images by focusing on the intensity changes, or edges, in the images. The similarity between the two images is determined by the presence of intensity changes at the same locations, regardless of the modalities of the images being compared.
NGF was originally developed for multi-modal applications like brain MR T1-to-T2 and PET-to-CT~\citep{haber2006intensity}.
However, it is now mostly used in learning-based registration models for lung CT registration~\citep{hering2019mlvirnet, hering2021cnn, mok2021conditional2}. 
This is because the complex structure of the lung, including bronchi, fissures, and vessels, can hinder accurate registration~\citep{hering2021cnn}. 
NGF focuses on edges rather than intensity values, making it a more suitable measure for this purpose.

\noindent\textbf{Recent Advancements.} There have been many efforts to improve upon or propose new loss functions due to some limitations of the aforementioned similarity measures.
{\citet{czolbe2021semantic,czolbe2023semantic}~leveraged a ConvNet feature extractor to obtain image features from the deformed and fixed images, and then computed the {LCC} between these features as a similarity measure.}
The benefit of this approach is that the features produced by the ConvNet feature extractor have less noise, resulting in a more consistent similarity measure in areas with noise which leads to a smoother transformation.
\citet{haskins2019learning}~were the first to propose using a ConvNet to learn a similarity measure for image registration.
However, this method relies on having ground truth target registration error for the training dataset to learn such a similarity measure.
{\citet{ronchetti2023disa} used a ConvNet followed by the inner product operation to approximate LC$^2$ similarity~\citep{fuerst2014automatic}. They showed that the trained network can be used for loss function in multi-modal registration.}
\citet{grzech2022variational}~went one step further and introduced a technique for learning a similarity measure using a variational Bayesian method. The method involves initializing the convolution kernels in the network architecture to model MSE and {LCC}, and then using variational inference to learn a similarity measure that optimizes the likelihood of the images in the dataset when aligning them to the atlas. Building on the success of adversarial networks in computer vision~\citep{makhzani2015adversarial, goodfellow2020generative}, researchers have developed a number of techniques for image registration that leverage adversarial training~\citep{fan2019adversarial, mahapatra2020training, luo2021deformable}. These methods can be used standalone or in conjunction with a traditional similarity measure.

\subsection{Deformation Regularizer}
\label{sec:def_reg}
A deformation regularizer, as the terminology implies, is used for DIR, with its usage being not necessary for rigid/affine transformations.
For DIR algorithms, producing smooth deformations is not only a desirable property but a necessary requirement: while diffeomorphic transformations may not be required for certain applications, smoothness remains imperative in almost all cases to avoid trivial solutions such as rearranging voxels~\citep{rohlfing2011image}, with which an almost perfect similarity measure can be achieved but result in unrealistic transformation~(also see Section~\ref{sec:Eval_Metric}).
The regularizer can be considered as a prior in a maximum a posteriori~(MAP) framework, while the similarity measure acts as the data likelihood (\emph{e.g.}, in the case of MSE, the data likelihood becomes a Gaussian likelihood). The diffusion regularizer is a commonly employed deformation regularizer, as demonstrated by its frequent appearance in Table~\ref{table:unsup_list}. This regularization computes the squared $\ell^2$-norm of the gradients of the displacement field, effectively penalizing the disparities between adjacent displacements. Other alternatives for regularization include using the $\ell^1$-norm instead of the $\ell^2$-norm to impart equal penalties on the neighboring disparities, or penalizing the second derivative of the displacements, commonly referred to as bending energy~\citep{rueckert1999nonrigid}. It is important to note that since bending energy and curvature-based regularizers penalize the second derivatives, thereby zeroing out any affine contributions, pre-affine alignment prior to the deformable registration step may not be necessary, as demonstrated in \citep{ding2022aladdin, fischer2003curvature}.
These conventional regularizers enforce an isotropic regularization on the displacement field~\citep{pace2013locally}.
As a result, they discourage discontinuities in the displacements in applications where sliding motion may occur in organs, such as registering exhale and inhale CT scans of the lung.
Historically, various improvements have been made to address this issue, including the isotropic Total Variation~(TV) regularization~\citep{vishnevskiy2016isotropic}, anisotropic diffusion regularization~\citep{pace2013locally}, and adaptive bilateral filtering-based regularization~\citep{papiez2014implicit}.
However, these regularization techniques have not been widely adopted in learning-based image registration.

\noindent\textbf{Recent Advancements.}
Enforcing spatial smoothness alone is insufficient to ensure the regularity of the transformations. A different strategy is to penalize the ``folding'' of voxels directly during training, in addition to applying the aforementioned regularizers to enforce smoothness in the deformation.
These foldings can be evaluated using local Jacobian determinants, where the magnitude of the Jacobian determinant indicates if the volume is expanding or shrinking near the voxel location.
A non-positive Jacobian determinant represents a locally non-invertible transformation. Several regularization methods based on local Jacobian determinants have been proposed to penalize such transformations~\citep{kuang2019faim, mok2020fast}.
Meanwhile, with the advent of deep learning, new methods have emerged that leverage the deep learning of deformation regularization from data.
One such method by  \citet{niethammer2019metric},~introduced a method that learns a spatially-varying deformation regularization using training data.
Spatially-varying regularization offers the advantage of accommodating variations in deformation that may be required for different regions within an image, such as the movement of the lungs in relation to other organs (\emph{e.g.}, rib cage) due to respiratory processes.
The technique proposed by Niethammer~\emph{et al.} involves training a registration network to produce not only a deformation field but also a set of weight maps, each of which corresponds to the weight of a Gaussian smoothing kernel in a multi-Gaussian kernel configuration.
The weighted multi-Gaussian kernel is then applied to the deformation field via convolution. To further impose spatial smoothness, an optimal mass transport~(OMT) loss function was introduced to encourage the network to assign larger weights to Gaussian kernels with larger variances.
While this method was developed for a time-stationary velocity field setting, \citet{shen2019region} later expanded upon it by incorporating it into a time-varying velocity field setting.
In this setup, a different set of weight maps are produced for each time point.
More recently, \citet{chen2023spr}~introduced a weighted diffusion regularizer that applies spatially-varying regularization to the deformation field.
The neural network generates a weight volume, assigning a unique regularization weight to each voxel and thus allows for spatially-varying levels of regularization strength.
As the diffusion regularizer is related to Gaussian smoothing, using spatially-varying strengths of diffusion regularization can be considered equivalent to employing a multi-Gaussian kernel, as originally proposed by~\citet{niethammer2019metric}.
This is because the convolution of multiple Gaussian kernels still results in a Gaussian kernel. To promote the overall smoothness of the deformation, they further applied a log loss to the weight volume, which encourages the maximum regularization strength when possible. 
In a different approach, \citet{wang2022deep}~employed a regression network to learn the optimal regularization parameter for an optimization-based method, specifically Flash~\citep{zhang2019fast}.
Flash is a geodesic shooting method in the Fourier space that requires only the initial velocity field to compute the time-dependent transformation.
Wang~\emph{et al.} generated ground truth optimal regularization parameters by assuming the prior of the initial velocity field given the regularization parameter as a multivariate Gaussian distribution.
Using gradient descent, they obtained the optimal regularization parameter for each image pair through MAP estimation.
A ConvNet regression encoder then estimates the optimal regularization parameter based on the image pair. 
This approach achieved registration performance comparable to Flash while significantly improving runtime and memory efficiency.
Alternatively, \citet{laves2019deformable}~were inspired by the deep image prior~\citep{ulyanov2018deep}. 
They used a randomly initialized ConvNet as a regularization prior.
They then fed a random image (\emph{i.e.}, a noise image) as input and the network gradually transformed it into a smooth deformation field through iterative optimization.
The deep image prior provided by the ConvNet enables the network to produce a smooth deformation in the early iterations, then gradually adds non-smooth high-frequency deformations.
As a result, early stopping is used for the network to generate a smooth deformation field without the need for explicitly encouraging smoothness in the loss function.

Transformations can also be implicitly regularized by imposing invertibility constraints.
This is achieved by using a symmetric consistency loss or cycle consistency loss.
Symmetric consistency typically uses a single DNN to output both the forward and reverse deformation fields, which transform the moving image to the fixed image and vice versa, respectively.
The similarity between the warped image and the target image is then calculated and backpropagated to update the network~\citep{mok2020fast, liu2022pc}.
Alternatively, a consistency loss can be calculated by composing the network-generated forward and backward deformation fields, and then comparing the outcome with the identity transformation~\citep{greer2021icon,tian2022gradicon}.
The underlying concept is that, theoretically, an invertible mapping should cancel itself when composed with its inverse.
Such an approach by itself imposes invertibility but does not explicitly enforce spatial smoothness over the deformation field.
\citet{greer2021icon}~demonstrated that incorporating such a loss within a DNN framework implicitly imposes spatial regularity {and inverse consistency} on the deformation field without necessitating an additional regularizer to enforce smoothness.
The authors showed that the errors of the DNN in computing the inverse, combined with the implicit bias of DNN favoring more regular outputs, enable such a consistency loss to entail a $H^1-$ or Sobolev-type regularization over the deformation field, thereby implicitly enforcing spatial smoothness.
Later, \citet{tian2022gradicon}~expanded on this regularizer and proposed to regularize deviations of the Jacobian of the composition from the identity matrix.
This improved regularizer led to faster convergence while offering greater flexibility, while maintaining an approximated diffeomorphic transformation.
{\citet{greer2023inverse}~introduced a simple method to ensure DNN-based registration models are inverse consistent by construction.
They began by defining the output of the registration network as a member of a specified Lie group that inherently supports inverse consistency.
Specifically, suppose a DNN $f$ generates a deformation field, this is mathematically represented as:
\begin{equation}
    f[I_m, I_f]:=\exp{(g(I_m,I_f))}.
\end{equation} The DNN ensures the generated deformation field retains inverse consistency when $g(I_f,I_m)=-g(I_m,I_f)$, as demonstrated by:
\begin{equation}
    f[I_m, I_f]\circ f[I_f, I_m]\equiv\exp{(g(I_m,I_f))}\circ\exp{(-g(I_m,I_f))}=id.
\end{equation} To ensure $g(I_f,I_m)$ meets this condition, the authors propose:
\begin{equation}
    g(I_f,I_m):=M_\theta(I_f,I_m)-M_\theta(I_m,I_f),
\end{equation}
where $M_\theta$ is an arbitrary DNN with parameters $\theta$. This approach structurally guarantees that the deformation field generated by the DNN maintains inverse consistency.
The authors recognize that their previous two-stage approach~\citep{greer2021icon, tian2022gradicon}, which initially estimates the deformation field at a coarse level and subsequently refines it with an additional field, does not guarantee inverse consistency due to the interleaving of inverses. To address this, they introduced a \texttt{TwoStepConsistent} operator that works on the square root of $\exp{(g(I_m,I_f))}$.
This operator can also be adapted for multi-step registration methods, which involve compositions as described later in Section~\ref{sec:multi_res_reg}.
The updated model has been evaluated on both synthetic and real-world brain MRI datasets and has shown significantly more precise inverse consistency compared to models that employ penalty-based approaches for ensuring inverse consistency~\citep{greer2021icon, tian2022gradicon}.}

On the other hand, cycle consistency employs two identical networks, where the first network generates a forward deformation field that deforms the moving image and the second network produces a reverse field that aims to warp the deformed image back to the original moving image~\citep{zhao2019unsupervised, kuang2019cycle, kim2021cyclemorph}.
Both consistency losses have been shown to improve the registration performance and provide regularization to the deformation field. However, since this regularization is not explicitly applied to the deformation fields, a separate deformation regularizer is often required in addition to the consistency loss.

{
Regularization terms in the objective function frequently necessitate tuning additional hyperparameters (e.g., $\lambda$ in Eqn. \ref{eqn:energy_func}) to effectively balance the trade-off between the image similarity measure and the regularity of the deformation field. In a recent study~\citep{dou2023gsmorph}, the authors observed that the similarity and regularity losses often conflict in terms of optimization direction. They proposed a layer-wise gradient surgery process that modifies the gradient of the total loss whenever the inner product of the gradients from these two components is less than zero. This gradient surgery process focuses solely on the direction of the regularization loss gradient, thereby eliminating the need to adjust its weighting.
}

\subsection{Auxiliary Anatomical Information}
\label{ss:anatomical_info}
The overlap of anatomical label maps of the fixed and transformed moving images is a widely used evaluation metric for image registration.
Hence, to improve registration performance on this metric, learning-based methods often apply the estimated deformation to the moving label map during training to compute an extra anatomy loss between the fixed and transformed moving label map.
Various loss functions used in image segmentation tasks, such as Dice loss, cross-entropy, and focal loss (see \citet{ma2021loss}~for a comprehensive review of such loss functions), can be borrowed as the choice of anatomy loss.
Despite the availability of different loss functions, Dice loss remains the most commonly used loss function in learning-based image registration, as evidenced by Table~\ref{table:unsup_list}.
This is likely because Dice loss is confined within the range of $[0,1]$, like {LCC}, which makes it easier to adjust hyperparameters when used in conjunction with {LCC}. 
{Auxiliary anatomical information can also be introduced in the registration network through multi-task learning~\citep{estienne2019u, qin2018joint}.
In particular, \citet{tan2023progressively} showed that such strategy is more effective than the direct application of anatomy loss.}

{\textbf{Recent Advancements.} 
        Building on the concept of integrating segmentation loss while
training the registration network, additional auxiliary data related
to anatomical prior knowledge can also be included during the training
process.} When anatomical landmarks are present in both the moving and fixed images, the transformation generated by the DNN can be applied to the landmarks of the moving image.
The resulting transformed landmarks can then be compared with the landmarks of the fixed image to create a loss.
This landmark supervision has been utilized in optimization-based registration methods to improve performance, as demonstrated in a number of studies~ \citep{ehrhardt2010automatic, polzin2013combining, ruhaak2017estimation, heinrich2015estimating, fischer2003combination}.
\citet{hering2021cnn}~were the first to incorporate landmark supervision into a DNN framework by comparing the MSE between the transformed and target landmarks, which resulted in a substantial improvement in the target registration error of the landmark.
Subsequently, \citep{heinrich2022voxelmorph++}~confirmed the superiority of landmark supervision on multiple benchmark datasets in their work.
It is worth mentioning that the landmarks can be generated automatically before or during the training stage without manual labeling using automatic landmark detection algorithms~\citep{heinrich2015estimating, ruhaak2017estimation, polzin2013combining}, making it straightforward to integrate into most learning-based registration frameworks.

The combination of anatomy loss and deformation regularization without an intensity-based similarity measure is also common, and in these cases, the anatomy loss serves as a modality-independent similarity measure~\citep{hu2018weakly, song2022cross, blendowski2020multimodal}.
However, the drawback of using anatomy loss without a similarity measure is clear: it does not penalize deformations in areas where anatomical labels are missing or ambiguous.
Thus, to achieve accurate and realistic deformations, the anatomical labels should be as detailed as possible, ideally with a unique label for each organ or structure.
However, obtaining such detailed labels is often challenging as anatomical label maps in medical imaging are usually manually delineated, which is a time-consuming and expensive process.

%% file: network_arch.tex
\begin{table*}[!h]
        \rowcolors{3}{white}{cyan!10}
        \centering
        \caption{{A summary of the anatomy, modality, and network infrastructure of the methods outlined in Table~\ref{table:unsup_list}.}}
        \label{table:infrastructure}
        \fontsize{8.5}{10}\selectfont
        \begin{tabular}{lc lc lc l}
        \\[-1em]\hspace*{-5em}\\[-0.8em]
        \toprule
        \rowcolor{white}\textbf{Method} && \textbf{Anatomy} && \textbf{Modality} && \textbf{Network Infrastructure}\\
        \cmidrule(lr){1-1}
        \cmidrule(lr){3-7}
AC-DMiR~\citep{khor2023anatomically} && Brain / Uterus && MRI && Transformer \\
ADMIR~\citep{tang2020admir}  && Brain && MRI && CNN \\
AMNet~\cite{che2023amnet} && Brain && MRI && CNN\\
Attention-Reg~\citep{song2022cross}  && Prostate && US / MRI && CNN (Self Attention) \\
BIRNet~\citep{fan2019birnet}  && Brain && MRI && CNN \\
Baum~\emph{et al.}~\citep{baum2022meta} && Prostate && MRI / Transrectal Ultrasound && CNN \\
CondLapIRN~\citep{mok2021conditional}  && Brain && MRI && CNN \\
CycleMorph~\citep{kim2021cyclemorph}  && Faces / Brain / Liver && Photographs / MRI / CT && CNN \\
de Vos~\emph{et al.}~\citep{de2020mutual}  && Breast && MRI && CNN \\
Deformer~\citep{chen2022deformer}  && Brain && MRI && CNN \\
DiffuseMorph~\citep{kim2022diffusemorph}  && Faces / Brain / Cardiac && Photographs / MRI && DDPM \\
DIRNet~\citep{de2017end}  && Cardiac && MRI && CNN \\ 
DLIR~\citep{de2019deep}  && Cardiac / Chest && MRI / CT && CNN \\ 
DNVF~\citep{han2023diffeomorphic}  && Brain && MRI && DNVF \\
DTN~\citep{zhang2021learning}  && Brain && MRI && CycleGAN-like \\
Dual-PRNet~\citep{hu2019dual}  && Brain && MRI && CNN / Siamese \\
Dual-PRNet++~\citep{kang2022dual}  && Brain && MRI && CNN / Siamese \\
FAIM~\citep{kuang2019faim}  && Brain && MRI && CNN \\
Fan~\emph{et al.}~\citep{fan2019adversarial}  && Brain / Pelvis && MRI / CT && GAN \\
Fourier-Net~\citep{jia2022fourier}  && Brain && MRI && CNN \\
FSDiffReg~\cite{qin2023fsdiffreg} && Cardiac && MRI && DDPM \\ 
GraformerDIR~\citep{yang2022graformerdir}  && Brain / Cardiac && MRI && CNN \\
Han~\emph{et al.}~\citep{han2022deformable}  && Brain && MRI / CT && CycleGAN \\
Hering~\emph{et al.}~\citep{hering2021cnn}  && Lung && CT && CNN \\
HyperMorph~\citep{hoopes2022hyper}  && Brain && MRI && CNN \\
IDIR~\citep{wolterink2022implicit} && Lung && CT && INRs \\
im2grid~\citep{liu2022coordinate}  && Brain && MRI && CNN \\
Krebs~\emph{et al.}~\citep{krebs2019learning}  && Cardiac && MRI && CNN \\
LapIRN~\citep{mok2020large}  && Brain && MRI && CNN \\
LKU-Net~\citep{jia2022u}  && Brain && MRI && CNN \\
Li~\emph{et al.}~\citep{li2018non}  && Brain && MRI && CNN \\
Liu~\emph{et al.}~\citep{liu2019probabilistic} && Brain && MRI && CNN / Siamese \\
MAIRNet~\cite{gao2024mairnet} && Pelvis / Spine && MRI / CT && CNN \\
MIDIR~\citep{qiu2021learning}  && Brain && MRI && CNN \\
ModeT~\citep{wang2023modet} && Brain && MRI && Transformer \\
MS-DIRNet~\citep{lei20204d}  && Abdomen && CT && CNN \\
MS-ODENet~\citep{xu2021multi}  && Brain && MRI && Neural-ODE \\
NICE-Trans~\cite{meng2023non} && Brain && MRI && Transformer \\
NODEO~\citep{wu2022nodeo}  && Brain && MRI && Neural-ODE \\
PC-Reg~\citep{yin2023pc} && Brain / Abdomen && MRI / CT && Transformer \\
PC-SwinMorph~\citep{liu2022pc}  && Brain && MRI && CNN \\
PDD-Net 2.5D~\citep{heinrich2020highly}  && Abdomen && CT && CNN \\
PDD-Net 3D~\citep{heinrich2019closing}  && Abdomen && CT && CNN \\
PIViT~\citep{ma2023pivit}  && Brain && MRI && Transformer \\
R2Net~\citep{joshi2023r2net}  && Brain / Lung / Cardiac && MRI / CT && Neural-ODE \\
SDHNet~\citep{zhou2023self}  && Brain / Liver && MRI / CT && CNN / Student -- Teacher\\
Shao~\emph{et al.}~\citep{shao2022multi}  && Abdomen && Endoscopic && CNN \\
SVF-R2Net~\citep{joshi2022diffeomorphic}  && Brain && MRI && CNN \\
SpineRegNet~\cite{zhao2023spineregnet} && Spine && MRI / CT && CNN \\
SYMNet~\citep{mok2020fast}  && Brain && MRI && CNN \\
SymTrans~\citep{ma2022symmetric}  && Brain && MRI && Transformer \\
SynthMorph~\citep{hoffmann2021synthmorph}  && Brain && MRI && CNN \\
TM-DCA~\citep{chen2023deform}  && Brain && MRI && Transformer \\
TM-TVF~\citep{chen2022unsupervised}  && Faces / Brain && Photographs / MRI && Transformer \\
TransMatch~\citep{chen2023transmatch} && Brain && MRI && Transformer \\
TransMorph~\citep{chen2022transmorph}  && Brain / Abdomen && MRI / CT && Transformer \\
ViT-V-Net~\citep{chen2021vitvnet}  && Brain && MRI && Transformer \\
VoxelMorph~\citep{balakrishnan2019voxelmorph}  && Brain && MRI && CNN \\
VoxelMorph-diff~\citep{dalca2019unsupervised}  && Brain && MRI && CNN \\
VoxelMorph++~\citep{heinrich2022voxelmorph++}  && Chest / Abdomen && CT && CNN \\
VR-Net~\citep{jia2021learning}  && Cardiac && CT && CNN \\
VTN~\citep{zhao2019unsupervised}  && Brain / Liver && MRI / CT && CNN \\
XMorpher~\citep{shi2022xmorpher}  && Brain / Cardiac && MRI / CT && Transformer \\
Zhang~\emph{et al.}~\citep{zhang2020diffeomorphic}  && Brain && MRI && CNN \\
\bottomrule
        %
        %
        %
        %
        %
        %
        \end{tabular}
        \end{table*}

The application of ConvNets has been the dominant trend in learning-based image registration since its inception.
Among different ConvNets architectures, the U-Net-like architectures~\citep{ronneberger2015u}, which were initially designed for image segmentation tasks, have played an important role.
Many noteworthy ConvNet-based registration models, including RegNet~\citep{sokooti2017nonrigid}, DIRNet~\citep{de2017end}, QuickSilver~\citep{yang2017quicksilver}{,} VoxelMorph~\citep{balakrishnan2019voxelmorph, dalca2019unsupervised}, VTN~\citep{zhao2019unsupervised}, DeepFlash~\citep{wang2020deepflash}, and CycleMorph~\citep{kim2021cyclemorph}, have demonstrated promising performance in various registration applications.
More recently, registration neural networks have witnessed notable advancements beyond the conventional ConvNet designs, owing to the progress of DNN architectures in computer vision and the development of architectures that are specifically tailored for registration tasks. 
Notably, models such as Transformers, diffusion models, and Neural ODEs are gaining increasing attention in the field of image registration.
{Table~\ref{table:infrastructure} summarizes the types of network architectures along with the associated anatomies and modalities of the methods outlined in Table~\ref{table:unsup_list}.
This section provides a comprehensive overview of the \textbf{recent advancements} in network architectures.}

\begin{figure}[!ht]
        \centering
        \includegraphics[width=0.45\textwidth]{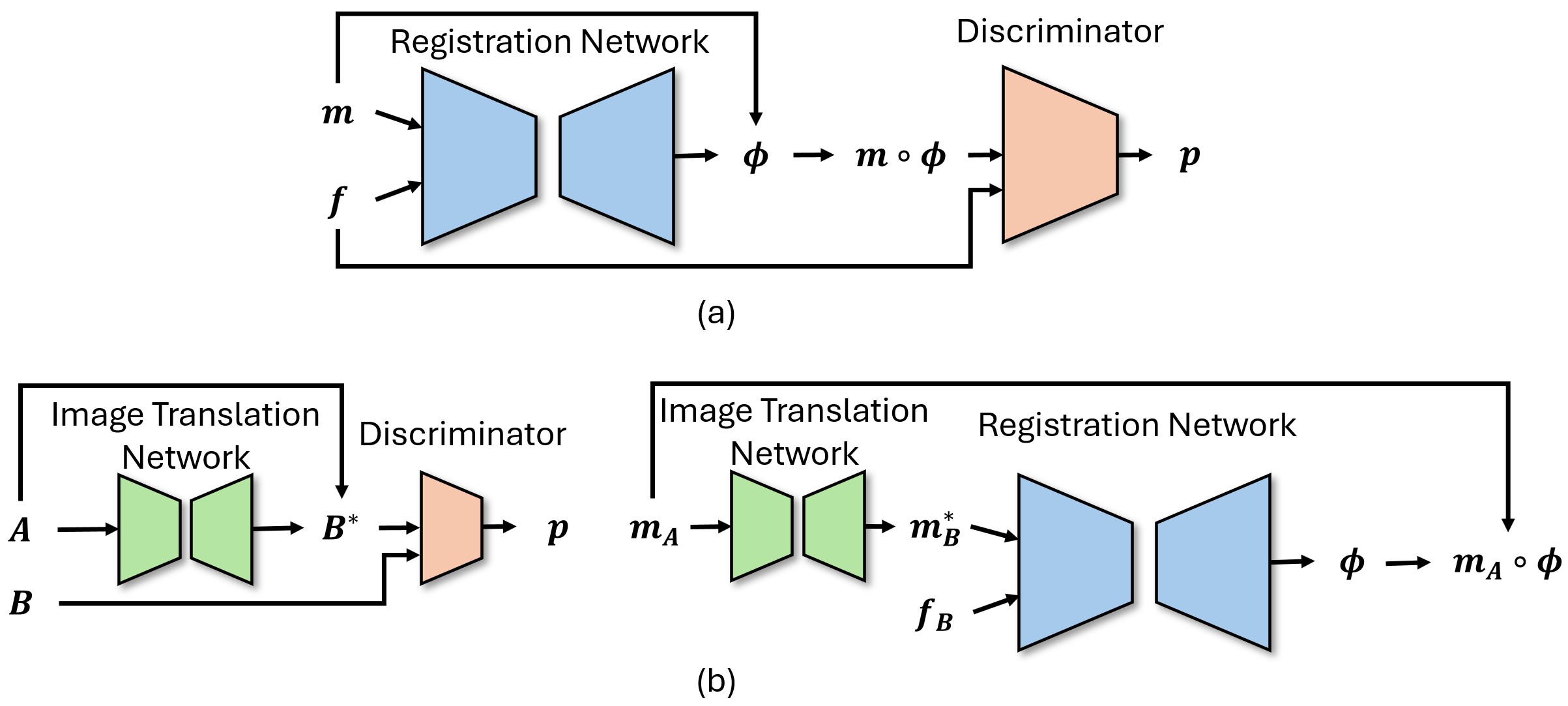}
        \caption{{Visual representation of adversarial
learning in medical image registration, with $\pmb{m}$ and $\pmb{f}$
denoting the moving and fixed images, respectively. Here, $\pmb{\phi}$
represents the deformation field, $\pmb{p}$ indicates a
pseudo-probability generated by the discriminator, and $\pmb{A}$ and
$\pmb{B}$ correspond to two different modalities. Panel (a)
demonstrates how adversarial learning serves as a metric for image
similarity, as similarly employed in \citet{fan2018adversarial}, \citet{yan2018adversarial}, \citet{mahapatra2018deformable, mahapatra2018joint}, and \citet{mahapatra2020training}. Panel~(b) shows the application of adversarial
learning to multi-modal image registration, synthesizing different
image modalities into the same modality for registration, as seen in
\citet{xu2020adversarial}, \citet{wei2019synthesis}, \citet{zhang2021learning}, and \citet{han2022deformable}.}}
        \label{f:network_adv}
        \end{figure}

\subsection{Adversarial Learning}
\label{sec:adv_learn}
The majority of adversarial learning applied to image registration relies on the foundational principles of generative adversarial networks~(GANs).
The concept of GANs is derived from a two-player zero-sum game involving a generator and a discriminator~\citep{goodfellow2020generative}.
The objective of the generator is to generate new samples by learning the data distribution, while the discriminator functions as a binary classifier, aiming to accurately distinguish between real and generated samples.
In the context of image registration, the registration network acts as the generator, producing a deformation field and subsequently warping the moving image.
Meanwhile, the discriminator functions as an image similarity measure, distinguishing between the warped image and the fixed image.
This offers the advantage of alleviating the need for an explicit similarity measure, making the approach adaptable to both mono- and multi-modality applications.

{Figure~\ref{f:network_adv} summarizes the designs of adversarial-based DNNs used for medical image registration.} In early applications of adversarial learning to image registration, \citet{fan2018adversarial}~and \citet{yan2018adversarial}~adhered to the aforementioned approach.
The former utilized the generator to produce a deformation field, while the latter employed a ConvNet encoder to generate affine transformation parameters.
Subsequently, a binary discriminator served as a similarity measure between the transformed and fixed images.
In a similar vein, Mahapatra~\emph{et al.}~\citep{mahapatra2018deformable, mahapatra2018joint, mahapatra2020training} applied adversarial learning to multi-modal image registration, with the additional implementation of CycleGAN~\citep{zhu2017unpaired, qin2019unsupervised} to further ensure the inverse consistency of the generated deformation field. 
\citet{elmahdy2019adversarial}~proposed incorporating anatomical label maps into a Wasserstein-GAN~(WGAN) to enhance the segmentation performance of the registration network.
Their generator was a U-Net-based network that generated a deformation field, which warped both the moving image and the associated anatomical label map.
The discriminator's role was to evaluate the alignment between the warped and fixed image, as well as the warped and fixed label maps.
In their approach, image and anatomical similarity measures were still employed, while the discriminator served as an additional measure of the alignment.
Similar approaches can be found in \citet{duan2019adversarial},  \citet{li2019adversarial}, and \citet{luo2021deformable}, where the authors used the discriminator in conjunction with image similarity measures as additional alignment indicators.
In another study, \citet{fan2019adversarial}~proposed a GAN-based registration framework applicable to both mono- and multi-modality registration.
Their generator was also a registration network based on U-Net, with the discriminator serving as the sole measure of image alignment.
However, the definition of positive pairs sent to the discriminator deviated from previous methods.
Ideally, in mono-modality registration, a positive pair would consist of identical images, but this strict requirement is impractical.
Given this observation, the authors proposed that the positive pair comprise the fixed image and an alpha-blended image created from the fixed and moving images.
For multi-modality registration, a positive pair consisted of pre-aligned multi-modal images from the same patient. 
The method was evaluated on mono-modal brain MRI registration and multi-modal pelvic MR and CT registration tasks, demonstrating favorable performance compared to the state-of-the-art at the time.

Given the promising results GANs have demonstrated in image translation, i.e., synthesizing one image modality into another, researchers have made efforts to leverage their capabilities in addressing multi-modal image registration{~\citep{xu2020adversarial, wei2019synthesis, zheng2021symreg, han2022deformable, zhong2023united}}.
This approach involves first synthesizing multi-modal images into the same modality and then applying a registration network to perform the image registration task. In \citep{xu2020adversarial}, the authors tackled the challenge of multi-modal registration of CT and MR images using a CycleGAN-based approach to translate CT images into MR images.
To ensure that the translated images maintained anatomical consistency with the original images, the authors introduced additional loss functions, including MIND and identity loss, alongside the standard CycleGAN loss. 
They then employed a three-stage registration framework to align the original and translated images.
In the first stage, a U-Net-based registration network learned the multi-modal registration between CT and MR images.
In the second stage, a network with the same architecture learned the mono-modal registration between the translated CT and the target MR images.
Finally, the deformation fields created by both registration networks were fused using a convolutional layer to produce the final deformation field.
A similar concept was presented in \citet{wei2019synthesis},~where mutual information was used instead of MIND to enforce structural consistency.
\citet{zheng2021symreg}~integrated an image translation network within a GAN-based image registration framework, where the modality of the moving image was first translated to the modality of the target image before a registration network was applied to register the two images.
The discriminator in this approach acted as an image similarity metric for both the registration and image translation networks.
Additionally, this approach employed a symmetric pipeline that reversed the order of the moving and fixed images, ensuring symmetric consistency in the resulting synthesized and deformation images.
More recently, \citet{han2022deformable}~proposed tackling the multi-modal registration between CT and MR images using a dual-channel framework.
Within each channel, an imaging modality was transformed into a target modality using a probabilistic CycleGAN, which was then followed by a registration network that predicted the deformation in the target modality.
The deformation fields from both channels were then fused, taking advantage of the uncertainty weighting generated by the synthesis networks.
This proposed dual-channel framework can be trained end-to-end, resulting in improved registration accuracy and faster runtime compared to baseline methods.

Adversarial learning has also been employed for knowledge distillation, enabling the transfer of information from a larger teacher network to a smaller student network (\emph{i.e.}, in terms of the number of parameters).
\citet{tran2022light}~aimed to compress the size of a registration network by transferring information from a computationally expensive VTN~\citep{zhao2019unsupervised} to a smaller registration network with only one-tenth of its parameters.
The training process for the student network involved calculating a correlation-based image similarity measure~\citep{zhao2019unsupervised} between the warped image generated by the student network and the fixed image. 
Meanwhile, a discriminator was used to differentiate the deformation field created by the student network and the pre-trained teacher network.
After training, the teacher network was discarded, and only the lightweight student network was used for inference. 
Despite having only one-tenth of the network parameters, the lightweight registration network demonstrated comparable performance to baseline learning-based methods with larger parameter sizes, in terms of both anatomical overlaps and deformation smoothness.

\begin{figure}[!ht]
        \centering
        \includegraphics[width=0.3\textwidth]{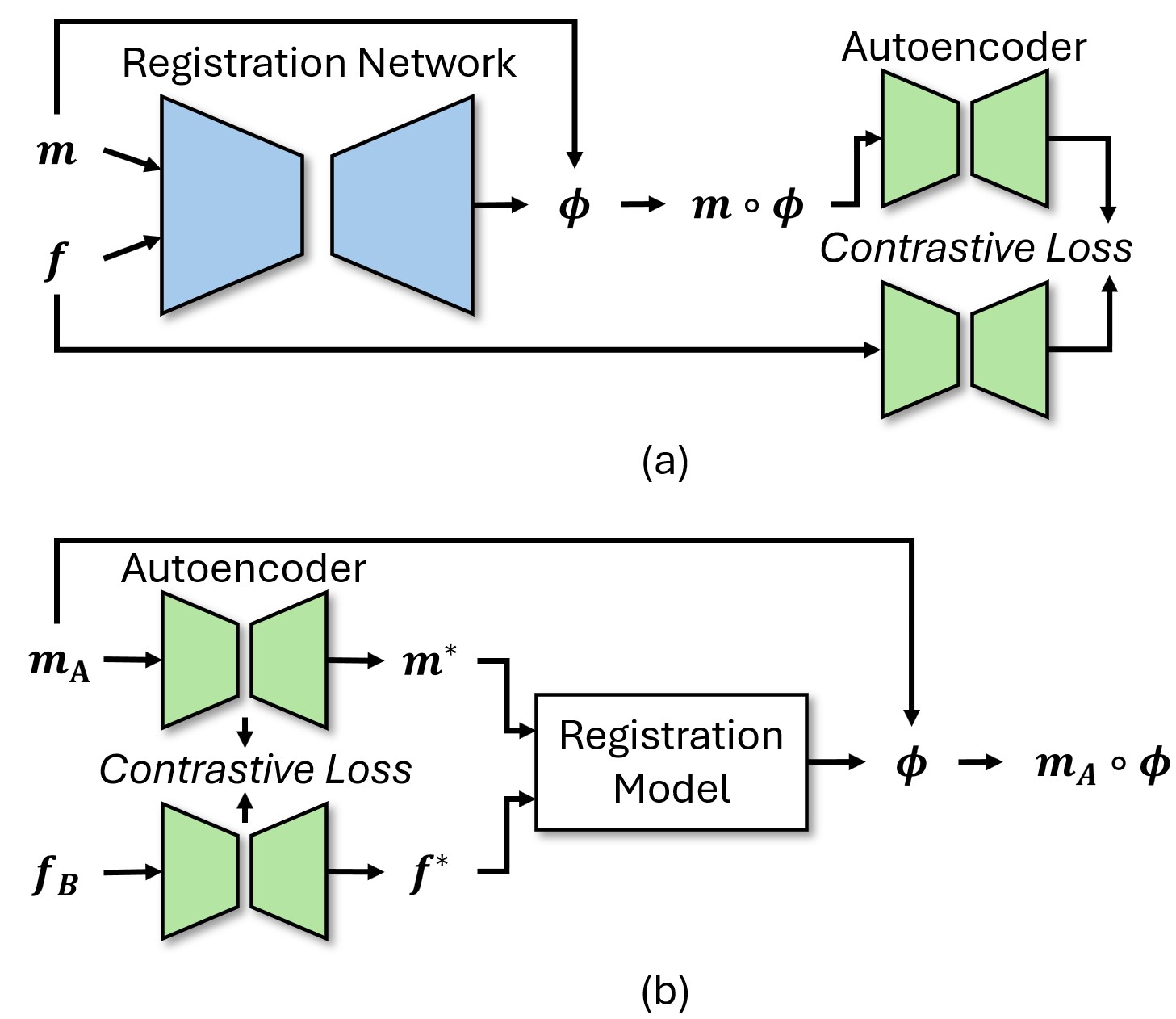}
        \caption{{Visual representation of contrastive
learning in medical image registration, with $\pmb{m}$ and $\pmb{f}$,
respectively, denoting the moving and fixed images, $\pmb{\phi}$
denoting the deformation field, and $\pmb{A}$ and $\pmb{B}$
corresponding to two different modalities. Panel~(a) illustrates the
application of contrastive learning as a similarity metric for
comparing the deformed moving image to the fixed image, as seen in~\citet{hu2019towards} and~\citet{dey2022contrareg}. Panel~(b) depicts the use of
contrastive learning to transform images from different modalities
into a unified feature representation, upon which the registration
model operates, as similarly employed in \citet{pielawski2020comir}, \citet{wetzer2023can}, and~\citet{casamitjana2021synth}.}}
        \label{f:network_contrast}
        \end{figure}

\subsection{Contrastive Learning}
\label{sec:contrast_learn}
The principle of contrastive learning enables DNNs to learn by comparing various examples instead of focusing on single data points independently.
This comparison process typically involves examining positive pairs of similar inputs and negative pairs of dissimilar inputs.
For a comprehensive understanding of this concept and a detailed overview of the evolution of contrastive learning, we recommend interested readers refer to~\citet{le2020contrastive}.
{Figure~\ref{f:network_contrast} provides a broad illustration of the contrastive learning designs used in medical image registration applications.} In the context of image registration, contrastive learning could be particularly beneficial as an alternative to using explicit image similarity metrics, which can be challenging to optimize due to their task-specific nature.
For example, different similarity metrics may be preferred for lung CT registration versus brain MRI registration or multi-modal versus mono-modal registration tasks.
Whereas, contrastive learning empowers the DNN to determine whether two images are registered or not without relying on a specific image similarity metric {(e.g., Fig.~\ref{f:network_contrast} (a))}, making it a more versatile approach for handling different registration tasks. 

\citet{hu2019towards}~were the pioneers in applying contrastive learning to multi-modal affine registration, concentrating on the inter-patient alignment of 2D CT and MR scans for patients with Nasopharyngeal Cancer.
Their method involved using an automatic keypoint detecting algorithm to identify keypoints in the CT and MR scans.
Subsequently, they extracted a patch centered on each keypoint and employed a Siamese network to minimize the contrastive loss, which minimized the distance between corresponding keypoints and maximized the distance between non-corresponding keypoints.
In the testing phase, after establishing correspondences between all keypoints in the CT and MR scans, the optimal affine transformation parameter was determined by means of least-squares fitting.
In another study, \citet{pielawski2020comir}~applied contrastive learning to transform multi-modal images into similar contrastive representations with equivariant properties.
Their method used two independent U-Nets to learn the representations for each modality such that the InfoNCE-based~\citep{oord2018representation} loss between the learned representations is minimized.
This minimization can be understood as maximizing the mutual information between the two learned representations.
Finally, conventional affine registration methods were used to align the learned representations as if they had undergone a mono-modal registration task.
\citet{wetzer2023can}~later investigated the contrastive learning approach proposed in \citet{pielawski2020comir}~to determine whether applying contrastive learning supervisions to the U-Nets' intermediate layers could improve multi-modal image registration performance.
However, they concluded that the best representations for the evaluated registration task were achieved when the contrastive loss was applied only to the features of the final layers.
\citet{casamitjana2021synth}~proposed a contrastive learning-based approach for multi-modal deformable registration.
They introduced a synthesis-by-registration method, where they trained a registration network for mono-modal registration on the target modality domain, and then froze the network's weight for training an image synthesis network using a loss function that leverages the registration network.
The image synthesis network's ability to accurately translate the moving image into the target modality directly influenced the performance of the registration network.
To enhance synthesis performance and ensure geometric consistency, a PatchNCE-based~\citep{park2020contrastive} contrastive loss was used, maximizing the mutual information between pre- and post-synthesis images at the patch level.
This method demonstrated promising results in multi-modal brain MRI registration applications, outperforming both MI-based registration and other image synthesis-based registration methods.
\citet{dey2022contrareg}~also addressed the multi-modal registration task using contrastive loss.
In their method, feature-extracting autoencoders were first pre-trained for each modality to derive modality-specific features.
These autoencoders were then used on the deformed moving image and the fixed image to extract features for a PatchNCE-based~\citep{park2020contrastive} contrastive loss.
In order to optimize contrastive learning, a single positive pair was sampled, corresponding to the multi-scale feature patches of the same spatial location across both modalities, while multiple negative pairs were sampled, corresponding to the feature patches of different spatial locations. 

Until now, the methods based on contrastive learning have been centered on multi-modal image registration. 
However, \citet{liu2020contrastive}~proposed the integration of contrastive learning in the intermediate stages of the network architecture for mono-modal brain MRI registration.
In their method, two identical ConvNet encoders of shared weights were applied to the moving and fixed images, each followed by a fully-connected layer to project ConvNet extracted features onto a latent space where the contrastive loss is applied.
The positive pair for computing the contrastive loss consists of the unregistered moving and fixed image pair, while any other pair apart from the current image pair under registration is considered a negative pair. 
In an extension of their work, \citet{liu2022pc}~proposed to compute the contrastive loss in a similar way, but between patches of the moving and fixed images. 
However, it is important to note that the positive pair used in these two methods contained structural dissimilarities as it was the unregistered image pair, as opposed to the registered images used in the methods mentioned earlier. 
The authors argued that this was because the image contents, including the number of brain structures, were consistent for brain registration.
Nonetheless, further research is needed to fully uncover the potential of these methods.

\begin{figure}[!ht]
        \centering
        \includegraphics[width=0.4\textwidth]{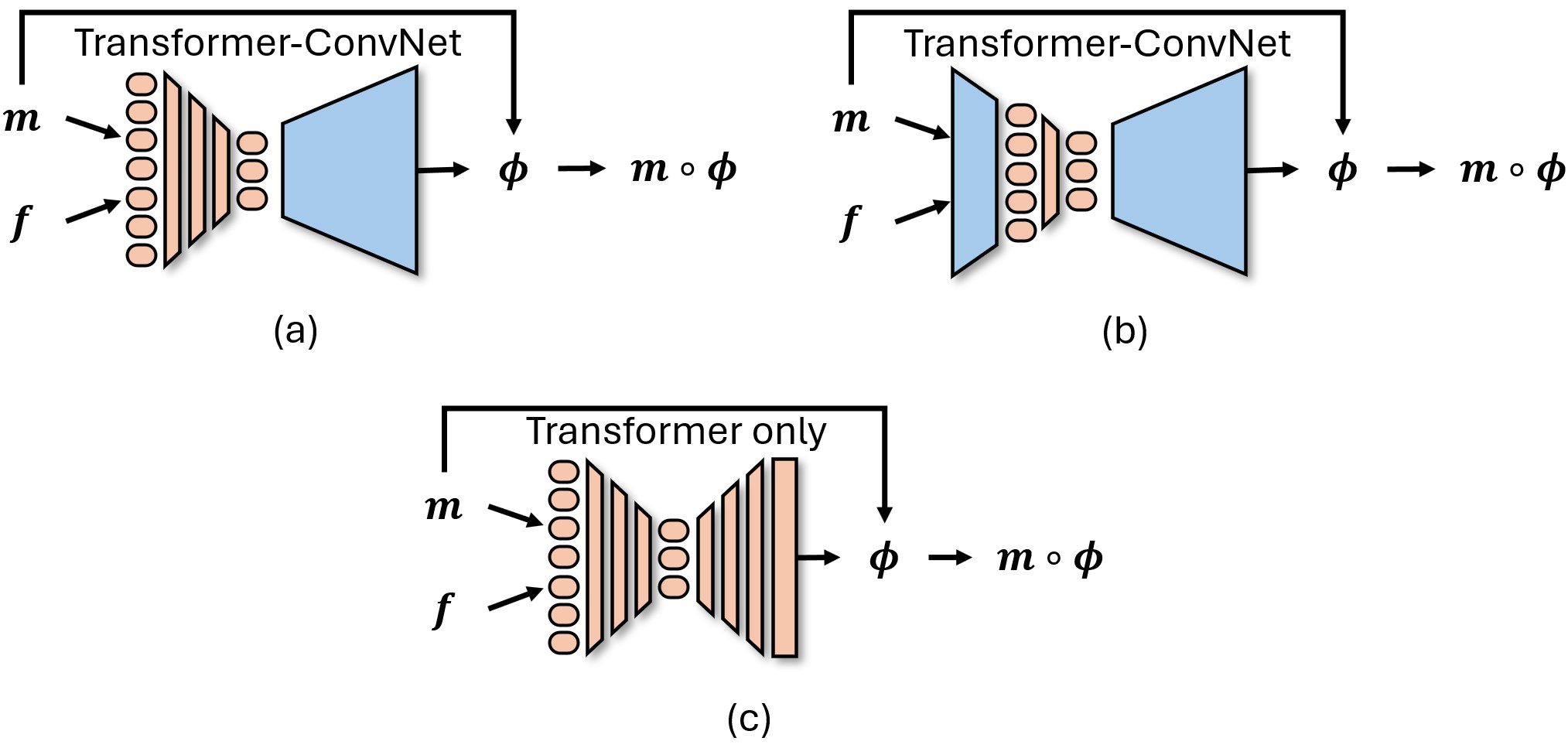}
        \caption{{Graphical representation of
Transformers used in medical image registration, with $\pmb{m}$ and
$\pmb{f}$ indicating the moving and fixed images, respectively, and
$\pmb{\phi}$ representing the deformation field. Panel~(a) displays a
Transformer-ConvNet hybrid architecture, where Transformers function
as encoders for extracting features, coupled with a ConvNet decoder
for generating the deformation field. A similar framework is adopted
in~\citet{chen2021vitvnet, chen2022transmorph, chen2023spr}, and~\citet{zhang2021learning}. Panel~(b) shows a design where a Transformer is
sandwiched between ConvNets, with the initial ConvNet acting as a
feature extractor and the Transformer applies self- or cross-attention
between image features, and the subsequent ConvNet decoder producing
the deformation field. This configuration is adopted
in~\citet{zhang2021learning}, \citet{chen2022deformer}, and~\citet{song2022cross}.
Panel~(c) illustrates the architecture where Transformers serve as
both the encoder and decoder, a design adopted
in~\citet{shi2022xmorpher}.}}
        \label{f:network_transformer}
        \end{figure}

\subsection{Transformers}
One of the key factors in designing ConvNets is the size of the receptive fields.
While incorporating consecutive convolutional layers and pooling operations can increase the theoretical receptive fields of ConvNets, its effective receptive fields are still limited~\citep{luo2016understanding}.
This makes them less effective at capturing long-range spatial correspondence, which is important to image registration since it aims to identify the correspondence between different parts of the images.
In contrast, Transformers are widely acknowledged for their superior ability to capture long-range dependencies and achieve exceptional performance when trained on large datasets~\citep{li2023transforming}.
Transformers differ from ConvNets in that they employ the self-attention mechanism, in which each local part of an image is compared in relation to the other parts, guiding the network on where to focus. 
Originally developed for natural language processing tasks~\citep{vaswani2017attention}, Transformers have recently become prevalent in various computer vision applications~\citep{dosovitskiy2021an, liu2021swin, liu2022video, chen2021crossvit, yuan2021tokens, dong2022cswin, carion2020end}. 
Inspired by their success, many Transformer-based models have been proposed and have demonstrated promising performance in medical imaging applications. 
{Figure~\ref{f:network_transformer} provides a graphical illustration of the Transformer-based DNNs used in medical image registration.} 
For a comprehensive review of the current Transformer-based models in medical imaging {in general}, interested readers are directed to a review paper by~\citet{li2023transforming}.
Despite their potential, Transformers have certain drawbacks, such as larger computational complexity and a lack of inductive bias when compared to ConvNets, hindering the training process.
To address these shortcomings, Transformers are commonly used in conjunction with ConvNets in medical image registration applications.
{Initially, the field saw the introduction of self-attention-based Transformers for these tasks.
More recently, cross-attention-based Transformers emerged, demonstrating potential advantages.
Given that registration involves identifying correspondences between images, cross-attention, which focuses on the features of one image in relation to another, may enhance this identification process.
We start by examining self-attention-based Transformers and then proceed to discuss the developments in cross-attention-based Transformers specifically designed for medical image registration.}
\paragraph{Self-attention}
\citet{chen2021vitvnet}~were the first to adopt Transformers for registration-based tasks. 
They proposed ViT-V-Net, which employs a ConvNet for extracting high-level features, followed by a Vision Transformer~(ViT)~\citep{dosovitskiy2020image} and a ConvNet decoder to generate a dense displacement field.
Subsequently, they proposed TransMorph~\citep{chen2022transmorph}, which employs a Swin Transformer~\citep{liu2021swin} in the encoder, replacing the ConvNet feature extractor and ViT.
TransMorph is capable of both affine registration and deformable registration.
The study provided empirical evidence that Transformer-based models have larger effective receptive fields than baseline ConvNets.
In inter-subject and atlas-to-subject brain MRI registration, as well as XCAT-to-CT abdomen registration applications, TransMorph achieved significantly improved registration performance when compared to top-performing traditional and ConvNet-based registration models.
\citet{zhang2021learning}~proposed DTN, which consists of two encoder branches with identical architecture.
Each branch contains a ConvNet feature extractor and a ViT.
In DTN, the moving and fixed images are first fed consecutively into one encoder branch, then concatenated and sent to the other branch.
The encoder outputs are then concatenated and sent to a ConvNet decoder to produce a deformation field.
\citet{mok2022affine}~introduced a Transformer encoder, C2FViT, specifically designed to tackle the affine registration problem.
Their Transformer architecture was inspired by ViT, but with augmented patch embedding and feed-forward layers to introduce locality into the model.
C2FViT adopts a coarse-to-fine strategy with an image pyramid for affine registration.
The registration process is carried out in multiple stages of ViTs with identical architectures, each corresponding to a different resolution of the fixed and moving images.
The affine parameters are estimated in each stage, and the moving image is affine-transformed using the parameters from the previous stage to refine the registration progressively. 
C2FViT was evaluated on several benchmark datasets and demonstrated superior performance compared to multiple ConvNet-based and traditional affine registration methods. 
\citet{chen2022deformer}~proposed a Deformer module, which leverages the attention mechanism on feature maps produced by a ConvNet encoder.
The authors argued that the Deformer module facilitated the image-to-spatial transformation mapping process by estimating the displacement vector prediction as a weighted sum of multiple bases.
Employing a coarse-to-fine strategy, the proposed model outperformed both the ConvNet and Transformer models in the comparative analysis. {\citet{yin2023pc} introduced PC-Reg, which uses a Swin Transformer similar to that in TransMorph \citep{chen2022transmorph} but incorporates modifications in patch partitioning through two convolutional residual blocks. The processed features are subsequently directed to a specially designed ConvNet decoder tasked with predicting the a deformation field. The decoder estimates the final deformation field using a coarse-to-fine approach, progressively upsampling the lower-resolution field and employing a separate module at each scale to refine the field by estimating the residuals. This refinement module draws inspiration from the predictor-corrector method commonly used in numerical methods for differential equations. PC-Reg shows promising results, performing competitively in brain MRI and abdominal CT registration tasks.}
\paragraph{Cross-attention}
\citet{song2022cross}~introduced Attention-Reg, a model that adopts cross-attention to correlate features extracted from multi-modal input images by a ConvNet encoder. 
To expedite the training process, they applied a contrastive pre-training strategy to the ConvNet feature extractor, allowing for the extraction of similar features from different modality images. 
The Dice loss was used as the multi-modal similarity measure, and they developed both rigid and deformable variations of the model.
The results showed that Attention-Reg performed favorably against several learning-based rigid and deformable registration models.
Similarly,  \citet{shi2022xmorpher}~introduced XMorpher, a full Transformer architecture featuring dual parallel feature extractors that exchange information via a cross-attention mechanism. 
The cross-attention module developed in their study is based on a Swin Transformer, where attention is computed between base windows of one image and searching windows of another image with differing sizes. 
This cross-attention mechanism exhibited improved performance compared to self-attention-based Transformers and ConvNet models. 
\citet{chen2023deform}~made further improvements to the cross-attention technique used in XMorpher. 
They proposed a novel deformable cross-attention module that enables tokens to be sampled from regions beyond the conventional rectangular window, while also reducing computational complexity. A lightweight ConvNet was introduced to deform the sampling window in a reference.
The attention is then computed between the tokens sampled from the deformed window in the reference and those sampled from a rectangular window in a base. This enables tokens sampled from a larger reference region to guide the network on where to focus within each local window in the base. 
The proposed network includes two encoding paths.
In one path, the moving and fixed images are used as the base and reference, respectively.
In the other path, the roles of the base and reference are switched, with the moving image used as the reference and the fixed image used as the base.
A ConvNet decoder then fuses the features extracted from the two encoders to generate a deformation field. 
Their method was evaluated on brain MRI registration tasks, and it performed favorably against self-attention, cross-attention, and ConvNet-based models. 
{Adopting a similar strategy of incorporating cross-attention for image registration, \citet{chen2023transmatch} developed TransMatch. This model employs a Transformer encoder with dual-stream modules, each alternating between windowed self-attention and cross-attention to enhance the matching of image features through dot-product attention. This process aims to establish more accurate spatial correspondences both between and within images. A ConvNet decoder subsequently processes the concatenated features from both streams to estimate a deformation field. TransMatch was evaluated on mono-modal brain MRI image registration tasks, where it outperformed both traditional ConvNets and self-attention-based Transformers.}
\citet{liu2022coordinate}~proposed im2grid, a model that uses cross-attention to explicitly guide the neural network in comprehending the coordinate system for image registration, which is usually learned implicitly from data.
Their approach uses ConvNet encoders to independently extract hierarchical features from the fixed and moving images.
Subsequently, their proposed coordinate translator block computes a softmax score function by comparing the extracted fixed image feature at a voxel location with the features of the moving image within a search window. 
Spatial correspondence between the voxel location in the fixed and moving images is established by linearly combining the coordinates of all voxel locations weighted by the score function.
Their approach is implemented as cross-attention with coordinates as one of the inputs.
This model was evaluated on inter-patient brain MRI registration tasks using publicly available datasets and demonstrated superior performance compared to the comparative ConvNets and Transformer-based models. 
{\citet{wang2023modet},~inspired by the im2grid introduced in~\citet{liu2022coordinate}, aimed to calculate matching scores to weight deformations but also enhanced the the weighting process by noting that feature maps at coarse-level resolutions typically exhibit different motion modes.
Specifically, the authors developed a motion decomposition Transformer, named ModeT.
This Transformer decomposes the feature maps of the moving and fixed images into multi-head feature representations using a linear projection layer followed by LayerNorm.
Cross-attention is then applied between the feature at each spatial location in the fixed image and the features of the neighboring locations in the moving image, employing a technique similar to that used in~\citet{liu2022coordinate}.
This process yields a series of deformation fields, with the count equaling the number of heads used. Subsequently, a competitive weighting module, which consists of a shallow ConvNet ending with a SoftMax activation function, is applied to the sequence of deformation fields. These fields are then aggregated to form a deformation field at each resolution.
Finally, the final deformation field at the full resolution is generated by sequentially composing the upsampled deformation fields from each resolution. This model was tested on brain MRI image registration tasks and showed superior performance compared to several state-of-the-art methods at the time.
In~\citet{khor2023anatomically}, the authors proposed AC-DMiR for joint learning of registration and segmentation. 
Specifically, AC-DMiR adopts two Transformer-based backbone networks, similar to the one used in TransMorph~\citep{chen2022transmorph}, to generate initial predictions: a deformation field for registration and segmentation for the warped moving image.
Subsequently, the feature maps from the registration and segmentation networks are coupled via cross-task attention modules.
These modules employ cross-attention to focus on features from one task to another, followed by self-attention and a decoder to estimate the final deformation field.
This integration allows the correlation of anatomical variations and alignment information from the segmentation network with the registration network, thereby enhancing the estimation of the deformation.
When compared to traditional ConvNet-based and other Transformer-based methods, AC-DMiR shows enhanced registration accuracy in terms of Dice coefficients and comparable deformation regularity on both Brain MRI and Uterus MRI image registration tasks.}

The mechanisms of Transformers have inspired various ConvNet designs in computer vision, leading to a debate on whether Transformers could replace ConvNets for image-related tasks~\citep{li2023transforming}.
ConvNet models such as ConvNeXt~\citep{liu2022convnet} and RepLKNet~\citep{ding2022scaling} have built upon Transformer concepts and demonstrated performance comparable to Transformers.
Inspired by these models,  \citet{jia2022u}~proposed a U-Net with increased kernel sizes to expand the effective receptive field of the U-Net.
Their method compared favorable against several Transformer-based registration methods. 
Currently, ConvNets still possess inherent advantages over Transformers, such as their invariance to input image sizes and the incorporation of inductive bias due to the nature of the convolution operation.
Therefore, there has been a growing interest in advancing ConvNets using Transformer concepts in computer vision.
It is anticipated that further research in this area will lead to improved ConvNet architectures for medical image registration applications.

\begin{figure}[!ht]
        \centering
        \includegraphics[width=0.35\textwidth]{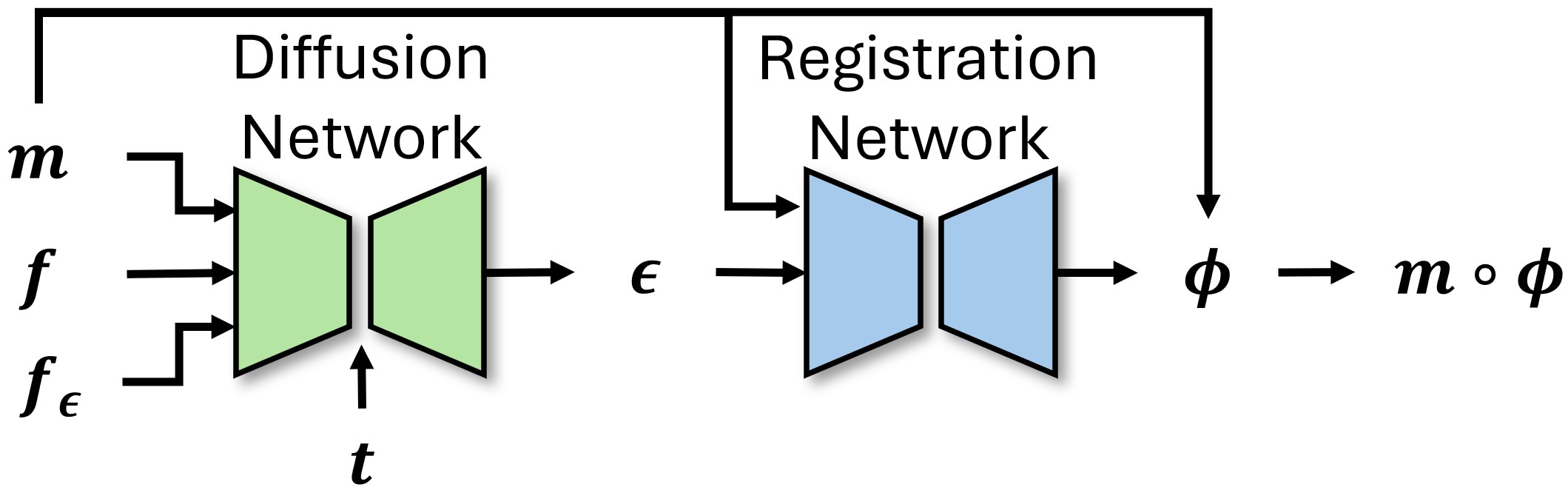}
        \caption{{Visual depiction of the diffusion
model in medical image registration, where $\pmb{m}$ and $\pmb{f}$ are
the moving and fixed images, respectively, $\pmb{\phi}$ is the
deformation field, $\pmb{\epsilon}$ symbolizes the score function, and
$\pmb{t}$ represents an imaginary time step in the diffusion process.
This approach adopts a diffusion model to create a conditional score
function associated with the time step. This function, in conjunction
with the moving image, facilitates the generation of continuous
deformations going from the moving image to the fixed image. This
approach is adopted in~\citet{kim2022diffusemorph}.}}
        \label{f:network_diffusion}
        \end{figure}

\subsection{Diffusion Models}
In recent years, diffusion models~\citep{sohl2015deep, ho2020denoising} have garnered significant research interest in computer vision.
Initially designed for generative tasks, such as image synthesis, inpainting, and super-resolution, diffusion models have now been widely explored in various applications in the field of medical image analysis~(see~\citet{kazerouni2022diffusion} for a survey). 
In contrast to other generative models like GANs and VAEs, which are either confined to data with limited variability or generating low-quality samples~\citep{ho2020denoising, kazerouni2022diffusion}, diffusion models have no such restrictions, making them an attractive alternative.
The goal of diffusion models is to use the known forward process of gradual diffusion of information caused by noise to learn the reverse process of recovery of information from noise.
The forward process is similar to the behavior of particles in thermodynamics, where particles spread~(\emph{i.e.}, diffuse) from areas of high concentration to those of low concentration~\citep{kirkwood1960jcp, sohl2015deep}.
The existing diffusion models use iterative steps of diffusion, which can include up to several thousand steps, to carry out the diffusion process. As a result, inference with these models, which requires the reverse diffusion process, is time-consuming.
{To date, only a few studies have incorporated diffusion models into medical image registration~\citep{kim2022diffusemorph,qin2023fsdiffreg}, and Fig.~\ref{f:network_diffusion} provides a graphical representation of these diffusion model-based models.}
In~\citep{kim2022diffusemorph}, the authors proposed DiffuseMorph, which involves a diffusion network and a deformation network.
The diffusion network learns a conditional score function~(\emph{i.e.}, the added noise), while the deformation network uses the latent feature in the reverse diffusion process to estimate the deformation field.
The registration process of DiffuseMorph is a one-step procedure as the fixed image is the target image at the end of the reverse diffusion process~(\emph{i.e.}, $t=0$), and it is already given. 
As a result, there is no need for time-consuming reverse diffusion steps to synthesize a target image from the moving image.
Furthermore, DiffuseMorph offers the added capability of producing continuous deformations through the interpolation of the learned latent space.
The method demonstrated promising results when compared to several ConvNet-based methods on a publicly available Cardiac MRI dataset and a human facial expression dataset.
However, since their forward process adopts the strategy of adding Gaussian noise to the fixed image, their diffusion network learns a conditional score function for the fixed image. {This score function is then used in the registration network, rather than directly inputting the fixed image, allowing the score function to carry semantic information. This results in a method that differs from the conventional diffusion models.} {In \citep{qin2023fsdiffreg}, the authors proposed an alternative diffusion model based on \citep{kim2022diffusemorph} for medical image registration, where the score function is used as a weighting map for the image similarity measure. However, this formulation deviates from conventional diffusion models, where the score function is typically modeled as Gaussian noise.}

\subsection{Neural ODEs}
\label{sec:nodes}
Inspired by Euler's method for discretizing the derivative of ordinary differential equations~(ODEs) into discrete time step updates, \citet{chen2018neural} proposed a new family of DNN models called Neural ODEs. 
In their method, DNN elements that progressively update their input (\emph{e.g.}, residual connections, or recurrent networks) are interpreted as updates of time steps in Euler's method.
Consequently, a chain of these elements in a neural network is essentially a solution of the ODE with Euler's method of the form:
\begin{equation}
    \frac{dh(t)}{dt} = f_\theta(h(t), t),
\end{equation}
and
\begin{equation}
    h(t+1)=h(t)+f_\theta(h(t), t),
\end{equation}
where $h(t)$ represents the $t$-th element, which may be a residual block or a network.
The final output at $t=T$ can be computed by integrating  $f$ over the time interval $[0, T]$, which is evaluated by a numerical solver taking many small time steps, thus approximating a neural network with infinite depth.

The first application of the NeuralODE framework for medical image registration was introduced by~\citet{xu2021multi}.
They proposed MS-ODENet, which parameterizes $h$ at the final time point $T$ (\emph{i.e.}, $h(T)$) as the deformation field that warps the moving image to the fixed image, and $\frac{dh(t)}{dt}$ as the small increment of deformation produced by a network at state $t$ from the preceding state $h(t-1)$.
To alleviate the computational burden of numerical solvers and accelerate the runtime, they proposed solving ODEs at different resolutions in a coarse-to-fine manner.
However, the loss function, consisting of a similarity measure and a deformation regularizer, is applied only to the final deformation field $h(T)$.
Similarly,~\citet{wu2022nodeo} proposed NODEO, which formulated $h(t)$ as the voxel movement at time $t$ and the trajectory of the movement as the solution to the ODE.
Drawing inspiration from dynamical systems, they expressed the ODE as $\frac{dh(t)}{dt} = \mathcal{K}v_\theta(h(t), t)$, where $\mathcal{K}$ is a Gaussian smoothing kernel, $v_\theta$ denotes the velocity of the voxel movement produced by a neural network, and the initial condition $h(0)$ is an identity.
It is noteworthy that this formulation bears similarities to LDDMM~\citep{beg2005computing}, an influential optimization-based method that considers image registration as an energy-minimizing flow of particles over time. 
In contrast to MS-ODENet~\citep{xu2021multi}, which applies loss solely to the deformation at $t=T$, NODEO optimizes image similarity at each $t$ while minimizing the energy of the flow and encouraging spatial smoothness and regularity of the velocity fields through the Gaussian kernel, diffusion regularizer, and Jacobian determinant loss. 
The authors compared NODEO to various widely-used traditional methods and a ConvNet model on brain MRI registration tasks.
It demonstrated superior registration performance measured by Dice while attaining diffeomorphic registration.
{\citet{joshi2023r2net}~introduced R2Net, which integrates both time-varying and time-stationary velocity fields using a neural ODE approach. This method employs Lipschitz continuous residual networks to estimate smooth intermediate velocity fields from an initial momentum, which is predicted by a U-Net-based ConvNet. The underlying theory relies on the Cauchy-Lipschitz theorem, positing that Lipschitz continuous integration ensures a well-defined mapping within the space of diffeomorphisms. The Lipschitz continuous residual networks comprise $N$ blocks, each corresponding to a time step in the integration process. In the time-stationary configuration, these $N$ blocks share weights, contrasting with the time-varying setting where the weights are not shared for each block. R2Net was evaluated against both optimization-based and learning-based diffeomorphic registration methods. It showed competitive performance, producing notably smoother deformations compared to the other methods.}

\begin{figure}[!ht]
        \centering
        \includegraphics[width=0.42\textwidth]{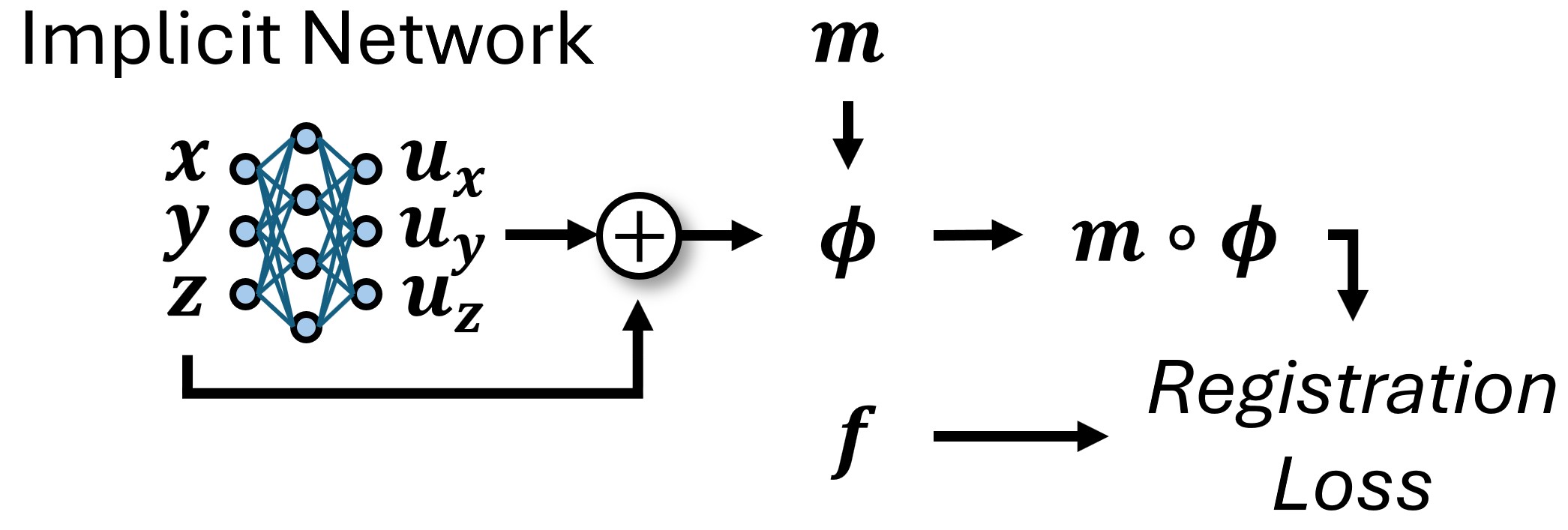}
        \caption{{Graphical depiction of implicit
neural representations in medical image registration, with $\pmb{m}$
and $\pmb{f}$ denoting the moving and fixed images respectively, and
$\pmb{\phi}$ denotes the deformation field. The set $\{\pmb{x},
\pmb{y}, \pmb{z}\}$ represents the grid in three distinct spatial
dimensions, while the associated $\pmb{u}$'s indicate the displacement
vectors. Implicit neural representations are typically employed within
a pair-wise optimization framework. In this setup, the DNN (typically
an MLP) receives coordinates from an image grid as input and outputs
the corresponding displacement vector for those coordinates. These
vectors are then iteratively refined through the pair-wise
optimization of the registration objective function. Similar
frameworks are found in~\citet{han2023diffeomorphic}, \citet{wolterink2022implicit}, \citet{byra2023implicit}, and~\citet{van2024deformable}.}}
        \label{f:network_implicit}
        \end{figure}

\subsection{Implicit Neural Representations}
Image registration can be formulated as an implicit problem of the form:
\begin{equation}
\label{eqn:imp_func}
    \mathcal{C}(\pmb{x}, \psi)=0,\ \ \psi:\pmb{x}\rightarrow \psi(\pmb{x}),
\end{equation}
where $\pmb{x}\in\mathbb{R}^{2,3}$ is the 2D or 3D spatial coordinate (\emph{i.e.}, from an integer grid), and $\psi$ represents a neural network that maps each coordinate $\pmb{x}$ to a value of interest, subject to the constraint $\mathcal{C}$.
In the context of image registration, $\psi$ typically maps the coordinate $\pmb{x}$ to its deformation $\psi(\pmb{x})$, while $\mathcal{C}$ comprises a similarity measure and a deformation regularizer.
The neural network $\psi$ can be considered as an implicit function of $\pmb{x}$, defined by the relation modeled by $\mathcal{C}$ (Eqn.~\ref{eqn:imp_func}).
This concept is commonly referred to as \textit{implicit neural representations} {(INRs)} in computer vision~\citep{sitzmann2020implicit, mescheder2019occupancy, niemeyer2019occupancy, mildenhall2021nerf}.
Although $\pmb{x}$'s used during training are discrete, the implicit function $\psi(\pmb{x})$ parameterized by a neural network{, often a multi-layer perceptron (MLP),} is a continuous and differentiable function.
As a result, {INRs} provide a more compact representation of a continuous function and facilitate smooth manipulation of that function. {Note that the input to the MLP is an image grid, which remains invariant across different image datasets. Consequently, INRs are commonly used in pairwise optimization-based registration~\citep{han2023diffeomorphic,wolterink2022implicit, byra2023implicit,van2024deformable}, as graphically illustrated in Fig.~\ref{f:network_implicit}.}

\citet{han2023diffeomorphic}~proposed to parameterize a continuous deformation field using a MLP introduced in \citep{sitzmann2020implicit}, given an integer grid representing the spatial coordinates of the voxels.
The MLP thus serves as the implicit function of the integer grid.
Since the MLP is not conditioned on the images and the only input is the coordinates that are deterministic for all images of the same resolution, optimization of the MLP is carried out iteratively and pair-wise for each image pair (similar to how the traditional registration methods are performed).
To further improve the registration performance, the authors proposed a cascade framework that combines the benefits of learning-based registration DNNs with the optimization-based {INRs} provided by the MLP.
Within this framework, the learning-based DNN predicts an initial deformation field, while the MLP produces the residual deformation that refines the initial deformation field, leading to an enhanced overall registration performance.
{Leveraging the capability of INRs to represent continuous image grids, \citet{wolterink2022implicit} introduced an implicit DIR model that facilitates continuous differentiation of the deformation. 
This feature is crucial for calculating deformation regularization terms, such as the first derivative in diffusion regularizer and the second derivative in bending energy. This approach marks a significant shift from conventional DL-based registration models, which typically predict a discrete deformation field and hence rely on finite differences for differentiation. The authors explored various activation functions within the MLP framework and found that periodic activation functions enable the network to represent high-frequency signals necessary for capturing small local deformations. In contrast, ReLU activation functions tend to favor low-frequency functions. The proposed model showed promising results on the DIR-LAB dataset, surpassing DL-based methods. However, it demonstrated slightly inferior performance compared to traditional optimization-based methods on this dataset. 
Building on this foundation, \citet{byra2023implicit} further refined the model by proposing the decomposition of a moving image into a residual image and a support image. The residual image accounts for image artifacts and texture patterns that differ from the fixed image, whereas the support image aids in enhancing registration performance. These images are estimated using two implicit networks designed to minimize an exclusion loss, which encourages the gradient structures of these networks to be decorrelated, as suggested by~\citet{gandelsman2019double}. Image registration is then conducted by jointly minimizing a reconstruction loss, ensuring that the sum of the support and residual images reconstructs the moving image, along with two image dissimilarity measures applied to the deformed moving and support images against the fixed image, a deformation regularizer, and the exclusion loss. This method was evaluated on \textit{in situ} hybridization image registration, where it demonstrated improved performance over both the earlier proposed INRs-based and CNN-based models, as well as traditional optimization-based methods.
In~\citet{van2024deformable}, the author used IRN to parameterize the temporal motion of the intestine.
Instead of using spatial coordinates from an integer grid, they align the coordinate system with the tangent space of the anticipated dominant motion, which conforms to the orientation of the intestines.
The authors contend that this adjustment simplifies the complexity of the IRN and leads to better registration performance.
}
{It is important to note to recognize that these aforementioned INRs-based registration methods, much like traditional approaches, suffer from a common limitation: the optimization is conducted on a pairwise basis without learning from a larger dataset. As a result, these methods are unable to take advantage of the guidance offered by anatomical label maps when such maps are unavailable during the inference process.}
Meanwhile, \citet{sun2022topology}~applied {INRs} to a task of organ shape registration.
Their approach was based on the idea of DeepSDF \citep{park2019deepsdf}, where an auto-decoder maps a latent code representing a unique organ shape and the 3D coordinates of a sampled point to a signed distance function~(SDF).
The value of an SDF determines whether the point lies inside ($<0$), outside ($>0$), or on the surface ($=0$) of the shape, consequently providing an implicit description of the organ shapes.
The resulting SDF is a continuous function, and the auto-decoder serves as the implicit neural representation of the discrete coordinates.
To register points from different organ shapes, the authors modeled the trajectory of the point movement in space as the solution to an ODE, akin to the formulation proposed in NODEO~\citep{wu2022nodeo}.
In this formulation, the time derivative corresponds to the velocity of the point movement at time $t$.
The authors solved this ODE using a NeuralODE solver (as briefly discussed in section \ref{sec:nodes}), resulting in a diffeomorphic mapping between shapes. 

\begin{figure}[!ht]
        \centering
        \includegraphics[width=0.75\columnwidth]{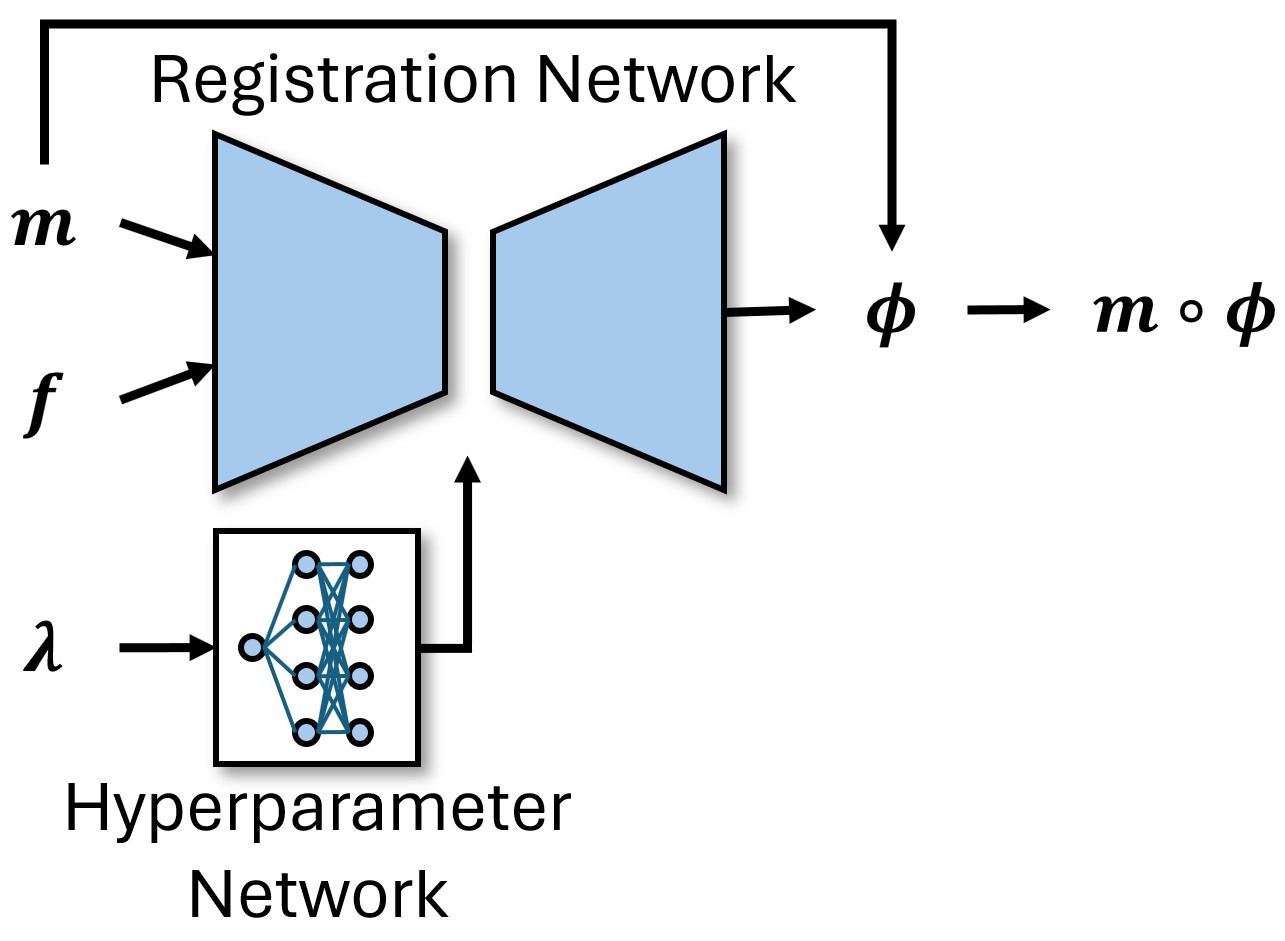}
        \caption{{Illustration of hyperparameter
conditioning applied in medical image registration, with $\pmb{m}$ and
$\pmb{f}$ denoting the moving and fixed images, respectively, and
$\pmb{\phi}$ indicating the deformation field. Here, $\pmb{\lambda}$
is the hyperparameter for tuning the loss function, commonly
associated with the regularization hyperparameter to control the
smoothness of the deformation field. The purpose of this strategy is
to facilitate the adjustment of hyperparameters at test time,
circumventing the need for time-intensive retraining. Hyperparameters
can either be integrated directly into the network architecture~\citep{mok2021conditional, chen2023spr} or input into a separate DNN
that then produces the parameters for the registration network~\citep{hoopes2022hyper}.}}
        \label{f:network_hyper}
        \end{figure}

\subsection{Hyperparameter Conditioning}
Inspired by HyperNetworks~\citep{ha2017hypernetworks} and Hyperparameter Optimization~\citep{franceschi2018bilevel}, recent research has introduced methods that integrate hyperparameters directly into the architecture of the registration DNNs{, which is graphically depicted in Fig.~\ref{f:network_hyper}.} This allows for the capturing of a wide range of hyperparameters within a single training process, consequently speeding up the hyperparameter tuning process without requiring multiple networks to be trained from scratch for each hyperparameter value. In the training process of these methods, a distinct hyperparameter value is randomly selected, and the network generates a deformation field associated with that value. Subsequently, the registration loss is calculated using the same hyperparameter value, which is then used to update the network parameters. The hyperparameter being conditioned typically relates to the weight of the deformation regularizer, which affects the smoothness of the deformation produced by the network.

\citet{hoopes2022hyper}~introduced HyperMorph, which is based on the concept of HyperNetworks~\citep{ha2017hypernetworks}. HyperMorph comprises two ConvNets: a hypernetwork and a U-Net-like registration network (\textit{i.e.}VoxelMorph~\citep{balakrishnan2019voxelmorph}).
The hypernetwork estimates the weights of the U-Net based on the provided hyperparameter value for the diffusion regularizer, while the U-Net generates a deformation field to warp the moving image.
In each training step, the hyperparameter value is randomly sampled from a uniform distribution, and the loss is computed using the same sampled value.
After training, the best-performing hyperparameter value is acquired using gradient descent.
In this process, the network weights are fixed, and an optimizer iteratively updates the hyperparameter based on a target objective function (commonly the Dice score) applied to a validation dataset.
In a parallel work, \citet{mok2021conditional}~proposed conditioning the regularization hyperparameter through conditional instance normalization~\citep{dumoulin2017a}. 
In this approach, the feature map statistics within the regularization network are normalized and shifted according to two affine parameters.
These affine parameters are generated by a lightweight mapping network, which takes the sampled hyperparameter value as input. 
Later, \citet{chen2023spr}~expanded the conditional instance normalization to a conditional layer normalization for application in Transformer-based registration models. The training processes in both~\citep{mok2021conditional, chen2023spr} are similar to the one used in HyperMorph, where the hyperparameter value is sampled from a uniform distribution and then employed for loss computation. However, it is worth noting that \citet{mok2021conditional} and \citet{chen2023spr} obtain the best-performing hyperparameter value through a grid search, whereas HyperMorph acquires it via gradient descent.

\subsection{Anatomy-aware Networks}
{Given the widespread success of DNN methods in medical image segmentation, anatomical label maps can now be relatively easily obtained using a segmentation network. As a result, some registration models have started to incorporate this prior anatomical information into their design to enhance registration accuracy, moving beyond simply using them as components of a loss function during training.}

{In~\citet{su2023nonuniformly}, the authors proposed using anatomical label maps to extract a set of nonuniformly spaced control points, sampled along the contours of anatomical structures. These control points allow for more precise deformation of anatomical structures, particularly at their boundaries, compared to uniformly distributed control points on image grids. The method involves extracting intensity features, which are generated by a ConvNet encoder, alongside spatial features, both of which are aligned with the positions of the sampled control points. In another approach, }to facilitate a spatially discontinuous deformation, which is important for many registration applications as delineated in Section~\ref{sec:loss}, \citet{chen2021deep2}~proposed an alternative approach. 
Rather than employing a discontinuity-permitted deformation regularization (as briefly mentioned in Section~\ref{sec:loss}), the authors proposed using anatomical label maps to segregate the moving and fixed images into different regions of interest and subsequently generate deformation fields for each region using multiple registration networks.
These deformation fields are then combined to yield a final deformation via addition. However, this method has an immediate drawback in necessitating the anatomical label maps throughout both the training and inference stages.
When label maps are not available, this method becomes infeasible.

\subsection{Correlation Layer}
Optical flow is the name given by the computer vision community to image registration.
In learning-based optical flow, it is common to employ a correlation layer~\citep{dosovitskiy2015flownet} to aid neural networks in pinpointing explicit correspondences between points in images.
This involves computing the correlation between the neighboring features of a spatial location in the moving image and the neighboring features of a range of spatial locations in the fixed image.
The correlation is computed between two feature patches centered at $\pmb{x}_m$ and $\pmb{x}_f$ in the moving and fixed images, respectively, using the following equation:
\begin{equation}
\label{eqn:corr_layer}
c(\pmb{x}_m, \pmb{x}_f)=\sum_{\pmb{o}\in[-k, k]} \langle F_m(\pmb{x}_m + \pmb{o}), F_f(\pmb{x}_f + \pmb{o}) \rangle,
\end{equation}
where $F_m$ and $F_f$ denote the feature patches of the moving and fixed images, respectively, and $k$ defines the patch size.
The selection of locations $\pmb{x}_m$ and $\pmb{x}_f$ is based on a maximum displacement $d$, meaning that for each $\pmb{x}_m$, the range of $\pmb{x}_f$ is limited to the locations that are at most $d$ distance away.
The output of the correlation layer is a set of correlation values that represent the correlation between one feature patch in the moving image and another feature patch in the fixed image.
The output has a size of $H\times W\times D\times d$, where $H\times W\times D$ represents the size of the feature maps. 

Although the concept of directing networks with explicit correspondences between voxels or patches has been employed in computer vision since 2015, it was only recently embraced in medical image registration.
This delay can be attributed to the potential computational challenges introduced by Eqn.~\ref{eqn:corr_layer}.
Since medical images are typically volumetric, the search space for each voxel location would be in a 3D volume, quickly becoming unmanageable as the search distance $d$ grows.
\citep{heinrich2019closing}~was the first to implement a correlation layer in their network design by introducing the PDD-net.
Instead of calculating the scalar product between two features as done in Eqn.~\ref{eqn:corr_layer}, PDD-net computes the correlation as the mean squared error between feature patches centered at each control point in the moving and fixed images:
\begin{equation}
\label{eqn:corr_layer_mse}
c(\pmb{x}_m, \pmb{x}_f) = \sum_{\pmb{o}\in[-k, k]} \Vert F_m(\pmb{x}_m + \pmb{o}) - F_f(\pmb{x}_f + \pmb{o})\Vert^2.
\end{equation}
Moreover, in their correlation layer, the search distance is represented by a 3D matrix, $\pmb{d}^3$, with each element in the vector, $\pmb{d}$, defining a discrete displacement distance from the current center of the feature patch.
This correlation layer produces a 6D matrix, where the first three dimensions outline the shape of the feature maps, and the final three dimensions describe the shape of the search space.
Due to the sparsity of the control points in comparison to the image size, the computational burden of this correlation layer remains relatively low. The correlation layer is applied to features independently extracted from the moving and fixed images using a ConvNet that incorporated deformable convolutional layers as introduced in~\citet{heinrich2019obelisk}.
Subsequently, min-convolutions and mean-field inference are employed to spatially smooth the dissimilarities produced by the correlation layer.
A softmax operation is then applied to the 6D matrix, converting the dissimilarities into pseudo-probabilities. 
The displacement field is subsequently generated by multiplying the probabilities with the displacement distance in $\pmb{d}^3$, resulting in a weighted average of these probabilistic estimates for the 3D displacement field.
The deformation field is then trilinearly interpolated to align with the image resolution. 
\citet{heinrich2020highly}~later extended this approach by proposing a 2.5D approximation of the quantized 3D displacement, significantly reducing the memory burden of the original Pdd-net.
Instead of creating a 6D dissimilarity matrix, they generated three 5D matrices (i.e., the 2.5D dissimilarity matrices), with each matrix representing the dissimilarities computed for two out of the three dimensions.
The 2.5D probabilities produced at the end of the network are interpolated to 3D using B-splines.
To minimize the error during the conversion from 2.5D to 3D, a two-step instance normalization is applied for each pair of test scans using gradient descent.
More recently,~\citet{heinrich2022voxelmorph++} further expanded the concept of probabilistic displacement and incorporated keypoint supervision into VoxelMorph~\citep{balakrishnan2019voxelmorph} through the introduction of VoxelMorph++.
They advanced VoxelMorph in two respects: probabilistic displacement via heatmap prediction and multi-channel instance optimization using one-hot embeddings of the anatomical label maps generated by a segmentation network.
In their model, high-level features are initially extracted from the VoxelMorph decoder, and feature vectors are then sampled at given keypoint locations.
These feature vectors are converted into larger heatmap patches through a convolution block followed by a softmax operation.
Consequently, each heatmap represents the probabilistic displacements of the corresponding keypoint.
The final displacement field is generated as the sum of the displacements weighted by the heatmap.
During the testing phase for each image pair, an instance optimization strategy~\citep{siebert2022fast} refines the displacement field using the supervision provided by the anatomical labels generated from a segmentation network.
The methods discussed in this subsection were evaluated on abdomen and lung CT datasets, where large deformations are necessary for accurate registration.
The architectures proposed in these methods proved to be efficient and demonstrated superior performance compared to traditional methods and learning-based networks that only generate dense displacement fields.

\begin{figure}[!ht]
        \centering
        \includegraphics[width=0.9\columnwidth]{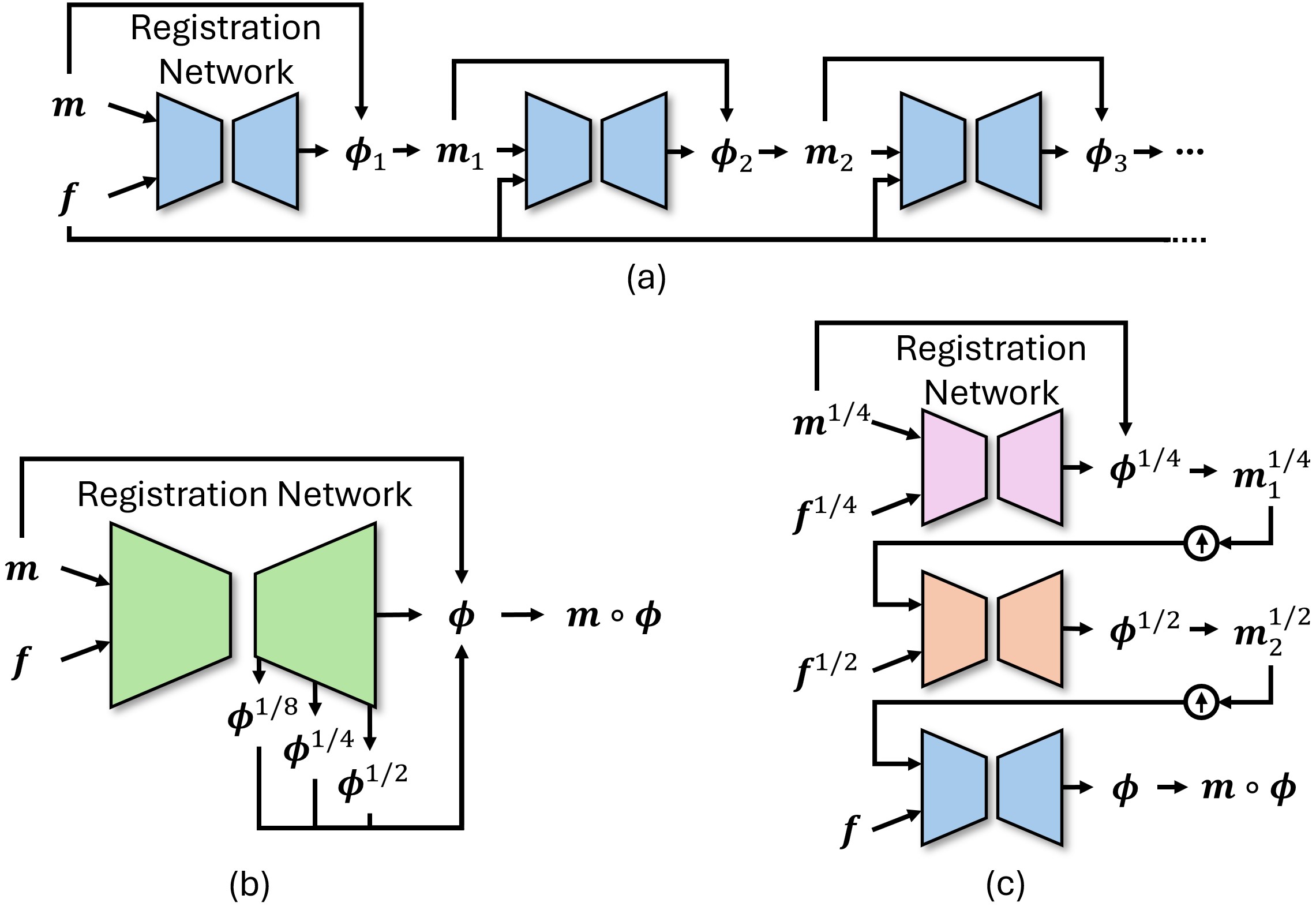}
        \caption{{This illustration depicts progressive and multi-scale registration network designs employed in medical image registration, where $\pmb{m}$ and $\pmb{f}$ represent the moving and fixed images, respectively, and $\pmb{\phi}$ denotes the deformation field. The subscripts indicate the intermediate deformed moving images at respective stages, while the superscripts specify the resolution fraction of the image or the deformation field. Panel~(a) outlines the framework for progressive image registration, e.g., VTN~\citep{zhao2019unsupervised}. Panels~(b) and (c) demonstrate two distinct approaches to multi-scale registration employed in learning-based methods. The former incorporates a multi-scale aggregation of deformation fields within a single network (e.g., im2grid~\citep{liu2022coordinate}), whereas the latter uses an independent network for each scale (e.g., DLIR~\citep{de2019deep} and LapIRN~\citep{mok2020large, mok2021conditional}). The designs for progressive and multi-scale registration are intended to emulate the optimization strategy commonly used in traditional methods, which typically involve iterative and multi-scale updates of the deformation field.}}
        \label{f:network_prg_ms}
        \end{figure}

\subsection{Progressive and Pyramid Registration}
\label{sec:multi_res_reg}
Recent research has also demonstrated that employing a network to progressively warp a moving image towards a fixed image, or performing registration through a multi-scale image pyramid employing a coarse-to-fine technique, may significantly improve registration performance.
{Figure~\ref{f:network_prg_ms} graphically summarizes these two types of architectures, in which panel~(a) illustrates the progressive registration framework, while panels~(b) and (c)~depict the multi-scale registration frameworks.}
\citet{zhao2019unsupervised}~introduced the VTN, which leverages cascade registration networks to align moving images with fixed images.
Drawing inspiration from FlowNet2.0~\citep{ilg2017flownet}, each subnetwork is responsible for aligning the current moving image with the fixed image, with the resulting warped image sent into the subsequent subnetwork as the new moving image. The final deformation field is the composition of the intermediate deformation fields produced by the subnetworks.
This approach has been shown effective in handling large displacements.
{Similar trends of adopting a progressive registration framework are evident in the development of image registration models, which have shown enhanced registration performance \citep{jia2021learning, chen2022unsupervised, lara2023deep}.} \citet{chen2022unsupervised}~proposed a method for progressive image registration within a single network.
Their method employs multiple convolution blocks in the decoding stage, each responsible for aligning the current moving image to the fixed image.
{In~\citet{jia2021learning}, the authors introduced VR-Net, an end-to-end learning framework that unfolds the iterative optimization process. First, the nonlinear objective function (i.e., Eqn.~\ref{eqn:energy_func}) is linearized with respect to the deformation by applying a first-order Taylor series expansion. Subsequently, an auxiliary variable is introduced to decouple the objective function into two convex problems. This approach mirrors earlier methods that used quadratic relaxation applied to the variational problem~\citep{chambolle2004algorithm, steinbrucker2009large, heinrich2014non}, where one sub-problem focuses on optimizing similarity and the other concentrates on the regularization of the deformation. VR-Net employs a sequence of network blocks that iteratively updates and refines the deformation field, with each block refining the output from the previous one, effectively mirroring the iterative update process for the decoupled objective function. In~\citet{lara2023deep}, the authors adopt a similar strategy to that of VTN~\citep{zhao2019unsupervised} by employing a sequence of intermediate registration networks that progressively warp the moving image towards the fixed image. The authors augmented the VTN framework for dynamic myocardial perfusion CT registration by introducing a contrast loss specifically designed to focus on the dynamic changes in contrast agent concentration within the heart. The proposed method surpasses several established optimization-based registration methods for this particular task.}
%
%

Concurrently, there have been efforts to apply progressive registration using a multi-scale image pyramid approach. 
Given the widespread adoption of hourglass-shaped network architectures in image registration, convolution blocks within the decoder generate deformation fields at multiple resolutions in a coarse-to-fine manner.
These deformation fields at different resolutions are subsequently upsampled and composited to form the final deformation field. Notable methods that adopt this scheme include \citep{jiang2020multi, kang2022dual, liu2022coordinate, lv2022joint}.
In addition to network architecture, the coarse-to-fine training scheme has also been adopted in learning-based image registration.
Taking inspiration from conventional registration methods that often employ multiple stages with varying resolutions, \citet{de2019deep}~pioneered a multi-scale training strategy for deformable image registration.
Their approach involves sequentially training the ConvNet in each stage for a specific image resolution by optimizing the image similarity measure.
Notably, a B-spline framework is adopted thus alleviating the need for a deformation regularizer.
During training, the weights of the preceding ConvNets are held fixed, and after training, the registration is performed through a single pass of input images to the multi-stage ConvNets.
\citet{eppenhof2019progressively}~proposed a novel progressive and multi-scale training scheme for learning-based image registration.
Instead of training a large network on the registration task all at once, they first train smaller versions of the network on lower-resolution images.
The resolution of the training images is then gradually increased, and additional convolutional layers are added to increase the network size.
{A similar approach is employed by~\citet{berg2023employing}.}
\citet{mok2020large}~proposed LapIRN, which adopts a similar pyramid training scheme.
However, unlike the previous training approach, which progressively increases the image resolution and network size of the same network, LapIRN employs three different networks, each producing a deformation field for a specific resolution.
Each network is equipped with a skip connection that propagates feature embeddings from a lower-resolution network to a higher-resolution network.
The networks are trained in a coarse-to-fine manner, with each network producing a deformation field that refines the upsampled deformation field from the previous resolution.
However, using multiple networks to generate a pyramid of deformation fields can be computationally inefficient and increase the network size, which can hinder training.
To address this issue, \citet{hu2020self}~proposed a self-recursive contextual network that employs a single feature extractor to produce features at different resolutions.
Then, a weight-sharing deformation generator and receptive module are then used to recursively generate and refine deformation fields in a coarse-to-fine manner.
Since the network weights are shared between resolutions, this method reduces the computational burden and the size of the network, resulting in more efficient training.
\citet{zhou2023self}~proposed a novel network architecture to leverage progressive registration at both single and multi-scale resolution.
The proposed method iteratively refines the deformation field generated from the previous iteration, with each iteration composing deformation fields of various resolutions to form the new deformation field. 
{\citet{meng2023non}~introduced NICE-Trans, a novel approach to address both affine and deformable registration simultaneously using a coarse-to-fine framework. The model features dual-path ConvNet-based encoders that independently extract features from the moving and fixed images across multiple scales. A Transformer-based decoder is employed to estimate affine parameters at its bottleneck, along with multiple displacement fields at various resolutions of the decoder. The final displacement field is derived by summing the upsampled affine grids and the multi-scale displacements. NICE-Trans has shown superior performance compared to previous DL-based DIR methods, which often rely on optimization-based methods for affine registration as a preprocessing step or require a separate affine network to pre-align the images prior to the DIR network.}

{In~\citet{ma2023pivit}, the authors developed a single Transformer-based network architecture for image registration that integrates both progressive and multi-scale strategies. The architecture features a dual-stream ConvNet decoder that independently extracts multi-scale features from both moving and fixed images. Recognizing that large deformations are typically captured at coarser scales, the network employs a series of Swin Transformer-based blocks~\citep{liu2021swin} at the bottleneck scale. These blocks progressively refine the deformation fields through successive additions, where each field is updated based on the output from the previous block, which takes in the deformed moving features together with the fixed image features. Subsequently, additional blocks that include upsampling and a convolution layer further refine the deformation field from the bottleneck to the full scale. The model was rigorously validated to identify the optimal number of progressive updates, with findings indicating that 2 to 3 updates are adequate to achieve satisfactory registration accuracy for the brain MRI and liver CT registration tasks it was evaluated on. For these tasks, the proposed method surpassed several contemporary registration models, including those based on ConvNets and Transformers.}

The registration methods discussed in this subsection have demonstrated the efficacy of decomposing the registration process into multiple steps, where each step refines the deformation fields from the previous step.
These approaches have consistently shown significant performance gains while enforcing a smoother deformation field for image registration tasks compared to using a single network to generate a deformation field all at once.

%% file: uncertainty.tex
DNNs are capable of learning complex representations. However, their predictions are typically deterministic and assumed to be accurate, which is usually not the case.
Estimating the uncertainty is important for evaluating what the models learn from the data and helps reduce risk in decision-making based on the model prediction.
In medical image analysis, uncertainty estimation has been widely used in tasks such as image segmentation, image classification, and image registration.
For example, registration uncertainty empowers surgeons to evaluate the surgical risk tied to the registration model's prediction, thereby avoiding undesirable consequences.
Prior to the deep learning-based registration, traditional registration uncertainty is based on the framework of probabilistic registration, where the probabilistic distribution of the transformation parameters is estimated. 

{In this section, we focus on reviewing the \textbf{recent advancements} in estimating registration uncertainty through deep learning methods, though many concepts have been drawn from traditional registration uncertainty estimation methods.}
We start with the general framework for estimating uncertainty using deep learning methods.
Next, we formally define the different types of registration uncertainty and elaborate on how the uncertainty estimation methods are used in learning-based registration.
Finally, we provide a summary of the methods used for evaluating the quality of uncertainty estimation.

\subsection{Bayesian Deep Learning}
\begin{figure}
\centering
\includegraphics[width=0.35\textwidth]{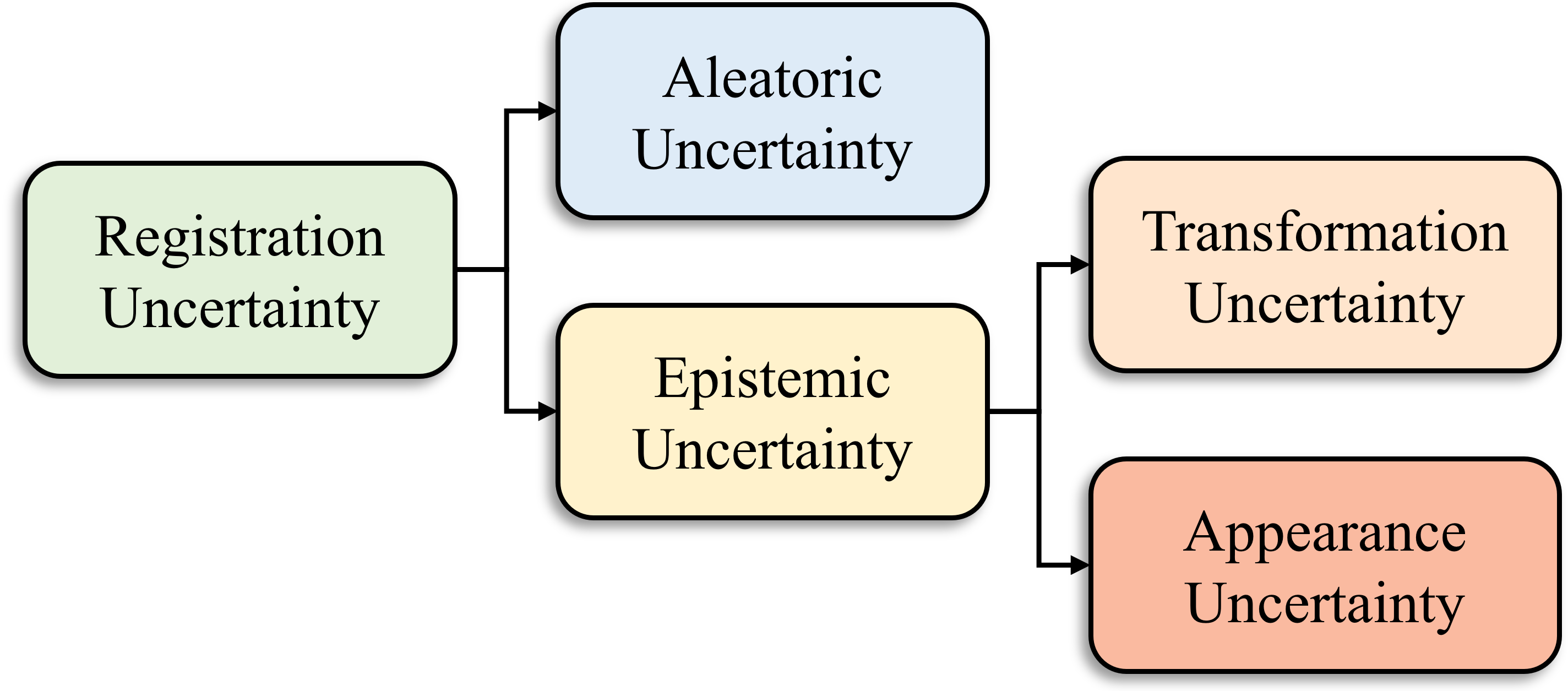}
\caption{Various types of registration uncertainty can be estimated using DNNs.} \label{fig:reg_uncert}
\end{figure}
As shown in Fig.~\ref{fig:reg_uncert}, in general, uncertainty can be categorized into two types: aleatoric and epistemic uncertainty~\citep{kendall2017uncertainties}.
Aleatoric uncertainty, also known as data uncertainty or inherent uncertainty, refers to the inherent randomness or variability present in observed data.
It can be thought of as the variability of the data given the underlying true data generation model due to factors such as measurement errors, sensor noise, or the intrinsic stochastic nature of the data generation process.
Epistemic uncertainty, also known as model uncertainty or knowledge uncertainty, refers to variability present in the model structure, model parameters, and model assumptions.
It arises due to our limited knowledge or understanding of the underlying model.
Aleatoric uncertainty may be reduced by improving the data quality, while epistemic uncertainty may be mitigated by improving model selection, refining parameter estimation, or acquiring additional relevant information.

To predict aleatoric and epistemic uncertainty using deep learning, we train a model $W$ using data set $D$ that takes an input $x$ to generate an output $y(x, W)$ and a variance prediction $\sigma^2(x,W)$. 
Aleatoric uncertainty describes the uncertainty that is inherent in the training data $D$.
It is expressed as follows:
\begin{equation}
\label{eq:ua}
    u_a = E_{p(W|D)}\left[ \sigma^2(x,W) \right],
\end{equation}
where $E$ represents taking the average of $\sigma^2(x,W)$ over the distribution $p(W|D)$.
Epistemic uncertainty describes the uncertainty of the model $W$.
It is represented as follows:
\begin{equation}
\label{eq:ue}
    u_e = V_{p(W|D)} \left[ y(x,W) \right],
\end{equation}
where $V$ represents taking the variance of $y(x,W)$ over the distribution $p(W|D)$.
Directly computing aleatoric uncertainty using Eqn.~\ref{eq:ua} and epistemic uncertainty using Eqn.~\ref{eq:ue} is usually impractical, as it requires the integration of high dimensional numerical functions.
Instead, these uncertainties are approximated from a set of outputs by using the model weights $W$ sampled from the posterior distribution $p(W|D)$.  

In theory, the posterior distribution $p(W|D)$ can be obtained through Bayes' rule,
\begin{equation}
    p(W|D)=\frac{p(D|W)p(W)}{p(D)},
\end{equation}
where $p(W)$ is an assumed prior. 
However, it is not feasible to obtain $p(D)$ in the denominator due to the intractable integral.
As an alternative, variational inference is used to approximate $p(W|D)$ as a distribution $q_{\theta}(W)$ with parameter $\theta$ by minimizing the Kullback-Leibler~(KL) divergence between them.
This process can be simplified as follows:
\begin{equation}
\label{eq:theta_weight}
    \hat{\theta} = \argmin_{\theta}D_{KL}\Big[q_{\theta}(W)\Vert p(W)\Big] - E_{q_{\theta}} \Big[ \log{p(D|W)} \Big],
\end{equation}
where $D_{KL}$ represents KL divergence and $E_{q_{\theta}}$ represents taking the average over the distribution $q_{\theta}(W)$.
With this, the aleatoric uncertainty in Eqn.~\ref{eq:ua} and the epistemic uncertainty in Eqn.~\ref{eq:ue} can be approximated by sampling $W$ from $q_{\hat{\theta}}(W)$.

Many sampling methods can be used for uncertainty estimation, including Monte Carlo dropout, bootstrap, and snapshot techniques. 
The Monte Carlo dropout sampling method operates under the assumption that $q_{\theta}(W)$ follows a Bernoulli distribution~\citep{gal2016dropout}.
It leverages dropout layers during the testing phase to perform multiple forward inferences.
This method is widely used in learning-based registration models, likely due to its straightforward implementation~\citep{yang2016fast, yang2017quicksilver, madsen2020closest, chen2022transmorph, xu2022double}.
Bootstrap sampling, a traditional method, involves training the registration model multiple times on independent training sets to produce multiple inferences~\citep{kybic2009bootstrap}.
Snapshot sampling uses the cyclic learning rate in one training process for perturbing the model to converge to multiple different local minimums~\citep{huang2017snapshot}.
It has shown that snapshot sampling performs better uncertainty estimation than other methods for the medical image registration use case~\citep{gong2022uncertainty}.

\subsection{Registration Uncertainty Estimation for DNN}
Both aleatoric and epistemic uncertainty are present in image registration. 
Aleatoric uncertainty in image registration may arise from factors such as image noise, image artifacts, lack of image features or image contrast, and natural anatomical variation between images.
There may be two types of aleatoric uncertainty in image registration.
One is that given the underlying true deformation field, two aligned images may not be exactly the same due to different image noise, image artifacts or natural anatomical variation between images.
Another type is that multiple deformation fields may align two images with similar performance due to lack of image features or image contrast.
On the other hand, epistemic uncertainty represents the limitations inherent in the modeling process. In image registration, this relates to the limited ability of the model to precisely capture the complex deformation field. This form of uncertainty can be attributed to factors like inadequate training data, choices in model architecture, or the inherent complexity posed by the inverse problem of estimating the deformation fields.
The following subsections provide details on each type of uncertainty.

\subsubsection{Aleatoric Uncertainty}
In medical image registration, the output of the network is usually a deformation field $\phi(I_f, I_m, W)$ as a function of the fixed image $I_f$, the moving image $I_m$ and the model $W$.
To help the model estimate the aleatoric uncertainty inherent from data, the model needs to predict a variance $\sigma(I_f, I_m, W)$ of the output.
Assuming the deformation field $\phi$ follows a voxel-wise Gaussian distribution, the model $W$ can be trained by minimizing the loss function,
\begin{equation}
\label{eq:loss}
    \mathcal{L} = \sum_{p} \frac{(\phi(p) - \phi^*(p))^2}{\sigma^2(p)} + \log{\sigma^2(p)},
\end{equation}
where $\phi^*$ is the ground truth for the deformation field. 
However, in unsupervised learning-based image registration, the ground truth deformation $\phi^*$ is unavailable, and the loss may be calculated in the image domain (\emph{i.e.}, in the form of image similarity measure) rather than directly comparing the deformation fields as in Eqn.~\ref{eq:loss}.
To overcome this issue, the aleatoric uncertainty for image registration is frequently estimated using a variational inference strategy, which optimizes a global neural network to produce {posterior} of deformation fields~\citep{dalca2019unsupervised, grzech2020image, grzech2022variational, wei2021recurrent, smolders2022deformable, croquet2021unsupervised, krebs2019learning}.
This approach circumvents the direct computation of the intractable posterior $p(\phi|I_f; I_m)$ by introducing an approximate posterior $q_{\theta}(\phi)$, where the parameter $\theta$ can be predicted by a network based on the inputs $I_f$ and $I_m$.
The KL divergence between the two posteriors is then minimized, which maximizes the evidence lower bound~(ELBO):
\begin{equation}
    \hat{\theta} = \argmin_{\theta} D_{KL} \Big[ q_{\theta}(\phi)\Vert p(\phi) \Big] - E_{q_{\theta}} \Big[ \log p(I_f|\phi;I_m) \Big],
\end{equation}
where $D_{KL}$ represents KL divergence and $E_{q_{\theta}}$ represents taking the average over the distribution $q_{\theta}(\phi)$, {and $p(\phi)$ represents the prior of the deformation field}.
Here, {the prior $p(\phi)$ is usually modeled using
a graph Laplacian matrix for smoothness
regularization~\citep{dalca2019unsupervised, grzech2020image,
grzech2022variational}}, and the approximate posterior $q_{\theta}(\phi)$ is frequently modeled as a multivariate normal distribution  (\textit{i.e.}, 
$\phi\sim\mathcal{N}(\mu_{\phi},\sigma_{\phi}^2)$). 
{A more complex prior can be designed to close the gap between the real posterior and the variational posterior~\citep{xu2023importance}.}
{To reduce the dimensionality of the covariance matrix $\sigma_{\phi}^2$ and simplify its optimization, a diagonal matirx~\citep{dalca2019unsupervised} or a sum of a diagonal and a low rank matrices~\citep{grzech2020image, grzech2022variational} are used.}
Moreover, the conditional probability $p(I_f|\phi; I_m)$ typically takes the Gaussian form (\textit{i.e.}, $I_f\sim\mathcal{N}(I_m \circ \phi, \sigma_I^2)$)~{\citep{dalca2019unsupervised}, or
can be implemented as a neural network for learning a more complex
form~\citep{grzech2022variational}}.
In practical applications, the mean $\mu_{\phi}$ and the standard deviation $\sigma_{\phi}$ of the deformation field can be {optimized directly as parameters using
backpropagation~\citep{grzech2020image, grzech2022variational}, or can be} predicted by the registration network, in a similar manner to a variational autoencoder~{\citep{dalca2019unsupervised}.
The obtained} variance $\sigma^2_{\phi}$ represents the aleatoric uncertainty associated with the deformation field, given the input images $I_f$ and $I_m$.

\subsubsection{Epistemic Uncertainty}
As illustrated in Fig. \ref{fig:reg_uncert}, the epistemic uncertainty in registration can be divided into two different measures \citep{luo2019applicability, xu2022double, chen2022transmorph}: transformation uncertainty and appearance uncertainty.
These measures refer to the uncertainty in generating the transformation and the plausibility of the transformation, respectively~\citep{bierbrier2022estimating}.
The former quantifies the uncertainty in the deformation space and tends to be large when the model is uncertain about establishing specific correspondences, such as when registering regions with piecewise constant intensity.
In contrast, the latter is often based on the assumption that high image similarity indicates correct alignment.
Consequently, this uncertainty would be large when the appearance differences between the warped and fixed images are significant. 

Transformation uncertainty can be described as the variance of the sampled deformation fields, which derives from Eqn.~\ref{eq:ue} by stochastic sampling:
\begin{equation}
\label{eq:ue_trans}
    u_{e,trans} = \frac{1}{N} \sum_{i=1}^{N} ( \phi_i - \frac{1}{N} \sum_{j=1}^{N} \phi_j ) ^2,
\end{equation}
where $\phi_i$ is generated by using the model weight $W_i$ sampled from the estimated variational distribution $q_{\hat{\theta}}(W)$ with the parameter $\hat{\theta}$ optimized by Eqn.~\ref{eq:theta_weight}, and $N$ is the total sampling number.
Appearance uncertainty is expressed as the variance of the warped images, which are created using the sampled deformation fields~\citep{luo2019applicability, xu2022double}:
\begin{equation}
\label{eq:uappea}
    u_{e,appea} = \frac{1}{N} \sum_{i=1}^{N} ( I_m \circ \phi_i - \frac{1}{N} \sum_{j=1}^{N} I_m \circ \phi_j ) ^2,
\end{equation}
where $\phi_i$ is generated by using the sampled $W_i$ as the
model, $N$ is the total sampling number and $I_m$ is the moving image. 
However, it should be noted that the uncertainty estimated using this equation for appearance uncertainty can be biased due to overfitting, as shown in~\citet{chen2022transmorph}. 
To correct this, the authors suggest using the following formulation instead:
\begin{equation}
    u_{e,appea} = \frac{1}{N} \sum_{i=1}^{N} ( I_m \circ \phi_i - I_f ) ^2,
\end{equation}
where the predictive mean is replaced by the fixed image $I_f$.

{It is important to note that while the presence of aleatoric and epistemic uncertainties in the image registration process, they may not provide useful insights for label propagation, where image registration is applied to warp between label maps. In such scenarios, segmentation uncertainty estimates could be more relevant. In \citep{chen2024registration}, the authors demonstrated that registration uncertainty estimates, which are primarily based on image intensity information rather than the underlying labels, do not correlate well with errors in label propagation. They proposed a plug-in-and-play DNN method that estimates both aleatoric and epistemic uncertainties in segmentation, in addition to registration uncertainties, to more effectively identify potential errors in using image registration for segmentation.}

\subsection{Uncertainty Evaluation in Registration}
One significant challenge in uncertainty estimation lies in its evaluation due to the absence of ground truth, especially in unsupervised learning-based registration.
To access the quality of uncertainty, sparsification plots are usually used for voxel-wise uncertainty evaluation~\citep{mac2012learning, wannenwetsch2017probflow, ilg2018uncertainty}.

Sparsification plots demonstrate how the registration error changes by gradually removing voxels ranked by the uncertainty measure. 
It is anticipated that removing a voxel with higher uncertainty will result in a greater reduction in registration error, and the opposite holds true for voxels with lower uncertainty.
If all voxels are arranged in descending order of uncertainty, and the uncertainty ranking matches the actual registration error ranking, the accumulated registration error under the sparsification plot will be small.
Therefore, the area under sparsification plots is also used as an evaluation metric to gauge the quality of the uncertainty estimation.

%
%


%

%% file: evaluation.tex
Manual correspondences are usually regarded as the gold standard for evaluating the performance of a registration algorithm~\citep{peter2021uncertainty}.
Landmark correspondences are the most frequently used type, although surfaces or lines may also serve as manual correspondences.
The evaluation of registration performance using landmark correspondences is relatively simple for rigid and affine transformations, since these transformations can be expressed as matrix multiplication and the ground truth transformation can be determined through several pairs of manual landmark correspondences.
In contrast, determining the parameters of deformable transformations requires dense manual landmark correspondences, which are typically not obtainable.
Even in cases where manual landmark correspondences are available, they are often restricted to highly selective intensity features~\citep{castillo2009framework} and neglect regions with homogeneous intensities.
Therefore, validating deformable registration algorithms is still considered a non-trivial task~\citep{viergever2016survey}.
In current literature, the performance of deformable registration algorithms is most commonly evaluated in terms of accuracy and regularity.
\subsection{Accuracy Measures}
When the manual landmark correspondences are available, the accuracy of the transformation can be evaluated by target registration error~(TRE),
\begin{equation}
    \text{TRE}_{\text{forward}} = \sum_{i=1}^{N} ||T_{\text{forward}}(l_m^i) - l_f^i||_k,
    \label{eqn:tre_forward}
\end{equation}
where $T_{\text{forward}}$ is the estimated forward transformation that takes the moving image to the fixed image; $l_m^i$ and $l_f^i$ are the $i^{\,\text{th}}$ pair of landmarks in the moving and fixed image, and $k\in\{1,2\}$ denoting either the $\ell^1$-norm or $\ell^2$-norm.
Both $l_m^i$ and $l_f^i$ as well as the warped moving landmark $T(l_m^i)$ can be non-integer locations.
Note that we used the forward transformation $T_{\text{forward}}$ in Eqn.~\ref{eqn:tre_forward}, but it is more common in practice to estimate the transformation $T_{\text{backward}}$ that maps the fixed image to the moving image.
In order to generate the warped image, $T_{\text{backward}}^{-1}$ can be applied in place of $T_{\text{forward}}$.
Both $T_{\text{forward}}$ and $T_{\text{backward}}$ are mappings from integer locations to non-integer locations.
The difference between these two schemes is manifested when rendering the warped image.
When $T_{\text{forward}}$ is applied to $I_m$, integer locations are mapped to non-integer locations, which necessitates interpolating scattered data~\citep{crum2007methods,zhuang2008atlas}.
On the other hand, $T_{\text{backward}}^{-1}$ maps non-integer locations back to integer locations.
Thus, rendering the warped image only requires interpolating the moving image, which is defined on a regular grid.
For algorithms that only output $T_{\text{backward}}$, TRE can be computed as
\begin{equation}
    \text{TRE}_{\text{backward}} = \sum_{i=1}^{N} || l_m^i - T_{\text{backward}}(l_f^i) ||_k.
    \label{eqn:tre_backward}
\end{equation}

Landmark correspondences can also be generated using artificial deformations~\citep{bauer2021generation,sdika2008fast,ger2017accuracy}. Different from manual landmark correspondences, artificial deformation can produce dense correspondences that are not limited to regions with highly selective intensity features.
However, the performance of algorithms on artificial deformation may not accurately reflect their actual performance due to the discrepancy between the artificial and real deformations~\citep{obeidat2016comparison, pluim2016truth}.
To overcome this issue, many works have been focused on generating deformations that are more akin to those observed in practical applications.
For instance, \citet{lobachev2021evaluating}~proposed a pipeline for simulating sectioning-induced deformation fields.
\citet{vlachopoulos2015selecting}~used a thin-plate kernel spline model to simulate lung deformations arising from respiration. Biomechanical simulation~\citep{fu2021deformable, teske2017construction} and phantoms~\citep{wu2019characterization,ayyalusamy2021performance} are other techniques used to generate artificial deformations.

In situations where manual landmark correspondences are not available, surrogate measures are used to evaluate accuracy.
The most straightforward measures of this kind include absolute intensity differences and the root-mean-square intensity difference between the warped image and the fixed image.
Other similarity measures such as mutual information, structural similarity index~(SSIM) can also be used.
When anatomic labels are available, evaluating the overlaps between the warped and fixed label images is a popular technique. The Dice coefficient and Jaccard Index are examples of such measures.
However, \citet{rohlfing2011image}~demonstrated that by simply reordering the voxels from the moving image based on the intensity values ranking without any geometrical constraints, one can achieve significantly better performance compared with the state-of-the-art registration algorithms in most of the surrogate measures.
They concluded that surrogate measures might still be useful to detect inaccurate registrations but many times they do not provide sufficient positive evidence for accurate registrations.
Only the overlap of sufficiently local labels among the surrogate measures was found to distinguish between reasonable and poor registrations in their experiments.

Label surface distances from segmentation maps offers an alternative to overlap measures.
\citet{dalca2019unsupervised}~converted segmentation maps into signed distance functions to approximate the distance between the fixed and warped surfaces.
They also showed that incorporating a similar surface distance loss during network training enhanced the surface alignment of anatomical structures.
\citet{cheng2020cortical}~used the mean minimum distance~(MMD), computed as the average Euclidean distance between manually defined surface points of anatomical structures and their corresponding nearest points on the warped surface, to measure the discrepancy between the surfaces.
Additionally, the Hausdorff distance has been extensively used~\citep{hering2022learn2reg}.

Previous studies \citep{lotfi2013improving, sokooti2016accuracy} have explored the use of machine learning algorithms for quantifying registration errors.
Compared to manual landmark correspondences, those methods provide dense error estimations that can be easily visualized.
More recently, several deep learning techniques have been employed, offering a speed advantage over traditional machine learning algorithms, especially when a graphical processing unit~(GPU) is available~\citep{sokooti2021hierarchical}.
Most of these methods were trained to predict the registration errors between a fixed and a warped image inputs.
During training, artificial deformations are used to produce the warped image and the accuracy of these methods were validated using manual landmark correspondences~\citep{eppenhof2018error,sokooti2021hierarchical}.
Additionally, these techniques can also be applied to inter-modality registration tasks by incorporating an extra image synthesis step~\citep{bierbrier2023towards}.

\subsection{Regularity Measures}
Given the challenge of acquiring dense manual landmark correspondences and the aforementioned limitation of surrogate measures, the regularity of the transformations is often used alongside accuracy measures to obtain a more comprehensive understanding of the transformations. 
The underlying assumption is that accurate transformations should be spatially smooth.
Particularly, transformations that fold the space result in physically un-realistic anatomy structures, which usually indicate errors. 
For continuous transformations, their Jacobian determinant $|J|$ must be positive everywhere to avoid folding of space.
This concept is extended to digital transformations where the number or the percentage of voxels with non-positive Jacobian determinant $|J|\leq0$ are reported to measure the irregularity~\citep{meng2022enhancing, liu2022coordinate, mok2022unsupervised, dey2022contrareg, chen2022deformer, mok2020fast, jia2021learning, wu2022nodeo}.
For a 3D transformation $T(x,y,z) = [T_x, T_y, T_z]$, the Jacobian is defined as
\begin{equation}
    J = \begin{vmatrix}
        \frac{\partial T_x}{\partial x} & \frac{\partial T_x}{\partial y} & \frac{\partial T_x}{\partial z} \\[4pt]
        \frac{\partial T_y}{\partial x} & \frac{\partial T_y}{\partial y} & \frac{\partial T_y}{\partial z} \\[4pt]
        \frac{\partial T_z}{\partial x} & \frac{\partial T_z}{\partial y} & \frac{\partial T_z}{\partial z} \\
        \end{vmatrix}.
\end{equation}

\begin{figure}[!tb]
    \centering
    \includegraphics[width=0.38\textwidth]{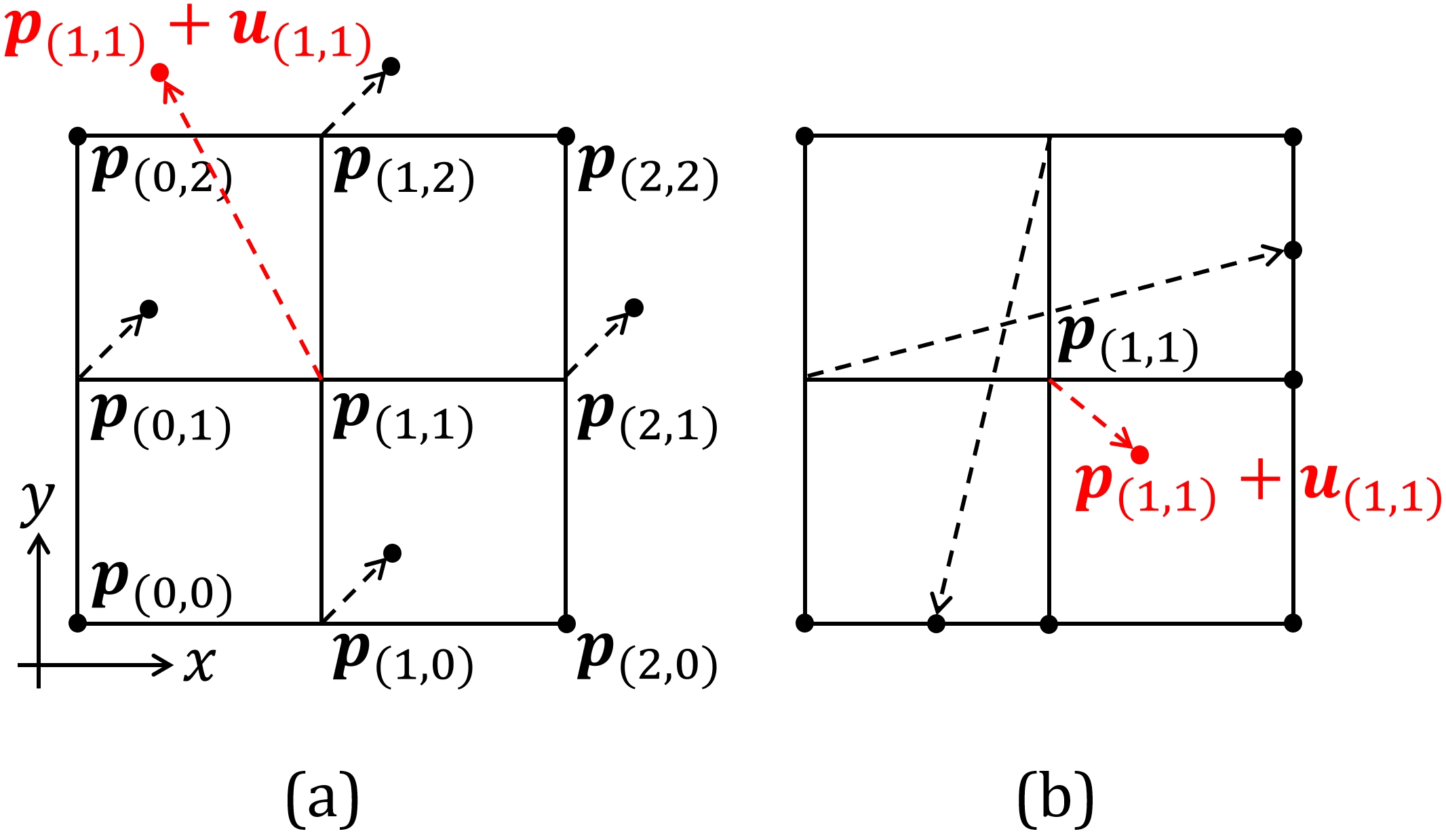}
    \caption{Examples of the checkerboard problem~(a) and the self-intersection problem~(b) for the central difference approximated Jacobian determinant $|J|$ on a $3\times3$ grid.
    The transformations are visualized as a displacement field and the displacement of the center pixel is highlighted in red.
    In~(a), the central difference approximated $|J|$ for the center pixel equals one but the displacement of the center pixel is \emph{not} involved in the computation. 
    Even if the center pixel moves outside the field of view, the central difference approximated $|J|$ still equals one.
    In~(b), the transformation around the center pixel already introduced folding in space regardless of the displacement of the center pixel but the central difference approximated $|J|$ is positive.}
    \label{fig:central_difference}
\end{figure}

\citet{ashburner1999high} considered the transformations to be locally affine, and the Jacobian determinant could be computed using singular value decomposition.
More generally, the Jacobian of a dense nonlinear transformation is estimated through numerical approximation using finite difference methods.
In a recent study,~\citet{liu2022finite} showed that when approximating the Jacobian using forward or backward difference, it is implicitly assumed that the digital transformations are linearly interpolated on a {tetrahedra} mesh grid.
{They found that the Jacobian determinant, when approximated using central difference, exhibits checkerboard and self-intersection problems.
Consequently, it consistently underestimates non-diffeomorphic spaces in a transformation.
In Fig.~4 of~\citet{liu2022finite}, the authors show examples of the checkerboard problem and the self-intersection problem in 2D. 
In both cases, the central difference approximated Jacobian determinants are positive, but the underlying transformations introduce folding of space~(under the assumption that the digital transformations are piecewise linear).
To address these issues, \citet{liu2022finite}~proposed the
non-diffeomorphic volume measurement, which uses a combination of forward and backward differences to measure the {folding implied by a displacement field based on the volume of folded tetrahedra in the mesh grid.}}
Importantly, \citet{liu2022finite}~showed that the Jacobian determinant, when approximated using central difference, results in the checkerboard and self-intersection problems. Consequently, it consistently underestimates non-diffeomorphic spaces.
Figure~\ref{fig:central_difference} shows examples of the checkerboard problem and the self-intersection problem in 2D.
In both cases, the central difference approximated Jacobian determinants are positive, but the underlying transformations introduce folding of space~(under the assumption that the digital transformations are piecewise linear).
They conclude that for a 2D transformation, four unique finite difference approximations of $|J|$'s must be positive to ensure the entire domain is invertible and free of folding; in 3D, ten unique finite differences approximations of $|J|$'s are required to be positive.
Note that their method is closely related to simplex counting~\citep{holland2011nonlinear} used in deformation-based volumetric change estimation.
Because of the issues associated with central difference approximation of $|J|$'s,~\citet{liu2022finite} recommend using non-diffeomorphic volume to accurately reflect the non-diffeomorphism introduced by transformations.

The logarithm of the Jacobian determinant is also an important measure, especially for applications where it requires the volume of the underlying anatomy to be preserved~\citep{rohlfing2003volume,jian2022weakly}. The logarithm is used to symmetrically weight local expansion and compression~\citep{rohlfing2003volume,lange2020symmetric}.
In recent works such as \citep{hering2022learn2reg,chen2022transmorph, dey2022contrareg,chen2022deformer,song2022cross}, the standard deviation of the logarithm of the Jacobian determinant has been used to quantify the smoothness of the displacement field. Additionally, the statistical distribution of the logarithm of the Jacobian determinant can be used as a visualization tool to reveal differences between registration algorithms~\citep{leow2007statistical, lange2020symmetric}.

Similar to many surrogate measures, the regularity of the transformations can detect inaccurate transformations, but by itself, it is insufficient to provide adequate positive evidence for accurate transformations. For example, the identity transformation would be deemed a perfectly regularized transformation, but it would not provide a meaningful registration.

%% file: datasets.tex
\begin{sidewaystable*}
        \rowcolors{2}{white}{cyan!10}
        \caption{{A summary of the publicly available benchmark dataset for medical image registration.}}
        \label{table:dataset_list}
        \fontsize{7.45}{8.95}\selectfont
        \begin{tabularx}{0.99\textwidth}{lc lc lc lc X}
        \toprule
        \rowcolor{white}
        \textbf{Dataset} && \textbf{Anatomy} && \textbf{Cohort Type} && \textbf{Modality} && \textbf{Highlights}\\
        \cmidrule(lr){1-1}
        \cmidrule(lr){3-9}
        IXI\textsuperscript{a}  && Brain && Healthy Controls && T1w, T2w, PDw MRI && Nearly 600 MRI images with cortical and subcortical label maps from prior studies~\citep{liu2024vector,chen2022transmorph,hoopes2022synthstrip}. \\
        LUMIR~\citep{L2R24}  && Brain && Healthy Controls && T1w MRI && Part of Learn2Reg 2024~\citep{L2R24}, using the OpenBHB dataset~\citep{dufumier2022openbhb}; 4,014 MRIs from ten public datasets with label maps and landmarks. \\
        LPBA40~\citep{shattuck2008construction}  && Brain && Healthy Controls && T1w MRI && 40 MRI scans affine-transformed to a common atlas with 50 manually delineated brain structures. \\
        Mindboggle~\citep{klein2012101} && Brain && Healthy Controls && T1w MRI && 101 MRIs affine-aligned to an atlas with 106 manually delineated brain structures.\\
        OASIS~\citep{marcus2007open, hoopes2022learning} && Brain && Alzheimer’s disease && T1w MRI && 416 MRIs from OASIS-1~\citep{marcus2007open} with label maps generated using FreeSurfer and SAMSEG, used in Learn2Reg 2021~\citep{hering2022learn2reg}.\\
        BraTS-Reg~\citep{baheti2021brain} && Brain && Glioma && T1w, T1ce, T2w, FLAIR MRI && 140 training, 20 validation, and 50 testing cases with manual landmarks across baseline and follow-up scans.\\
        CuRIOUS~\citep{hering2022learn2reg} && Brain && Glioma && T1w, T2-FLAIR MRI, 3D US && Part of Learn2Reg 2020, 22 subjects with pre-op MRI, and intra-op 3D US with annotated landmarks from EASY-RESECT~\citep{xiao2017re}.\\
        ReMIND2Reg~\citep{juvekar2024remind}  && Brain && Tumor resection && T1w, T2w MRI, 3D US && Part of Learn2Reg 2024~\citep{L2R24}, 104 intra-operative US, 98 T1ce, and 67 T2 MRIs from 104 patients, with manual landmarks.\\
        Hippocampus-MR~\citep{hering2022learn2reg} && Brain && Non-affective psychosis && T1w MRI && Part of Learn2Reg 2020, 394 MR scans of the hippocampus region with manually tracings for evaluation.\\
        DIR-Lab~\citep{castillo2013reference,castillo2009four}  && Lung && COPD, cancer && Breath-hold and 4DCT && 20 CTs (COPDgene and 4DCT subsets) with 7,000+ manually paired landmarks for evaluating deformable registration.  \\
        NLST~\citep{national2011reduced}  && Lung && Smokers && Spiral CT && 100 paired inhale-exhale CTs with lung masks and keypoints; 10 test images with manual landmarks for Learn2Reg 2022~\citep{l2r2022}.  \\
        Lung-CT~\citep{hering2022learn2reg}  && Lung && Healthy Controls && Inspiratory, expiratory CT && 30 paired lung CTs with lung masks and keypoints; evaluation with manual landmarks from vessels and airways for Learn2Reg 2021~\citep{hering2022learn2reg}.  \\
        EMPIRE10~\citep{murphy2011evaluation}  && Lung && Healthy Controls && Inspiratory, expiratory CT && 30 lung CT pairs with 100 manual landmarks for each, covering different scan types to evaluate registration methods.  \\
        Thorax-CBCT~\citep{hugo2016data}  && Lung && Cancer Patients && CT, CBCT && 18 paired CTs from TCIA-4D-Lung with manual organ and target delineations for interventional registration in Learn2Reg 2023~\citep{L2R2023}.  \\
        Lung250M-4B~\citep{falta2024lung250m}  && Lung && Mixed && CT && 248 paired CTs from seven datasets with 4 billion voxels and 250M keypoints, providing ground truth displacements and nnUNet segmentations.\\
        ACDC~\citep{bernard2018deep} && Heart && Cardiac diseases && 4D cine-MRI && 150 subjects with manual LV, RV, and Myo segmentations at ED and ES phases for intra-patient registration.  \\
        M\&Ms~\citep{campello2021multi}  && Heart && Cardiac diseases && 4D cine-MRI && 375 subjects from multiple centers with LV, RV, and Myo segmentations at ED and ES phases for intra-patient registration.  \\
        MM-WHS~\citep{zhuang2019evaluation} && Heart && Cardiac diseases && CT, MRI && 120 cardiac scans (CT and MRI) from 60 subjects with 7 key heart structures manually annotated for mono- and multi-modal registration.\\
        Abdomen-CT-CT~\citep{hering2022learn2reg} && Abdomen && Cancer Patients && CT && Part of Learn2Reg 2020~\citep{hering2022learn2reg}, featuring 50 CT images with 13 manually labeled structures from~\citep{xu2016evaluation}.\\
        Abdomen-MR-CT~\citep{hering2022learn2reg} && Abdomen && Cancer Patients && CT, MR && Part of Learn2Reg 2021~\citep{hering2022learn2reg}, containing 16 CT/MR pairs with 4 labeled structures.\\
        ACROBAT~\citep{weitz2024acrobat} && Breast && Breast Cancer && Pathological images && 4,212 whole-slide-images from 1,152 breast cancer patients.\\
        ANHIR~\citep{borovec2020tmi} && Body-wide && Cancer tissue samples && Pathological images && 355 images with 18 different stains, resulting in 481 valid image registration pairs.\\
        COMULISglobe SHG-BF~\citep{L2R24} && Breast / Pancreas && Cancer tissue samples && Pathological images && Part of Learn2Reg 2024~\citep{L2R24}, featuring paired second-harmonic generation and bright field pathology images.\\
        COMULISglobe 3D-CLEM~\citep{L2R24}  && Cell && Mitochondria, nuclei && Microscopy && Part of Learn2Reg 2024~\citep{L2R24}, featuring 3 pre-processed microscopy datasets with manually annotated landmarks.\\
\bottomrule
    \end{tabularx}
    \vspace{1ex}
    
    \footnotesize \textsuperscript{a} https://brain-development.org/ixi-dataset/
\end{sidewaystable*}

{Due to the inherent black-box nature of DL, developing DL models often relies on empirical studies and author-reported evaluations. The reproducibility and replicability of these experiments have been central focuses since the rise of DL, and DL-based registration is no exception. A key method to promote reproducibility is evaluating models on publicly available datasets. To support this effort, we provide a list of benchmark datasets in Table~\ref{table:dataset_list} to serve as a reference for the medical image registration community. While segmentation label maps are frequently used as a surrogate measure for registration accuracy, theoretically, any dataset containing segmentation label maps for the same anatomical structures across patients or within the same patient could be used to evaluate registration performance. However, our focus here is specifically on datasets designed for, or have been widely used in, registration tasks. These datasets span anatomies such as the brain, lung, heart, abdomen, and pathological images, each presenting unique challenges.}

{For brain imaging, the challenge arises from the need for precise alignment of small structures such as the hippocampus, ventricles, and cortical sulci. Additionally, conditions like Alzheimer's disease lead to cortical thinning and atrophy, while tumors cause local and global distortions, making registration complex and requiring both fine and large-scale alignment across and within subjects.}
    
{In lung registration, one of the primary challenges is managing the large deformations caused by respiratory motion while also accounting for finer structural changes, such as those in vessels and airways. Evaluation is typically landmark-based, as anatomical label maps of lung structures are often too coarse to precisely measure registration accuracy. However, obtaining high-quality landmarks is both time-consuming and labor-intensive.}

{Cardiac image registration similarly faces the challenge of complex deformations due to both respiratory and cardiac motion. However, a unique difficulty is the limited availability of anatomical landmarks~\citep{makela2002review}, as the heart presents fewer distinct features compared to the lungs, complicating both alignment and evaluation.}
    
{Abdominal registration faces significant challenges due to substantial inter-subject variability in organ size, shape, and appearance, driven by factors such as age, gender, and disease. Soft tissue deformation adds another layer of complexity, constantly altering inter-organ relationships within the same subject~\citep{xu2016evaluation}. Compounding these issues, factors like body pose, respiratory cycle, and digestive status cause dynamic shifts in organ positioning. The multimodal nature of abdominal imaging, where anatomical (e.g., CT or MRI) and functional (e.g., PET or SPECT) data are often combined, further intensifies the complexity. Despite the advancements in modern hybrid scanners that perform co-registration, persistent errors remain~\citep{livieratos2015technical, roy2015quantitative}, underscoring the need for continued research in this area.}

{Pathological image registration also presents distinct challenges. Thin tissue sections, only a few micrometers thick, can deform easily during sample preparation, and parts of the tissue in whole-slide images (WSIs) of the same sample may be absent in other WSIs, particularly if sections are non-consecutive~\citep{weitz2024acrobat}. Additionally, tissue appearance can vary significantly depending on the staining method. The complexity of pathological features, combined with the gigapixel resolution of WSIs, demands robust registration methods that provide multi-resolution alignment while accommodating potential topological changes for accurate registration.}

{While Table \ref{table:dataset_list} lists datasets commonly used for benchmarking registration methods, the advent of automated segmentation tools (e.g., FreeSurfer~\citep{fischl2012freesurfer}, SynthSeg~\citep{billot2023synthseg}, SLANT~\citep{huo20193d}, TotalSegmentator~\citep{wasserthal2023totalsegmentator}) enables researchers to convert raw medical images from open-source platforms such as OpenNeuro~\citep{markiewicz2021openneuro} and the Cancer Imaging Archive~\citep{clark2013cancer}, as well as from large-scale publicly available datasets such as ADNI~\citep{jack2008alzheimer}, ABIDE~\citep{heinsfeld2018identification}, and autoPET~\citep{gatidis2023autopet} into benchmark datasets for image registration methods.}

%% file: application.tex
\subsection{Atlas Construction}
\label{sec:atlas}
In computational anatomy, atlases have been an essential tool for investigating the variability of human organs across populations and facilitating the segmentation of organs in individual patients. Typically, atlases are constructed through an iterative averaging process (\emph{i.e.}, \textit{procrustean averaging}~\citep{ma2008bayesian}) using a population of patient images~\citep{allassonniere2007towards, davis2004large, guimond2000average, joshi2004unbiased, ma2008bayesian, avants2010optimal}.
This procedure commences with the registration of images to a common frame of reference, followed by the computation of an average based on the registered images, which serves as the atlas for the current iteration.
The iteration cycle continues until convergence has been achieved, resulting in the final atlas.
However, these traditional methods tend to blur regions exhibiting high-frequency deformations~\citep{dey2021generative}.
This shortcoming arises from the averaging of intensities when constructing the atlas, which invariably results in the loss of high-frequency information essential for capturing anatomical details.
{Note that the sharpness of the atlas directly affects its ability to depict structural details, which then impacts the precision with which structures are aligned within the atlas space. An atlas that is blurry and lacks detail can result in the poor alignment of anatomical structures, particularly the smaller ones, across different subjects.} 

Recent advancements in learning-based registration have demonstrated significant improvements in the quality of constructed atlases while concurrently expediting the atlas construction process.
\citet{dalca2019learning}~pioneered the development of a brain atlas within a deep learning framework, in which an initial approximation of the atlas is derived from the mean of the brain images under study.
This atlas is then jointly optimized with a registration network, employing the VoxelMorph architecture to align the atlas with individual patient images.
Throughout the training process, both the atlas and the registration network weights are updated.
To promote an unbiased atlas and enhance spatial smoothness in the resulting deformation fields, the authors introduce a Gaussian-inspired prior.
This prior serves to penalize sharp deformation changes while simultaneously encouraging minimal average deformation across the entire dataset.
{This strategy resembles the maximum likelihood methods traditionally used in atlas
construction, where the atlas is derived from the data by considering it as the mean within a Gaussian model, with the data samples being the deformed images~\citep{glasbey2001penalized, joshi2004unbiased, allassonniere2007towards, ma2008bayesian}.
Moreover, for an atlas to be considered unbiased, the total/mean deformations from the atlas to all images must sum to zero.
This principle applies both in the context of small deformations~\citep{allassonniere2007towards,
guimond2000average, bhatia2004consistent}, where linear averaging is applicable, and in large deformation settings~\citep{joshi2004unbiased, avants2004geodesic, ma2008bayesian}, where averaging extends to more
general metric spaces, such as geodesics.
In essence, the problem of constructing an unbiased atlas is to estimate an image that requires the least deformation to closely match each of the input images.
The prior introduced by~\citep{dalca2019learning} is therefore specifically designed to ensure that the mean deformation approaches an identity, with the mean deformation being calculated through a moving average in their implementation.}
Moreover, patient demographic information is conditioned into the network architecture, facilitating the generation of conditional atlases that vary according to the specific attributes of different individuals.
This work has inspired a variety of applications. For instance,~\citet{cheng2020unbiased} establish continuous spatio-temporal cortical surface atlases for neonatal brains.
Similarly, both~\citet{zhao2021learning} and~\citet{bastiaansen2022towards} construct continuous spatio-temporal atlases for fetal and infant brains.
Zhao~\emph{et al.} developed a multi-scale spherical registration network featuring group-wise registration, while Bastiaansen~\emph{et al.} applied group-wise registration to volumetric ultrasound images.
Alternatively,~\citet{yu2020learning} constructed an unconditional and universal atlas while incorporating demographic information into the displacement field generation.
This approach explicitly models morphological changes related to attributes as a diffeomorphic deformation, which captures variations in shape and size.
Recognizing that the necessity for images to be affinely aligned in a preprocessing step as suggested in~\citep{dalca2019learning} could not adequately capture the dynamic size and shape development of fetal brain structures,~\citet{chen2021construction} proposed incorporating an affine network, conditioned on patient demographic data, to register the constructed atlas to individual patient images.
This approach preserves the dynamic size and shape variations of patients at different ages.
\citet{li2021cas} proposed integrating the segmentation produced by a segmentation network into the atlas construction method proposed in~\citep{dalca2019learning}.
This method enables the joint training of segmentation and registration networks while simultaneously constructing both image and segmentation atlases. 
Similarly,~\citet{sinclair2022atlas} embraced the concept of jointly training segmentation and registration networks.
They were motivated by the observation that segmentation networks often yield spurious voxel-wise predictions.
By warping the label map of the constructed atlas to match the segmentation prediction through learning-based diffeomorphic registration, the topology of the original anatomical structure can be preserved, thus avoiding the potential segmentation errors produced by the segmentation network.
Drawing inspiration from \citet{dalca2019learning} and \citet{shu2018deforming},~\citet{siebert2021learning} proposed using a shared encoder to extract features from input images, followed by two decoders.
One decoder generates an unconditional atlas, while the other produces deformation fields that warp the atlas to individual images.
To improve registration performance and enforce unbiased atlas construction, they introduced an inverse consistency and a bias reduction loss, in addition to the commonly seen similarity measure and deformation regularizer.
In a related study,~\citet{wu2022hybrid} proposed a closed-form update for constructing the atlas by leveraging pre-trained registration networks as a priori knowledge of the deformation field.
Their approach involves an alternating update process for both the deformation field, which warps the atlas, and the atlas itself.
This method results in an atlas construction framework that is independent of the registration model choice, offering flexibility in its application.

Researchers have explored various strategies to enhance the quality of the constructed atlases.
\citet{dey2021generative} improved the constructed atlas by incorporating adversarial learning, which improved both the sharpness and centrality of the resulting atlas.
In a similar vein,~\citet{he2021learning} aimed to improve the atlas' sharpness through adversarial learning and by integrating edge information derived from anatomical label maps.
Additionally,~\citet{pei2021learning} leveraged anatomical label maps to improve the quality of the constructed atlas by applying anatomical consistency supervision.
However,~\citet{ding2022aladdin} contended that the importance of atlas sharpness is secondary to the registration model's ability to align corresponding points between images in the atlas space.
Therefore, they focused on the registration model upon which the atlas construction is based and proposed using the constructed atlas as a bridge.
In their method, an image is first warped to align with an atlas and then further warped to match the target image using the registration network. This process facilitates a direct comparison between the warped image and the target image while enabling the construction and evaluation of the atlas without requiring segmentation of the atlas itself.
Inspired by implicit neural shape representations~\citep{mescheder2019occupancy}, ~\citet{yang2022implicitatlas} proposed constructing atlases of anatomical shapes using a continuous occupancy grid instead of representing them in a voxel-based manner.
Given the latent representation of the shape, this alternative approach constructs an atlas based on the linear combination of a learned template matrix.
Their method offers a novel perspective on atlas representation, diverging from traditional voxel-based representations.

{The aforementioned methods for constructing atlases through deep learning predominantly adopt a Bayesian framework~\citep{allassonniere2007towards, ma2008bayesian}, in which the atlases are derived as an outcome of the training process using the training
dataset.
Consequently, these methods do not allow for atlas
construction during test-time.
To navigate this limitation, group-wise registration methods come into play, allowing for the simultaneous alignment of multiple images to an implicit template.
An illustrative example is found in the work of~\citet{zhang2021groupregnet}, where they introduced a DNN-based group-wise registration framework.
This DNN takes in a group of images, aligning them to an implicit template that is represented by the intensity average of the warped images in the group.
\cite{he2023groupwise}~proposed a similar group-wise
registration method where the atlas is built using principal component analysis~(PCA) rather than an arithmetic mean. 
Furthermore, constructing an atlas at test-time is not feasible with this method, as the DNN needs the atlas as an input, which undergoes updates throughout the training phase. 
Other group-wise registration strategies have been introduced for atlas construction during test-time, leveraging variational autoencoders~(VAEs) to produce latent vectors from a collection of images.
These vectors facilitate the decoupling of the atlas from the images themselves.
One technique~\citep{siebert2021learning, gupta2021ar}, inspired by deforming autoencoders, uses these latent representations to create the atlas.
Another approach proposed in~\citet{he2023groupwise}
constructs the atlas by averaging these latent vectors arithmetically, then employing a registration network to enhance the atlas.
This is achieved by adjusting it to more closely align with the images in the group throughout the training process. 
However, it is important to recognize that these methods are limited by their reliance on the small deformation assumption, where deformations are modeled as the addition between the identity and a dense displacement field, or by
adopting a time-stationary velocity framework.
Such assumptions can lead to inaccuracies in representing the shape average of the group through intensity averages, which may not align with the geodesic mean in the curved shape space, as highlighted in~\citet{joshi2004unbiased} and~\citet{avants2004geodesic}.}

Advancements in learning-based atlas construction methods have facilitated the fast construction of high-quality atlases. The following subsection explores the application of the atlases and learning-based image registration in achieving the goal of image segmentation.
\subsection{Multi-atlas Segmentation}
Multi-atlas segmentation is a well-established registration-based segmentation technique in existence for several decades~\citep{rohlfing2003ipmi, iglesias2015multi}.
The typical approach involves registering atlas images or their patches to a target image and fusing the propagated atlas labels.
For deformable registration-based multi-atlas methods, the process of pairwise registration between atlas images and the target image can be computationally expensive and time-consuming. 
However, recent advancements in deep learning-based deformable registration algorithms provide a promising solution to address the speed issue and potentially improve the accuracy of registration, which can subsequently improve the accuracy of multi-atlas segmentation.
While many works have explored the use of deep networks to improve the fusion of multiple registered atlas images~\citep{zhu2020fcn, fang2019automatic, xie2019improving, xie2023deep, yang2018neural}, there are relatively few studies that incorporate deep learning-based deformable registration algorithms into their pipeline.

Ding~\emph{et al.}~\citep{ding2019votenet, ding2020votenet+} proposed VoteNet, which predicts a voxel probability of the agreement between registered atlas images and the segmentation target image.
They adopted Quicksilver~\citep{yang2017quicksilver} as their registration algorithm to speed up the pairwise registration process.
Their follow-up work~\citep{ding2021votenet++} experimented with improving the initial registration results from Quicksilver by incorporating a registration refinement step.
The results showed that registration accuracy is a critical factor in achieving accurate multi-atlas segmentation.
In~\citet{ding2022cross}, the authors addressed the challenging problem of cross-modality multi-atlas segmentation. They proposed a deep network that learns the bi-directional registration between atlas images and the target image, as well as a second network that estimates the weights for label fusion.
To account for the modality differences between the atlas and target images, they used Dice loss as a similarity measure to train their registration network and conditional entropy to train the fusion network.
For registering 3D first-trimester ultrasound images,~\citet{bastiaansen2022multi} proposed a two-stage network for learning an affine transformation.
They then applied the VoxelMorph architecture to perform deformable registration on the affinely aligned images. 
The segmentation of the target images was achieved by propagating the labels of the atlas images and combining them using majority voting.

The good performance of supervised training in image segmentation could be a reason for the relative lack of research on deep learning-based registration in multi-atlas segmentation.
Deep neural networks have demonstrated impressive results in supervised image segmentation tasks, making them a popular choice for many researchers. However, the performance of single atlas segmentation is often used to evaluate the accuracy of a registration algorithm, as discussed in Section~\ref{ss:anatomical_info}.
Due to the close relationship between registration and segmentation, there is an increasing interest in exploring the possibility of integrating the learning of segmentation and registration~\citep{sinclair2022atlas, khor2023anatomically, xu2019deepatlas}.
Overall, the use of deep learning-based registration in multi-atlas segmentation is still in its early stages, and there is a significant opportunity for further research.

\subsection{Uncertainty}
Accurate registration is critical for many medical image analysis applications, such as image-guided surgery, radiation therapy, and longitudinal studies. 
However, registration uncertainty can arise due to factors such as training data artifacts or predictive model variances.
To address this issue, incorporating registration uncertainty into medical image analysis can help guide the interpretation of the registration results and improve the reliability of various analysis tasks. 

In clinical decision-making, understanding registration uncertainty is critical for image-guided surgery and radiation therapy. The absence of proper registration uncertainty awareness may lead surgeons to presume a substantial registration error throughout the entire region based on a large error in a single location, resulting in the total disregard of registration. Furthermore, the lack of registration uncertainty may also cause surgeons to place unwarranted confidence in regions with inaccurate registration, resulting in potentially severe consequences. 

For image-guided surgery,~\citet{risholm2013bayesian} showed that the registration uncertainty increased at the site of resection using clinical data from neurosurgery for resection of brain tumors, which demonstrated the potential utility of registration uncertainty in recognizing the surgical regions and guiding surgery.
For radiation therapy,~\citet{risholm2011estimation} had previously presented a probabilistic framework to estimate the accumulated radiation dose and corresponding dose uncertainty delivered to significant anatomical structures during radiation therapy, such as the primary tumor and healthy surrounding organs.
The uncertainty in the estimated dose directly results from registration uncertainty in the deformation used to align daily cone-beam CT images with planning CT. 
The accumulated radiation dose is an important metric to monitor during treatment, potentially requiring treatment plan adaptation to conform to the current patient anatomy. 

A study by~\citet{nenoff2020deformable} employed six different deformable registration algorithms to analyze dose uncertainty in proton therapy and investigate their impact on dose accumulation for non-small cell lung cancer patients with inter-fractional anatomy variations.
The results show that dose degradation caused by anatomical changes was more pronounced than the uncertainty arising from using different deformable image registration algorithms for dose accumulation. 
However, accumulated dose variations between these algorithms can still be substantial, leading to additional dose uncertainty.

In longitudinal medical image analysis, registration is an essential step because it enables the comparison of measurements taken at different time points, which is necessary for correcting anatomical variability and tracking changes over time.
Registration uncertainty estimation can be beneficial for longitudinal image processing tasks, such as image smoothing, segmentation prior propagation, joint label fusion, and others. \citet{simpson2011longitudinal}~proposed an approach to calculate the deformable registration uncertainty using a probabilistic registration framework, integrating the uncertainty into spatially normalized statistics for adaptive image smoothing.
This method showed improved classification results in longitudinal MR brain images acquired from Alzheimer's Disease Neuroimaging Initiative compared to not smoothing or using a straightforward Gaussian filter kernel.

In summary, incorporating registration uncertainty into medical image registration can facilitate interpreting registration results and improve the reliability of various medical image analysis tasks. 
It is crucial for clinicians to understand registration uncertainty and its potential applications in clinical decision-making.
Further research is needed to explore other potential applications of registration uncertainty.

\subsection{Motion Estimation}

In the context of medical images, deep learning-based motion estimation has been closely associated with the unsupervised optical flow~\citep{jonschkowski2020matters, stone2021smurf,bian2022learning} and point tracking~\citep{lai2019self, harley2022particle, ranjan2019competitive, bian2022learning} techniques within the computer vision domain. However, the application of motion estimation in medical imaging presents unique challenges, including limited training data, heterogeneous patient data for testing, and special desired properties on the motion field, such as diffeomorphism~(to preserve anatomical relationships) and incompressibility~(to preserve anatomical integrity).
Deep learning-based registration has demonstrated successful outcomes in estimating motion for various organs, such as the human heart, brain, lungs, and tongue.
Registration-based motion estimation plays a significant role in enabling the assessment of changes in the position, shape, and size of organs over time. 
Multiple dynamic imaging modalities are used for motion estimation in medical imaging, including but not limited to cine images,  tagged-MRI~\citep{axel1989heart, axel1989mr}, and echocardiography.

Cine images are a temporal sequence of MR images captured in quick succession, allowing for the monitoring of organ movement and deformation over time.
Recent research~\citep{qin2018joint, morales2019implementation, meng2022mulvimotion, yu2020foal, qin2023generative, lopez2022warppinn, yu2020motion} has successfully applied deep learning-based registration techniques to cine images.
For example, FOAL~\citep{yu2020foal} proposed online optimization to mitigate distribution mismatch between the training and testing datasets for motion estimation, using meta-learning techniques to enable more efficient online optimization with fewer gradient descent steps and smaller data samples, which differs from instance-specific optimization~\citep{balakrishnan2019voxelmorph}. 
\citet{yu2020motion}~applied similarity and smoothness loss to multiple scales of motion fields (pyramid) using a deep supervision strategy. 
 
The relatively uniform signal within tissues from cine images and the lack of reliable, identifiable landmarks have motivated the exploration of additional regularization methods for estimating motion that is biologically plausible and clinically reliable.
 For example, Qin~\emph{et al.}~\citep{qin2023generative} trained a variational autoencoder-based generative model to capture the prior of biomechanically plausible deformations by reconstructing simulated deformations using finite element models. This prior is then used as regularization during the training of the registration network. 
 Lopez~\emph{et al.}~\citep{lopez2022warppinn} incorporated hyperelastic regularization terms into the framework of physics-informed neural networks~\citep{raissi2019physics} to estimate incompressible motion fields.

Tagged-MRI, on the other hand, employs a spatially modulated periodic pattern to magnetize tissue temporarily, producing transient tags in the image sequence that move with the tissue and capture motion information. It allows for tracking the motion of inner tissue where the region does not have contrast on cine images.
DeepTag~\citep{ye2021deeptag} takes raw 2D tagged images as input and estimates the incremental motion between two consecutive frames using a bi-directional registration network. Then it composes the incremental motion field to estimate motion between any two time frames. Harmonic phase images~\citep{osman1999cardiac} have been found to be more robust to tag fading and imaging artifacts during motion tracking than raw tagged images. DRIMET~\citep{bian2023deep} proposed a simple sinusoidal transformation on the harmonic phase images,  enabling end-to-end training for estimating a 3D dense motion field from tagged-MRI. It also incorporates a Jacobian determinant-based loss that penalizes \textit{symmetrically} for contraction and expansion to estimate a biologically-plaussible incompressible motion field. DRIMET shows promising results in terms of superior registration accuracy, a comparable degree of incompressibility, and faster speed over its traditional iterative-based counterparts~\citep{PVIRA,ilogdemons2011}.
{MomentaMorph~\citep{bian2023momentamorph} proposed a framework of momenta, shooting, and correction to overcome motion estimation issues in presence of large motion for time-series tagged-MRI. MomentaMorph first accumulates momenta in the tangent space, and then uses exponential mapping for shooting momenta to the group of diffeomorphism to avoid local optima, and finally finds the true optima with a correction step. The resutls show that MomentaMorph produces accuarate, dense, and diffeomorphic motion fields in presence of large motion.}

Numerous deep learning-based techniques have been devised to estimate 2D motion, and although this may be adequate for certain applications, tracking dense 3D motion is typically necessary or highly desirable when estimating the motion of biological structures.  
To address this issue, Meng~\emph{et al.}~\citep{meng2022mulvimotion} integrate features extracted from multi-view 2D cine CMR images captured in both short-axis and long-axis planes to learn a 3D motion field of the heart. The edge map of myocardial wall is used as a shape regularization of the estimated motion field. Alternatively, DRIMET~\citep{bian2023deep} uses sparsely acquired tagged images and interpolates them onto an isotropic grid with a resolution based on the in-plane resolution. This approach is based on the observation that the tag pattern changes slowly in the through-plane direction and therefore will not cause aliasing issues during sampling. By doing so, DRIMET is capable of tracking \textit{dense} 3D motion.

Recent studies have shown that joint learning of segmentation and motion estimation can be mutually beneficial~\citep{qin2018joint,ta2020semi,ahn2020unsupervised}. For instance, Qin~\emph{et al.}~\citep{qin2018joint} employ a dual-branch framework consisting of a segmentation branch and a motion estimation branch to simultaneously estimate motion and segmentation from a sequence of cardiac cine images. During training, a \textit{shared} feature encoder is learned under the premise that joint features can complement both tasks. In contrast, Ta~\emph{et al.}~\citep{ta2020semi} and Ahn~\citep{ahn2020unsupervised} adopt a task-level approach to jointly tackle motion estimation and segmentation in the context of estimating cardiac motion from echocardiography. Specifically, they warp the segmentation (of one time frame) using the estimated motion field and regularize the motion field by incorporating shape information obtained from the segmentation. This approach differs from previous studies which couple motion estimation and segmentation at the feature-level, and may offer a novel perspective on joint learning of these tasks.

In addition to MRIs and echocardiography, numerous deep learning-based algorithms have been developed for motion estimation with {4D-CT~\citep{fu2020lungregnet,ho2023unsupervised,wolterink2022implicit,fechter2020one,hering2021cnn,ji2022one,liang2023orrn}}. 4D-CT imaging captures images at different phases of respiratory or cardiac cycles, providing valuable insights for lung imaging applications, including radiation therapy planning and lung function assessment. DIR-LAB~\citep{castillo2009framework} is a widely-used dataset, containing 4D CT images of ten patients, to evaluate 4D-CT registration techniques, with the aim of registering inspiration images to expiration images. This task is challenging due to the superimposed motion of the heart and lungs, which is larger in scale than the small lung structures being studied.

LungRegNet~\citep{fu2020lungregnet} trains two separate networks to handle large lung motion. One network predicts large motion on a coarse scale, and the other network takes the coarsely warped image and fixed image as input to predict fine motion. In addition to similarity and smoothness losses, an adversarial loss is applied as extra regularization to prevent unrealistic deformed images. Hering \emph{et al.}~\citep{hering2021cnn} employs a coarse-to-fine multi-level optimization strategy. The deformations of coarse levels provide an initial guess for subsequent finer levels. Networks are trained progressively, with each handling one level and initialized with parameters from the previous level. It incorporates a penalty for volume change and utilizes an $l2$ loss function to match corresponding keypoints that are automatically detected. Ho \emph{et al.}~\citep{ho2023unsupervised} applied cycle-consistent training~\citep{kuang2019cycle} to reduce foldings using two networks. After the first network's forward pass, the warped and moving images are sent to the second network to predict inverse deformation, with a similarity loss applied to maximize the similarity between the moving images and inversely-deformed moving images. IDIR~\citep{wolterink2022implicit} use a multi-layer perceptron to represent the transformation function of coordinates and demonstrate the ability to incorporate the Jacobian regularizer, hyperelastic regularizer~\citep{burger2013hyperelastic}, and bending energy~\citep{rueckert1999nonrigid} into the framework. The resulting deformation is void of foldings and achieves a mean target registration error (TRE) of 1.07 mm on DIR-LAB datasets. However, this method requires more time compared to CNN-based approaches, prompting researchers to consider acceleration as a potential future direction. 

In order to accurately register previously unseen images outside of training datasets, the application of one-shot learning has been employed for the estimation of lung motion~\citep{fechter2020one,ji2022one}. Fechter \emph{et al.}~\citep{fechter2020one} concatenated images captured at different phases in the channel dimension in order to leverage temporal information. To minimize memory requirements, they partitioned images into non-overlapping patches and applied a boundary smoothness constraint on the transitions between patches. Additionally, they utilized a coarse-to-fine approach by constructing an image pyramid, where the estimated vector fields of finer scales were added to the upsampled vector fields of coarser scales. The proposed method showed a competitive performance without the need for training in advance. 

{Despite advancements, current motion estimation techniques in medical imaging face several key challenges. A primary challenge is accurately modeling large organ movement and deformation, such as lung expansion during respiration, tongue deformation during speech, or cardiac motions during the heartbeat cycle. These deformations are nonlinear and complex, making it difficult for algorithms to capture both coarse and fine-scale motions without introducing artifacts or unrealistic distortions. }

{The second key challenge is the high computational cost associated with capturing dense 3D motion over time, which is essential for a comprehensive understanding of organ dynamics. Transitioning from 2D to 3D significantly increases computational and memory requirements, especially when temporal information (more than two frames) needs to be considered. Incorporating biomechanical constraints like diffeomorphism and incompressibility into models further adds to the computational complexity, as these require sophisticated regularization and may involve iterative optimization processes, limiting their applicability in time-sensitive clinical settings. Approaches like integrating multi-view 2D data to infer 3D motion \citep{meng2022mulvimotion} and incrementally accumulating neighboring motions \citep{bian2023momentamorph} have been explored, but efficient and accurate 3D motion estimation over time remains an open problem. Accelerating these algorithms without sacrificing accuracy is an ongoing research endeavor.}

{The third challenge arises when image intensity or contrast changes over time, which disrupt traditional photometric consistency assumptions. For example, in tagged MRI, the contrast diminishes due to the process reaching steady-state and T1 relaxation, which makes widely-used image similarity losses, such as MSE, NCC, NGF, MI, and MIND, less effective at accurately tracking deforming tissue throughout the entire sequence \citep{bian2024registering}. Similarly, intraoperative imaging faces challenges due to the application of dyes and changes in lighting conditions~\citep{alam2018medical}, and the alignment in imaging-guided radiation therapy is complicated by tissue responses to radiation that can alter image intensity~\citep{jaffray2012image}. Addressing this challenge may involve a better understanding of the underlying physics to model and compensate for intensity variations. Additionally, borrowing experiences from inter-modality image similarity learning---which are robust to contrast differences---could provide valuable insights.}

\subsection{2D-3D Registration}
Recent progress in the field of interventional procedures for invasive treatment protocols has been associated with high precision in surgeries performed at a reasonable cost~\citep{pfandler2019technical, dlouhy2014surgical}. In these procedures, 2D-3D registration plays a significant role in determining the spatial relationship between the 3D anatomical structures and 2D images, such as X-Ray fluoroscopic images, ultrasound image frames, or endoscopic images. 2D-3D medical image registration primarily involves registering 2D interventional images to 3D pre-operative CT/MR images, \emph{i.e.}, to obtain the 3D geometric transformation that aligns with the 2D view available. 
Conventional 2D-3D registration methods involve iterative optimization methods with similarity metrics \citep{maes1997multimodality}~based on image intensity as the objective function. 
Due to the sparsity of spatial information derived from 2D images, the problem is non-convex, which may lead to convergence at a local minimum if the initial estimate is not sufficiently close to the correct one. 2D-3D registration is a problem with a minimum of six degrees of freedom which may also lead to registration ambiguity as the spatial information along each projection line is compressed to a single point in the 2D plane. This high-dimensional optimization problem increases the difficulty of determining the parameters associated with the depth of anatomical features in the 3D volume. 
Alternatively, deep-learning-based methods have gained popularity for this application as they do not require explicit functional mappings~\citep{unberath2021impact}.
In this discussion, we briefly highlight recent advancements in 2D-3D registration, while directing interested readers to \citep{unberath2021impact} for a comprehensive review of the influence of various learning-based methods in this area.

Common 2D-3D registration applications and examples include registration of 2D fluoroscopic/angiography images to 3D CT/MR images of pelvic, lung, or brain {region~\citep{Gu2020ExtendedCR, liao2019multiview, gao2020generalizing, gao2020fiducial, jaganathan2023self, huang2022novel, shrestha2023x}}, registering endoscopy images to CT/MR images~\citep{liu2020extremely, bobrow2022colonoscopy}, {registering histopathological image to CT~\citep{leroy2023structuregnet},} and registering 2D ultrasound (US) frames to 3D MR images to facilitate interventional procedures, such as liver tumor ablation{~\citep{Wei2021a, wang2023multimodal}} or prostate cancer biopsy~\citep{guo2022ultrasound}. 

In~\citep{Gu2020ExtendedCR, Wei2021a, huang2022novel}, the 2D-3D registration problem was modeled as a regression learning problem where the network is trained to directly predict the desired geometric parameters. These models are trained by completely relying on the data, \emph{i.e.}, it has little to no tie to the actual image formation physics involved. Specifically, in \citep{Gu2020ExtendedCR} a 2D X-Ray image is registered to a 3D CT volume using a ConvNet, which takes the X-Ray image and a digitally reconstructed radiograph (DRR) from the CT volume at some known pose as input. 
The ConvNet regresses a geodesic loss function over the geometric parameter space to estimate the relative pose between the fixed X-ray image and the DRR from the CT volume without the need for accurate pose initialization. {A similar approach is employed in~\citet{zhang2023patient}, which enhances the model by simulating patient-specific X-ray images from CT scans for self-supervised learning of X-ray to CT registration.
In contrast, \citet{aubert2022x} adopts a GAN-based image-to-image translation method to synthesize X-ray images into DRRs for the registration of spine X-ray images to 3D vertebra models.
The registration employs a non-DL-based regression technique to estimate the rigid transformation of the 3D model that produces a DRR most closely resembling the synthesized DRR, as determined by a specific similarity measure.}

In~\citep{Wei2021a}, Wei~\emph{et al.} propose a two-step registration process to determine the position and orientation of the ultrasound plane in the 3D MR volume data. In the first step, a ResNet-18 network is employed to determine the US probe orientation. Following this, a U-Net is used to regress a weighted dice loss function, which facilitates the determination of the orientation and position of the corresponding XY plane in the resampled 3D MR volume associated with the US frame.
In \citep{huang2022novel}, Huang~\emph{et al.} also implemented a two-step registration process for aligning 3D MR vessel wall images (VWI) with 2D Digital Subtraction Angiography (DSA) images. This approach encompasses a ConvNet regressor~\citep{miao2016cnn} that estimates the initial pose, followed by an instance-based centroid alignment, which serves to further minimize parameter estimation errors between the images. {Adopting a different two-step method, \citet{haouchine2023learning} addresses the registration of preoperative MRI scans to intraoperative RGB images by initially simulating the expected appearances of intraoperative images from various 3D poses of the MRI scan using an image synthesis network. This is followed by employing a regression network to determine the optimal 3D pose that best aligns the simulated 2D image with the intraoperative image.} {In addition to these two-step methods,~\citet{leroy2023structuregnet} proposed StructuRegNet for multi-model 2D-3D registration, which used generative model for style transfer between different imaging modalities prior to registration, and used deformable registration after cascaded plane selection for refining the final alignment, especially the out-of-plane distortions.}

{In~\citet{shrestha2023x}, Shrestha et al. proposed to regress the scene coordinate with respect to the CT volume directly from the X-ray image using a fully convolutional neural network. The authors applied the camera intrinsic matrix and extrinsic matrix to calculate the back-projection from a pixel on the X-ray to the CT volume, and used that as gound truth in the loss funciton for training the network. The results show that the proposed method performs well under partially visible structure and extreme view points.}

As an alternative to formulating registration as a regression problem that necessitates ground truth transformation parameters, several recent studies~\citep{liao2019multiview, gao2020generalizing, gao2020fiducial, jaganathan2023self, guo2022ultrasound} have explored framing it as an unsupervised optimization problem. In such a formulation, the cost function is determined by a similarity metric measured between the transformed and fixed images. Liao~\emph{et al.}~\citep{liao2019multiview} trained a network to track a set of points of interest (POIs) derived from the 3D CT volume in the 2D DRR and in the multi-view fluoroscopic 2D images (used as fixed images), enabling the network to learn the spatial correspondences between the POIs. In this method, a Siamese U-Net architecture is employed to extract features from the DRRs and fixed images, subsequently tracking the POIs within the extracted features. A triangulation layer is incorporated to pinpoint the locations of the tracked POIs within the fixed image in 3D space. Finally, the geometric transformation between the estimated locations of POIs derived from the fixed image and their true positions is determined analytically. In \citep{gao2020generalizing}, Gao~\emph{et al.} proposed a novel differential volume rendering transformer network combined with a feature extraction encoder to approximate the image similarity metric in a manner that renders the geometric parameter estimation as a convex problem with respect to the pose parameters. 

The examples and applications discussed thus far have primarily focused on rigid 2D-3D registration. However, non-rigid 2D-3D registration is essential in certain applications, such as cephalometry~\citep{Li2020Nonrigid} and lung tumor tracking in radiation therapy~\citep{foote2019real,dong20232d}. Cephalometry, for instance, involves formulating the problem as deformed 2D-3D registration with the objective of generating a 3D volumetric image from a 2D X-ray image using a 3D skull atlas. Li~\emph{et al.}~\citep{Li2020Nonrigid} developed a convolutional encoder that uniquely codes the cephalogram image into a volumetric image. The network is trained by minimizing the NCC between the synthesized DRR originating from the volumetric image and the 2D cephalogram. 

Numerous deep learning-based models and metrics have been developed to improve the performance of 2D-3D registration in specific applications, although these methods are specialized and not as versatile as traditional optimization methods. Nonetheless, Machine Learning/Deep Learning has been instrumental in tackling the persistent challenges associated with algorithmic approaches. These techniques have tackled a narrow optimal range of parameters, while also decreasing registration ambiguity. CNN-based approaches are also comparably fast. These factors encourage users to further improve learning-based 2D-3D registration pipeline.

%% file: future_perspective.tex
\begin{figure}[t]
\centering
\includegraphics[width=0.48\textwidth]{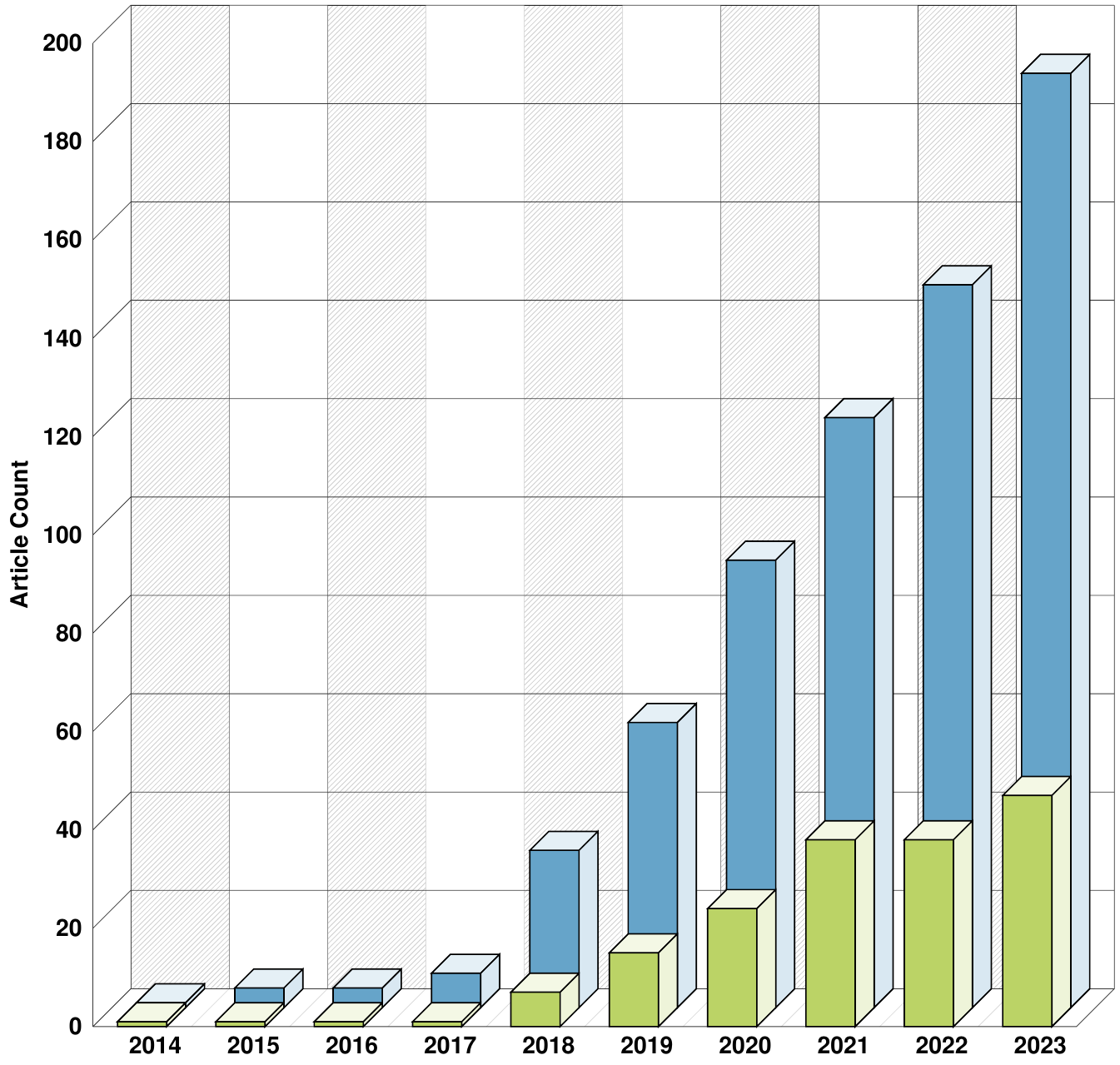}
\caption{{The figure shows the paper count statistics from PubMed for the years 2014 to 2023. The blue bars represent the count of papers that include keywords related to learning-based image registration in their title or abstract, while the green bars display the count of papers specifically related to unsupervised learning-based image registration.}} \label{fig:paper_stats}
\end{figure}

Over the past decade, learning-based registration models have been attracting increasing research interest. As illustrated in the left panel of Fig. \ref{fig:paper_stats}, there has been a growing trend in developing and applying these models since 2013. Unlike other medical image analysis tasks, such as segmentation or classification, which typically necessitate labor-intensive and time-consuming manual annotations to develop high-performing models, registration is inherently a self-supervised task. Traditional registration models have predominantly been unsupervised, requiring only moving and fixed images to execute registration. While traditional registration models are typically unsupervised, learning-based registration models initially began as a supervised process, generating ground truth deformation fields using traditional registration methods. However, these supervised models often could not surpass the performance of traditional methods. Instead, they often served as an intermediate step to expedite conventional approaches like geodetic shooting~\citep{shen2019region}, FLASH~\citep{wang2020deepflash}, etc. Despite traditional methods providing appealing deformation properties such as time-dependent diffeomorphic transformations, researchers have recently begun exploring the unsupervised nature of learning-based registration (as seen in the left panel of Fig. \ref{fig:paper_stats}). By training DNNs using loss functions adapted from traditional methods' objective functions, these methods aim to improve both registration accuracy and speed. Incorporating segmentation and landmark correspondences during training can further enhance registration accuracy, providing capabilities not achievable with traditional methods. {Moreover, with a powerful pretrained feature extractor such as self-supervised anatomical embedding~\citep{yan2022sam}, registration performance can be further improved~\citep{liu2021same, li2023samconvex}.} Given the rapid progress of deep learning and its growing adoption in medical applications, we anticipate an increasing focus on learning-based medical image registration.

In this section, we provide future perspectives and discuss potential avenues for advancing learning-based medical image registration. Our discussion will focus on the development of registration models, assessment of registration uncertainty, and prospective applications.

\subsection{Deep Learning-based Registration Models}
\subsubsection{Network Architecture}
Network architectures employed in image registration occasionally draw inspiration from other image analysis tasks, such as segmentation. For instance, VoxelMorph~\citep{balakrishnan2019voxelmorph}, CycleMorph~\citep{kim2021cyclemorph}, SYMNet~\citep{mok2020fast}, and DiffuseMorph~\citep{kim2022diffusemorph} all borrow U-Net-like architectures, originally developed for image segmentation. In such cases, they often generate deformation fields at a single resolution. In contrast, traditional registration algorithms have demonstrated the benefits of adopting a multi-resolution registration strategy, which decomposes deformations across multiple scales. This method not only improves registration performance but also imparts beneficial deformation properties, such as the ability to enforce larger deformations. This aspect can be particularly beneficial for lung or abdominal organ registration, where organ displacement between scans can be significant. As discussed in section \ref{sec:multi_res_reg}, there has been a growing interest in integrating multi-resolution strategies into network architecture, and these methods have consistently demonstrated notable performance improvements compared to using a single resolution alone. It is worth noting that this finding has parallels with observations in other image analysis fields, where adopting deep supervision can significantly boost performance~\citep{zhou2019unet++, isensee2021nnu}. 

The task of image registration is fundamentally distinct from other image processing tasks, as its primary objective is to capture the correspondences between images, with less emphasis on semantic information.
In this context, SynthMorph~\citep{hoffmann2021synthmorph} demonstrated that training an effective medical image registration network does not necessarily require actual medical images. Instead, random shapes or synthetic images can also be used as training datasets for registration networks.
{Similarly,~\citet{bigalke2023unsupervised} introduced a teacher-student model for unsupervised deformable registration. They also found that at the initial training stage, matching random features can produce reasonable deformation.}
Given the objective of establishing correspondences, network architectures that incorporate inductive biases focused on this capability may offer significant advantages. Architectures such as Transformers (particularly cross-attention Transformers), contrastive learning, Siamese networks, and correlation layers, which leverage comparisons between moving and fixed images, are of special interest for image registration. We expect to see an increasing number of studies incorporating these designs in the future, along with other advancements in deep learning applied to image registration.

{Optical flow estimation, a field closely related to image registration, shares significant conceptual and methodological overlap that has greatly influenced the development of medical image registration methods. Notable examples include the classic Demons algorithm~\citep{thirion1998image}, where the displacement-driven force} {is derived from an optical flow equation inspired by~\citep{horn1981determining, aggarwal1988computation}. Another example is the transformation of the variational optical flow energy function into a convex optimization problem through quadratic relaxation~\citep{steinbrucker2009large}, which has inspired various medical image registration methods~\citep{heinrich2013mrf,ha2018model} and even recent learning-based approaches~\citep{jia2021learning, li2023samconvex, siebert2021fast}. The development of FlowNet~\citep{dosovitskiy2015flownet,ilg2017flownet} for learning-based optical flow estimation, which integrates a correlation layer for explicit feature matching, has also influenced various learning-based medical image registration methods~\citep{heinrich2019closing, siebert2022learn, siebert2021fast}. Moreover, many motion estimation techniques designed for dynamic medical imaging draw from by optical flow methods, as discussed in Section 7.4. As the field of optical flow estimation continues to rapidly evolve, with its distinct perspectives on handling challenges such as large displacements~\citep{weinzaepfel2013deepflow, ilg2017flownet} and pretraining strategies~\citep{shi2023flowformer++}, it is poised to further inspire innovative learning-based registration methods for medical imaging. For readers interested in this intersection, we recommend a recent review article on optical flow by~\citet{zhai2021optical} for further reading.}


\subsubsection{Loss Function}
In unsupervised models, the image similarity measures predominantly used for mono-modal registration are MSE and NCC, as shown in Table \ref{table:unsup_list}. NCC is generally considered a better choice than MSE, as it is locally adaptive and less sensitive to local intensity variations~\citep{avants2008symmetric}. For multi-modal registration, MI has historically been the preferred choice~\citep{maes1997multimodality, wells1996multi}. However, to auto-differentiate MI using modern deep learning frameworks for end-to-end training, joint and marginal probabilities are often approximated using the Parzen window. A notable drawback of this approximation is the increased computational burden. In actual implementation, each voxel location expands to include a vector, with the elements in the vector representing the probability of the voxel belonging to each intensity bin. Increasing the number of intensity bins effectively results in an increased channel dimension, ultimately leading to a higher computational burden. Conversely, using a small number of bins often limits the registration performance. Recent learning-based methods have explored surrogates to tackle multi-modal registration problems. For instance, given the advantage of learning, anatomical loss functions like Dice can serve as a modality-independent loss function for training the registration network~\citep{hoffmann2021synthmorph}. The trained network can then be applied to images without requiring anatomical segmentation, offering an advantage that traditional methods cannot provide. Multi-modal registration can also be addressed using advanced learning methods, such as contrastive learning and adversarial learning, as discussed in sections \ref{sec:contrast_learn} and \ref{sec:adv_learn}. These methods guide the neural network in understanding similarities and dissimilarities between images across different modalities using paired data without requiring explicit multi-modal similarity measures. We expect future research to continue to develop more efficient and innovative approaches to tackle multi-modal registration.

Regarding the use of regularization in learning-based deformable registration, there is currently an inadequate emphasis on the development and application of spatially-varying regularization. Despite being a significant area of research historically~\citep{schnabel2016advances, schmah2013left,vialard2014spatially,stefanescu2004grid,tang2010reliability,pitiot2008geometrical,gerig2014spatially,simpson2015probabilistic,pace2013locally,myronenko2010intensity,papiez2014implicit,fu2018adaptive,risser2013piecewise,papiez2015liver}, spatially-varying regularization has been largely overshadowed by the rise of learning-based registration, with only a few studies addressing it within a deep learning framework framework~\citep{niethammer2019metric,shen2019region,chen2023spr,chen2021deep2}. As illustrated in Table \ref{table:unsup_list}, most methods opt for a simple spatially-invariant regularization, predominantly employing the diffusion regularizer. However, as outlined in section \ref{sec:def_reg}, spatially-varying regularization provides the advantage of accommodating spatially-varying deformations, preserving discontinuities, and facilitating sliding motion, all of which are essential for a variety of applications. Advancements in modeling spatially-varying regularization within or through deep learning frameworks are eagerly anticipated in the future.

\subsection{Registration Uncertainty}
Registration uncertainty in medical image analysis is an ongoing challenge and opportunity. On the one hand, advancements in deep learning have the potential to improve registration accuracy and reduce registration uncertainty by extracting features that are robust to noise and other artifacts. On the other hand, uncertainty can be estimated for use in interpreting the registration results and providing valuable information for clinical decision-making.

However, there are several limitations that restrict the further usage of uncertainty estimation in various applications. One significant limitation is the lack of ground truth for evaluating the quality of uncertainty estimation. Without ground truth, it is challenging to validate the accuracy of uncertainty estimation directly. Instead, most existing evaluation methods rely on indirect proofs such as sparsification analysis. This not only affects the reliability of uncertainty estimation, but also limits further developments for better uncertainty estimations. Another limitation is the computational complexity of estimating uncertainty, which can be time-consuming and may limit its usage in real-time clinical applications. Additionally, interpreting uncertainty estimates can be challenging for clinicians, as some statistical measures are not always straightforward. This can limit the adoption of uncertainty estimation in clinical decision-making, where clear and concise information is essential for making informed decisions.

To overcome these limitations, it may be helpful to develop improved evaluation methods that rely on direct validation rather than indirect proofs, such as the creation of synthetic data or the use of simulation frameworks where the ground truth is known. This could enhance the accuracy and reliability of uncertainty estimation. Moreover, new computational techniques and algorithms such as incorporating Markov Chain Monte Carlo into a multilevel framework \citep{schultz2018multilevel} and quantifying image registration uncertainty based on a low dimensional representation of geometric deformations \citep{wang2019registration} can reduce the computational workload. Last, efforts should be made to provide clinicians with more accessible and intuitive ways to interpret uncertainty estimates, such as visual aids or simpler statistic measures.

In addition to the limitations of uncertainty estimation in medical image analysis, there are also many potential applications of registration uncertainty that remain unexplored. One promising area of application is atlas-based segmentation, where registration is often used to align an atlas image to a target image for the purpose of segmenting anatomical structures. In this context, registration uncertainty can be used as a criterion for generating a soft segmentation mask, where the probability of each voxel belonging to a particular anatomical structure is weighted by the uncertainty estimate. Another potential application of registration uncertainty is multi-atlas-based segmentation, where multiple atlases are registered to a target image and combined to produce a final segmentation result. In this context, registration uncertainty can be used to weight different segmentation results, producing a more reliable segmentation. This approach could be particularly useful in cases where some atlases are more appropriate for a particular image than others. 

Overall, these limitations and potential applications of registration uncertainty in medical image analysis offer an exciting range of opportunities for future research, and it is likely that continued progress in this area will have a substantial impact on the field of medical image analysis and beyond.

\subsection{Towards Zero-shot Registration}
Classical registration algorithms, while potentially slower, are usually available for immediate use and provide end-users with the flexibility to choose the similarity measure and weighting of regularization terms that best meet their needs. 
In contrast, deep learning algorithms are susceptible to the domain shift problem, which arises when a trained network struggles to perform well when presented with input images from a different distribution than the training data.
Several sources of domain shift can arise in learning-based registration algorithms, such as changes in the input image modality, different populations of subjects, or variations in the direction of registration.
To address this challenge, researchers have explored several methods to improve the generalizability of registration networks.
For instance, SynthMorph~\citep{hoffmann2021synthmorph} uses synthetic images to force the network to learn contrast-invariant features, while HyperMorph~\citep{hoopes2021hypermorph} uses a hypernetwork to enable the adjustment of the regularization term during test time.
{Although these approaches have shown promising results, they have not yet been widely adopted, and further studies and validations are necessary to verify their generalizability and practical usability, particularly with clinical-quality data.}

Recent developments in zero-shot learning offer a promising avenue for further improving the generalizability of learning-based registration algorithms. In particular, Foundation models that are pretrained on a broad range of data have shown competitive or even superior zero-shot performance compared to prior supervised models in various tasks~\citep{kirillov2023segment, brown2020language}, without requiring specific training data for each new task. Leveraging these techniques can potentially reduce the time and resource requirements for developing deep learning registration algorithms in clinical pipelines, making the existing registration algorithms more accessible and useful to a wider range of users.

\subsection{Metamorphic Image Registration}
As outlined in Section \ref{sec:diff_reg}, diffeomorphic registration is a bijective mapping that preserves topology. In clinical scenarios, however, registration often involves the deformation of a healthy control or an atlas to fit patient images that may contain tumors or other anomalies. For example, longitudinal scans of the same patient with a tumor may need to be mapped to one another to facilitate the study of the tumor progression or response. Such situations violate the one-to-one mapping assumption of diffeomorphisms due to topological changes between scans. To address this challenge, alternative registration methods such as metamorphic registration models~\citep{brett2001spatial, sdika2009nonrigid, niethammer2011geometric, hong2012metamorphic, franccois2022weighted} have been proposed, which can accommodate changes in topology and appearance. For a mathematical definition of metamorphosis, readers can refer to~\citep{trouve2005metamorphoses, younes2010shapes}. However, these methods often require manual segmentation of the anomalies and are optimization-based, which can be time-consuming and computationally expensive, thereby limiting their practical adoption.

Recently proposed learning-based metamorphic registration methods~{\citep{wang2023metamorph,liu2023co,han2020deep,bone2020learning, maillard2022deep}} have been built upon a metamorphic framework~\citep{trouve2005metamorphoses, younes2010shapes}, which adds time-varying intensity variations on top of the diffeomorphic flow, thereby enabling topological changes over time. The registration networks learn to disentangle geometric and appearance changes and sometimes leverage available segmentation to constrain changes within a desired location (e.g., a tumor). With the success of learning-based image segmentation, the time-consuming manual segmentation previously required by classical methods to guide the metamorphosis can now be addressed using segmentation networks. Several recent approaches take advantage of segmentation networks through joint training or by integrating segmentation capabilities directly into the registration network{~\citep{han2020deep,liu2023co}}. Alternatively, some studies learn to disentangle appearance and shape changes directly from data without requiring to explicitly define a region~\citep{bone2020learning, maillard2022deep}. Despite these recent advancements, learning-based metamorphic registration is still in its infancy. Successfully capturing topological changes still greatly depends on the accuracy of the segmentation network. Meanwhile, the effective modeling of time-varying diffeomorphic flow using DNNs continues to be an area of ongoing research. Considering the practical potential of metamorphic registration, metamorphic registration represents an appealing direction for future investigation in learning-based registration research.

\subsubsection{Spatial-temporal Image Registration}
Learning-based image registration methods have primarily been centered around aligning just one pair of images. Yet, there is a crucial but underexplored need in medical imaging applications: tracking tissue motion across multiple frames. This is particularly relevant in modalities such as tagged/cine MRIs, 4D-CT, and echocardiography.

Addressing the challenge of motion tracking involves overcoming several challenges and taking into account key questions that require careful consideration. 
For instance, how can one ensure the preservation of desired properties, such as smoothness, diffeomorphism, and incompressibility, throughout temporally long-range tracking? Achieving accurate 4D tracking while maintaining a reasonable computational burden in terms of both temporal and spatial complexity poses another challenge. Moreover, how can varying input frame lengths be effectively managed? These questions demand further exploration and investigation in order to advance our understanding and capability in motion tracking. Encouragingly, recent advancements in computer vision, particularly in the context of natural video, have demonstrated promising results. Methods such as correspondence learning~\citep{jabri2020space, bian2022learning, zhang2023boosting, araslanov2021dense} and spatial-temporal representation learning~\citep{kim2019self, yao2020video, wang2019self, qian2021spatiotemporal, dave2022tclr} may offer valuable insights to tackle these challenges we face. By building upon these recent advancements, we may pave the way for more effective and efficient motion-tracking methods for medical imaging applications.

\subsubsection{Spatial Normalization}
Spatial normalization is the process of aligning medical images to a common atlas to reduce anatomical variations~\citep{ashburner1999nonlinear, friston1995spatial}. This process facilitates voxel-based analysis, allowing for comparisons between individual patients as well as between a patient and a larger population. Additionally, it is useful for mitigating anatomical differences that arise due to factors like motion across various image modalities from the same patient. However, a primary challenge of this approach is achieving accurate registration, often hindered by significant anatomical differences among patients. Moreover, traditionally constructed atlases tend to be of subpar quality, primarily because conventional atlas construction methods usually involve averaging, leading to significant blurring of anatomical features. However, as discussed in section \ref{sec:atlas}, deep learning is currently playing a transformative role. It is not only accelerating the atlas construction process but also significantly enhancing the quality of the atlases in aspects like contrast and sharpness. Historically, atlases have been largely used in brain research owing to the comparatively smaller anatomical differences among patients. But with the advancements introduced by deep learning, combined with learning-based registration models, there is potential to broaden their application beyond just the brain to include other body parts. Such an expansion can be invaluable in various medical imaging applications, from cancer treatment planning and monitoring tumor progression or therapy response, to the creation of patient-specific digital twins.